\documentclass[aps,prb,showpacs,amsmath,amssymb,aps,preprintnumbers,twocolumn,superscriptaddress,nofootinbib]{revtex4-2}
\usepackage{amsmath,amssymb}
\usepackage{bm}
\usepackage{tipa}
\usepackage{upgreek}
\usepackage{comment}
\usepackage{mathrsfs}
\usepackage{graphicx}
\usepackage{braket}
\usepackage{enumitem}
\usepackage{mathbbol}
\usepackage{booktabs}
\usepackage{gensymb}
\usepackage{subcaption}
\usepackage[normalem]{ulem}
\usepackage{color}
\usepackage{xcolor}
\usepackage[colorlinks,bookmarks=true,citecolor=blue,linkcolor=red,urlcolor=blue]{hyperref}
\renewcommand{\vec}[1]{\mathbf{#1}}
\usepackage{pifont}

\newcommand{\bk}{\bm{k}}
\newcommand{\bG}{\bm{G}}

\newcommand{\bp}{\bm{p}}

\newcommand{\ha}{\hat{a}}
\newcommand{\hb}{\hat{b}}
\newcommand{\hba}{\hat{\bm{a}}}

\newcommand{\halpha}{\hat{\alpha}}
\newcommand{\hbalpha}{\hat{\bm{\alpha}}}
\newcommand{\norm}[1]{\left\lVert#1\right\rVert}

\renewcommand{\comment}[1]{}

\begin{document}

\title{Putting a new spin on the incommensurate Kekul\'{e} spiral: from  spin-valley locking and  collective modes to fermiology and implications for superconductivity}

 \author{Ziwei Wang}
	\affiliation{Rudolf Peierls Centre for Theoretical Physics, Parks Road, Oxford, OX1 3PU, UK}

 	\author{Glenn Wagner}
	\affiliation{Institute for Theoretical Physics, ETH Zürich, 8093 Zürich, Switzerland}
 
 \author{Yves H. Kwan}
	\affiliation{Department of Physics, University of Texas at Dallas, Richardson, Texas 75080, USA.}
    \affiliation{Princeton Center for Theoretical Science, Princeton University, Princeton NJ 08544, USA}

\author{Nick Bultinck}
	\affiliation{Department of Physics, Ghent University, Krijgslaan 281, 9000 Gent, Belgium} 

	\author{Steven H. Simon}
	\affiliation{Rudolf Peierls Centre for Theoretical Physics, Parks Road, Oxford, OX1 3PU, UK}
	\author{S.A. Parameswaran}
	\affiliation{Rudolf Peierls Centre for Theoretical Physics, Parks Road, Oxford, OX1 3PU, UK}

\begin{abstract}
We revisit the global phase diagram of magic-angle twisted bilayer and [symmetric] trilayer graphene (MA-TBG/TSTG) in light of recent scanning tunneling microscopy (STM)  measurements on these materials. These experiments both confirmed the importance of strain in stabilizing the predicted incommensurate Kekul\'{e} spiral (IKS) order near filling $|\nu|=2$ of the weakly dispersive central bands in both systems, and suggested a key role for electron-phonon couplings and short-range Coulomb interactions in selecting between various competing orders at low strain in MA-TBG. Here, we show that such interactions {\it also} play a crucial role in selecting the spin structure of the strain-stabilized IKS state. This in turn influences the visibility of the IKS order in STM in a manner that allows us to infer their relative importance. We use this insight in conjunction with various other pieces of experimental data to build a more complete picture of the phase diagram, focusing on the spectrum of low-lying collective modes and the nature of the doped Fermi surfaces. We explore the broad phenomenological implications of these results for superconductivity.

\end{abstract}

\maketitle

\section{Introduction}
\label{sec:introduction}

Since the discovery of correlated insulating behaviour and proximate gate-tunable superconductivity in magic-angle twisted bilayer graphene (MA-TBG)~\cite{Cao:2018aa,Cao2018,Yankowitz2019,Lu2019ins}, understanding both phenomena and their relationship to each other has been a central objective of theoretical and experimental investigations of graphene-based moir\'e materials. Although many aspects of the superconducting state(s) remain debated, including fundamental questions about its gap structure and the underlying pairing mechanism, a measure of consensus has emerged as to the nature of the correlated insulators, especially near the most prominent insulators at a filling of $\nu=\pm2$ electrons per moir\'e unit cell. Pioneering scanning tunneling microscopy (STM) experiments~\cite{Nuckolls_2023} near this filling in MA-TBG reveal the presence of a Kekul\'e charge order on the microscopic graphene scale modulated at a single incommensurate wavevector on the longer moir\'e scale: a hallmark of the ``incommensurate Kekul\'e spiral'' (IKS) order first proposed theoretically in Ref.~\cite{Kwan2021}. The IKS state has since also been observed via similar STM studies~\cite{Kim_2023} in the closely-related setting of twisted symmetric trilayer graphene, again in good agreement with subsequent theoretical modeling~\cite{Wang2024}.

This progress  has reinforced the view  that adding interactions to the celebrated Bistritzer-MacDonald (BM) model of MA-TBG --- that extracts a description of electrons in the eight flat bands near charge neutrality by adding interlayer tunneling to a continuum description of the individual graphene layers ---  only gives a partial picture of the TBG phase diagram.  Most notably, this approach fails to explain why some experiments see an insulator \cite{Lu2019ins,Pierce:2021aa,Polshyn:2019aa,Serlin2020ins,sharpe2019ins,Wu:2021aa,Stepanov_2020,Stefanov2021} while others observe a semimetal \cite{Cao2018,Cao:2018aa,Cao2021a,Das:2021aa,Liu2021i,Park:2021aa,Rozen:2021aa,Saito_2020,Saito:2021aa,Uri:2020aa,Yankowitz2019,Zondiner:2020aa}  at charge neutrality ($\nu=0)$.  Furthermore, the insulating states predicted at $\nu=\pm2$  by the interacting BM model and its various solvable limits are characterized by a symmetry that forbids~\cite{Calugaru2022,JungPyo}  the Kekulé pattern detected at these fillings in STM on both twisted bilayer \cite{Nuckolls_2023} and trilayer \cite{Kim_2023} graphene. However, both features {\it can} be explained by incorporating  strain \cite{Parker2021,Kwan2021}, known to be present experimentally \cite{Kerelsky2019,Choi2021,Xie2019,Wong2020}, into the interacting BM model. Specifically, the presence of  heterostrain  --- a relative strain between the graphene layers --- strongly modifies the bandstructure and stabilizes the IKS phase at all nonzero integer fillings in conjunction with the emergence of a semimetal at $\nu=0$~\cite{Kwan2021}. 

\begin{figure}[t]
    \centering
    \includegraphics[width=\linewidth]{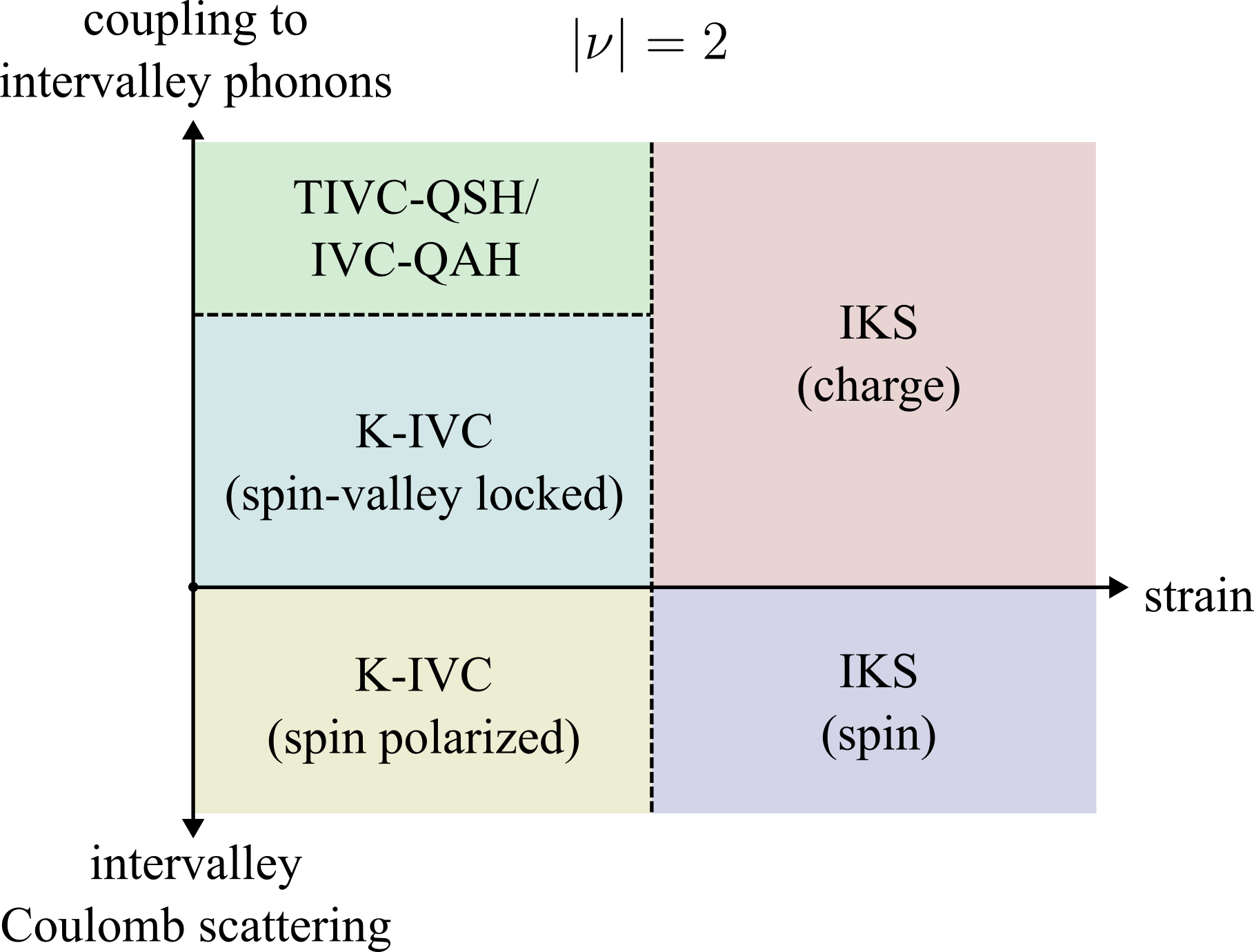}
    \caption{Schematic ground-state phase diagram of realistic TBG at $|\nu| = 2$. The TIVC-QSH/IVC-QAH state at zero to small strain was  identified in Ref.~\cite{Kwan2024}; the IKS states and their spin structure at moderate strain is the subject of this work. The distinction between the IKS (charge) and IKS (spin) states is explained in Sec.~\ref{subsec:nu2}.}
    \label{fig:phase_diagram}
\end{figure}

Further analysis of experiments suggests that the BM model has additional shortcomings, beyond its neglect of strain. First, STM data shows that a Kekulé charge pattern persists, although without a detectable moir\'e-scale modulation, at $|\nu|=2$ in ultralow-strain samples~\cite{Nuckolls_2023}, again in contradiction to predictions for the interacting BM model~\cite{Bultinck2020,Lian2021}. However, this observation can be reconciled with theory if the latter also incorporates the coupling of electrons to (a certain subset of) phonon modes that modify the competition between various candidate insulating states~\cite{Blason2022, Kwan2024,Shi_2025}. Specifically, Ref.~\cite{Kwan2024} found that electron-phonon coupling stabilizes a kind of intervalley-coherent state that preserves time-reversal symmetry and exhibits a quantum spin Hall effect, dubbed T-IVC-QSH. The state is energetically degenerate at mean-field level with a time-reversal-symmetry-breaking state that exhibits a quantized anomalous Hall effect, dubbed IVC-QAH. Crucially, both states are predicted to exhibit Kekul\'e charge density patterns, consistent with the aforementioned experiment, but can be distinguished by transport measurements. Second, the standard approach of incorporating long-range interactions often neglects intervalley Coulomb scattering, which is crucial in breaking degeneracies linked to the spin and valley structure of various correlated states.

Motivated by these discrepancies between experiments and the simplest theoretical models, and spurred by the insights offered by the observation of  the IKS state and the persistence of Kekul\'e charge order to low strain, in this work we construct a more complete picture of the Kekul\'e spiral order, and hence of the phase diagram of magic-angle moir\'e graphene. To do so, we incorporate the various modifications of the BM model mentioned above and use the resulting augmented microscopic model to directly study the competition between various broken symmetry orders both at integer and noninteger filling as well as the corresponding spectrum of collective modes. Our primary tool is numerical: we use   Hartree-Fock mean field simulations, both time-independent and (for analysing the stability of ground states and probing collective modes) time-dependent. We explain the rationale behind this choice below, but we remark here that it is a proven approach for integer $\nu$ (where in certain special limits it matches rigorous results) and remains one of the only techniques readily applicable  away from integer filling. We also  use  existing experimental results to inform our choice of parameters: for instance, as we will show below, the observability of the IKS in STM provides important clues to the spin structure, and thereby constrains the relative strength of a subset of the beyond-BM couplings.

Before proceeding, we place our work in context of previous studies. While the effect of strain has previously been studied in some detail both at integer and noninteger fillings~\cite{Parker2021,Kwan2021,Wagner2022,herzogarbeitman2025kekulespiralorderstrained}, we focus here on the less well-explored effects of the electron-phonon coupling  and related modifications to the BM model. The effect of the electron-phonon coupling at integer filling was studied previously in Ref.~\cite{Kwan2024}, which focused primarily on the zero strain case. To complement this analysis, in the present work we focus on the role of electron-phonon coupling  and intervalley Coulomb interactions in the experimentally more prevalent regime of substantial strain, sufficient to stabilize the $\nu=0$ semimetal and the $\nu\neq 0$ IKS state. [In an ongoing work~\cite{ongoing}, we explore how to model realistic particle-hole asymmetry, which has only received limited attention in many previous studies.] We therefore present a comprehensive  study of {\it realistic} MA-TBG at both integer and noninteger fillings, and explore the corresponding collective excitations and Fermi surfaces (in the latter case). Finally, we  examine, in a phenomenological manner, how IKS states may serve as parent states for superconductivity in the presence of these realistic perturbations.

Our main results may be summarized as follows. First, we explore the degenerate  manifolds of IKS ground states of the strained BM model at $|\nu| = 2, 3$, which are distinguished by their spin and valley structure. By incorporating perturbations that lift the degeneracy, we determine the specific  states selected from this manifold. Combining these results with the zero-strain phase diagram found in Ref.~\cite{Kwan2024}, we arrive at a schematic phase diagram of realistic TBG at $|\nu| = 2$ (Fig.~\ref{fig:phase_diagram}). Notably, we find that a specific IKS state that preserves the physical time-reversal symmetry and exhibits maximal Kekul\'e charge density pattern  --- in accord with the STM experiments~\cite{Nuckolls_2023} --- is  the favored ground state at $|\nu| = 2$ due to electron-phonon coupling. We also discover a nontrivial Zeeman response of this $|\nu| = 2$ IKS state. For $|\nu| = 3$, we note that two distinct effects of the electron-phonon coupling --- namely, inducing a valley anti-ferromagnetic Hund's coupling~\cite{Chatterjee2020} and strengthening charge density Kekul\'e patterns~\cite{Kwan2024} ---  have opposite effects. This may be understood as stemming separately from the the Fock and Hartree contributions of the effective electron-electron interaction induced upon integrating out phonons; we find that the anti-ferromagnetic tendency dominates.

Building on the knowledge of the symmetry of the IKS ground states under realistic perturbations, we numerically map out the spectrum of soft collective excitations about these states. We confirm that (a) the counting of linear and quadratic gapless Nambu-Goldstone modes matches general symmetry expectations~\cite{watanabeannualrev}, and (b) the number of gapped low energy collective modes agrees with the predictions of Ref.~\cite{khalafSoftModesMagic2020}. We extend the discussion to non-integer fillings, where we find the ground state symmetry to closely match that of the nearby integer fillings. As recent experimental study~\cite{kimResolvingIntervalleyGaps2025} has provided new evidence suggesting a link between IKS and superconductivity, we examine the pairing symmetry of superconductivity from (doped) IKS parent states; {\it inter alia}, we map the Fermi surfaces obtained by doping the correlated insulator at $|\nu|=2$, paying particular attention to the spin structure. Based on this analysis, we conjecture a two-regime picture for superconductivity between $-3<\nu<-2$, wherein the nature of the superconductor changes significantly at a critical value $\nu_c$. For $\nu < \nu_c$, we propose that the superconducting pairing symmetry is neither singlet nor triplet due to a spin-valley locked parent state, similar to that proposed in Ref.~\cite{lake_pairing_2022} but with the added feature of intervalley coherence. For $\nu > \nu_c$, the superconductor could be either singlet or triplet as the parent state preserves spin rotation symmetry. Furthermore, we emphasize that as the normal state in this regime preserves the physical time-reversal symmetry, the superconductor is protected by Anderson's theorem if it is $s$-wave --- a stronger statement than the  generalized Anderson's theorem proposed previously~\cite{zhou2024doubledomeunconventionalsuperconductivitytwisted} (including by four of the present authors), which relies on a modified time-reversal symmetry.

The remainder of this paper is organized as follows.
In Section~\ref{sec:method}, we introduce the Hamiltonian under discussion, including the terms due to electron-phonon coupling and non-local tunneling. In Section~\ref{sec:spinvalley}, we investigate the effects of electron-phonon coupling and other perturbations that break the symmetry of independent spin rotations in each valley on the IKS ground states at $\nu = \pm 2, \pm 3$. In Section~\ref{sec:nonint}, we consider the impact of electron-phonon coupling at non-integer filling factors. In Section~\ref{sec:collective}, we study the collective mode spectra of IKS states. In Section~\ref{sec:sc}, we consider the relevance of IKS states as parent states for superconductivity in twisted bilayer graphene. In Section~\ref{sec:conclusion}, we present our conclusions.

\section{Model and Methods}\label{sec:method}

Our starting point, as advertised, is the interacting BM Hamiltonian~\cite{Bistritzer2011} with strain, to  which we add various additional perturbations. Schematically, we have
\begin{equation}
    \hat{H} = \hat{H}_{\text{BM}} + \hat{H}_{\text{int}} + \hat{H}',\end{equation}
where $\hat{H}_{\text{BM}}$ is the non-interacting BM Hamiltonian with strain, $\hat{H}_{\text{int}}$ is the dual-gate screened Coulomb interaction, and $\hat{H}'$ includes perturbations such as electron-phonon coupling ($\hat{H}^{\text{inter}}_{\text{ph}}$ for coupling to intervalley phonons), intervalley-Coulomb scattering ($\hat{H}^{\text{inter}}_{\text{e-e}}$), and possible external fields. We discuss each of these in turn.

\subsection{Strained BM Model}
We begin by describing the BM Hamiltonian~\cite{Bistritzer2011} for TBG under uniaxial heterostrain~\cite{bi_designing_2019,Kwan2021,Parker2021}, corresponding to distortions along a single axis but with opposite signs in the two layers. Our discussion closely follows the conventions in Refs.~\cite{bi_designing_2019, Parker2021}.

First, consider strain of magnitude $\epsilon$ and direction $\varphi$, measured counter-clockwise from the $\hat{x}$-axis, in a single layer. The symmetric strain tensor is given by
\begin{equation}
    S(\epsilon, \varphi) = R^{-1}(\varphi) \begin{pmatrix}
        - \epsilon & 0 \\
        0 & \nu \epsilon
    \end{pmatrix}R(\varphi)=\begin{pmatrix}
    \epsilon_{xx} & \epsilon_{xy} \\ \epsilon_{xy} & \epsilon_{yy}
    \end{pmatrix}
\end{equation}
where $\nu = 0.16$ is the Poisson ratio. Including a twist angle $\theta$, the transformation matrix, which acts via $\bm{r} \to M^T\bm{r},\,\bm{g} \to M^{-1}\bm{g},$
is given for small $\epsilon,\theta$ by $M \simeq 1 + \mathcal{E}^T$, where
\begin{equation}
    \mathcal{E} \simeq \begin{pmatrix}
    \epsilon_{xx} & \epsilon_{xy} - \theta \\
    \epsilon_{xy} + \theta &\epsilon_{yy}
\end{pmatrix}.\end{equation}

To describe TBG (with  $l=1,2$ labelling the top and bottom layers) under uniaxial heterostrain, we consider layer-dependent transformations parameterized by
\begin{equation}
    \theta_1 = - \theta_2 = \frac{\theta}{2}, \quad \varphi_1 = \varphi_2 = \varphi,\quad  \epsilon_1 = - \epsilon_2 = \frac{\epsilon}{2}.
\end{equation}
Defining graphene reciprocal lattice vectors (RLVs)
\begin{equation}
    \bm{G}_{G1} = \frac{4\pi}{3a}(\frac{\sqrt{3}}{2}, \frac{1}{2}), \quad \bm{G}_{G2} = \frac{4\pi}{3a}(\frac{\sqrt{3}}{2}, - \frac{1}{2})\end{equation}
where $a=0.142\,\text{nm}$ is the C-C bond length, we then find that the moir\'e RLVs are given by
\begin{align}
    \bm{G}_1 &= (M^{-1}_1 - M^{-1}_2)(\bm{G}_{G2} - \bm{G}_{G1}),\nonumber\\
    \bm{G}_2 &= (M^{-1}_1 - M^{-1}_2)\bm{G}_{G1}.\end{align}

The kinetic term in valley $\tau$ (where $\tau = \pm 1$ for $K, K'$) is given by 
\begin{equation}
    \braket{\bm{k}, l |\hat{H}_{\text{BM}}|\bm{k}', l'} = \hbar v_F[M_l(\bm{k} - \tau \bm{A}_l) - \bm{K}_\tau] \cdot \begin{pmatrix}
    \tau\sigma_x \\ - \sigma_y
\end{pmatrix}\delta_{\bm{k}\bm{k}'}\delta_{ll'}\end{equation}
where $\bm{A} = \frac{\beta}{2a}(\epsilon_{xx} - \epsilon_{yy}, -2\epsilon_{xy})$
and $\bm{K}_\tau =  \frac{4\pi}{3\sqrt{3}a}(\tau, 0)$ is the original (i.e.~in the absence of rotation and strain) Dirac momentum in valley $\tau$. $\bm{k}$ is measured with respect to the global origin of momentum. We use the Dirac velocity $v_F=8.8\times 10^5\,\text{ms}^{-1}$. 

The (local) interlayer hopping term from layer 2 to layer 1 in valley $K$ is~\cite{Bistritzer2011}
\begin{equation}\braket{\bm{k},1|\hat{H}_{\text{BM}}|\bm{k}',2} = T_1\delta_{\bm{k} - \bm{k}',\bm{0}} + T_2\delta_{\bm{k} - \bm{k}', \bm{G}_1 + \bm{G}_2} + T_3\delta_{\bm{k} -\bm{k}', \bm{G}_2}\end{equation}
where the hopping matrices are given by
\begin{equation}
    T_j = \begin{pmatrix}
    w_\textrm{AA} & w_\textrm{AB}e^{\frac{2\pi i (j-1)}{3}}\\
    w_\textrm{AB}e^{-\frac{2\pi i (j-1)}{3}} & w_\textrm{AA}
\end{pmatrix}.
\end{equation}
The expressions for valley $K'$ can be found by time-reversal.

Unless otherwise stated, we consider twist angle $\theta=1.05^\circ$, inter-sublattice interlayer hopping $w_\textrm{AB}=110$~meV, intra-sublattice interlayer hopping $w_\textrm{AA}=80$~meV, strain magnitude $\epsilon=0.3\%$, and angle $\varphi = 0$ (i.e. strain along $\hat{x}$-axis).

\subsection{Screened Coulomb interaction}\label{subsec:Coulomb}

The interaction Hamiltonian is given by
\begin{equation}\label{eq:INT}
    \hat{H}_{\text{int}} = \frac{1}{2A}\sum_{\bm{q}}V(\bm{q}):\rho_{\bm{q}}\rho_{-\bm{q}}:
\end{equation}
where $\rho_{\bm{q}}$ is the projected density operator given by
\begin{equation}\rho_{\bm{q}} = \sum_{\bm{k} \in \text{BZ}, ab\tau s} \braket{u_{\bm{k}+\bm{q},\tau a}|u_{\bm{k},\tau b}}c^\dagger_{\bm{k} + \bm{q},\tau s a} c^{\phantom\dagger}_{\bm{k},\tau s b},
\end{equation}
where $a,b$ are band indices, $\tau$ is the valley index and $s$ is the spin index. We consider the dual-gate screened Coulomb potential $V(q)=e^2\tanh(qd)/(2\epsilon_0\epsilon_r q)$, with gate screening distance $d=25$\,nm and relative dielectric constant $\epsilon_r=10$. To specify the interaction term, we also need to select an `interaction scheme' to avoid double-counting interactions~\cite{Xie2020,parker2021fieldtunedzerofieldfractionalchern,Kwan2021,Kwan_2022,kwan2023mfci3}. The interaction scheme can be parameterized by a reference density matrix with respect to which we measure the deviations of $\rho$ that then enter the interaction term. In this work, we use the ``average" subtraction scheme~\cite{Lian2021}, where the reference density corresponds to half-filling of the central bands.

In the above formulation, we have only considered $\rho_{\bm{q}}$ with $|\bm{q}| \ll 1/a$, where $a$ is the graphene C-C bond length. As such, we have treated ``valley" as an internal degree of freedom and ignored momentum transfer of the order $|\bm{q}| \sim 1/a$, which corresponds to intervalley scattering. We will revisit this issue in Section~\ref{subsec:IVS}.

\subsection{Electron-phonon coupling}\label{subsec:EPC_model}
To motivate the inclusion of electron-phonon interactions, we remind the reader that one puzzle raised by the experimental observation of Kekulé patterns in twisted bilayer graphene in Ref.~\cite{Nuckolls_2023} is that  even low strain samples show a Kekulé pattern at $\nu=+2$. The interacting BM model predicts that the $\nu=+2$  ground state in this limit is the K-IVC phase \cite{Bultinck2020}, which is forbidden from exhibiting a Kekulé pattern in the charge sector for symmetry reasons \cite{JungPyo,Calugaru2022}. One route to resolving this discrepancy is to consider the coupling of the electron modes to a particular phonon mode, namely the graphene zone-corner $K$-phonon. Including the coupling to this phonon leads to a ``valley Jahn-Teller effect" \cite{Angeli2019,Blason2022}, which stabilizes a distinct phase, the so-called time-reversal symmetric IVC (T-IVC) state\footnote{There is also a time-reversal-breaking variant of the state, with the same energy under Hartree-Fock, that exhibits quantum anomalous Hall effect (therefore called IVC-QAH) and also Kekul\'e charge density pattern.}, which does exhibit a Kekulé charge pattern \cite{Kwan2024}. Added motivation for considering this specific phonon mode comes from recent measurements from quantum twisting microscope~\cite{birkbeck_quantum_2025} that sees strong coupling to the intervalley $K$-point optical phonons. Furthermore, the experiment found that the frequency and electron-phonon coupling strength of these modes depend only weakly on twist angles, justifying our approach of treating these phonons as arising from decoupled graphene monolayers. This mode has also been proposed as a route to superconducting pairing in MA-TBG~\cite{Wu2018,Liu_2024,liu_nodal_2025}, although this mechanism alone seems unlikely to yield a sufficiently high $T_c$ for a realistic electronic density of states \cite{Wagner2024}. 

At low strain, the phonons affect the competition between the strong coupling states \cite{Kwan2024,Wang2025}: in particular, they lower the energy of states which exhibit a Kekulé pattern and hence stabilizes the T-IVC relative to the K-IVC. The K-IVC and T-IVC both belong to the ``strong-coupling manifold of low-energy states'', identified by working in a special $\text{U}(4)\times \text{U}(4)$ symmetric limit, and then lifting degeneracies between competing orders through subleading anisotropy terms in the Hamiltonian \cite{Bultinck2020, Lian2021} that lower the symmetry to the more physical $\text{SU}(2)_K \times \text{SU}(2)_{K'} \times \text{U}(1)_v \times \text{U}(1)_c$ symmetry of interacting BM model, which further reduces to $\text{SU}(2)_s\times \text{U}(1)_v\times \text{U}(1)_c$ if intervalley-phonons or inter-valley scattering are included ($s, v, c$ correspond to spin, valley, and charge quantum numbers, and $\text{SU}(2)_{K(K')}$ refers to spin-rotation in $K(K')$ valley). The splitting between the strong coupling states due to the anisotropy terms is of the same order as the energy corrections due to the electron-$K$-phonon coupling \cite{Kwan2024}, so that the latter have a particularly prominent role in this limit. At higher strains, in contrast, there are no states in close energetic competition with the IKS state due to the large strain energy scale that favors the latter. Nevertheless, the IKS states on their own constitute a large degenerate manifold due to the symmetry of independent rotations of spins in each valley~\cite{Kwan2021, Wagner2022}; thus, even at high strain,  phonons play an important role in selecting out the ``true'' spin structure of the correlated insulators at integer $\nu$. Furthermore, phonons can affect the competition between closely competing metallic states at non-integer fillings at both high and low strain.

Accordingly, we consider the effects of the in-plane transverse optical phonons with symmetry irreps $A_1$ and $B_1$ at the graphene zone corners~\cite{Basko2008,Chatterjee2020,Wu2018,Angeli2019,Blason2022,Kwan2024}. Note that we do not consider the zone-center phonons or the low-energy moir\'e phonons or phasons, in the spirit of incorporating the minimal ingredients to capture the normal state phase structure; however, the latter may be important to include when exploring possible routes to superconductivity.

With these preliminaries, the phonon Hamiltonian is given by 
\begin{equation}
    \hat{H}_{\text{ph}} = \hbar\omega \sum_{l\alpha\bm{q}}\hat{a}^\dagger_{l\alpha}(\bm{q})\hat{a}^{\phantom\dagger}_{l\alpha}(\bm{q}),
\end{equation}
where $l$ labels the layer, $\alpha=1,2$ labels the degenerate modes, and the phonon dispersion is approximated as constant, given by $\hbar\omega \simeq 160$~meV. The intervalley phonons couple to electrons as~\cite{Wu2018}
\begin{equation}\label{eq:H_EPC}
    \hat{H}_{\text{EPC}} = \mathcal{F}\sum_{l\alpha}\int d\bm{r} \, \, \hat{\psi}^\dagger_l(\bm{r})\, [\hat{u}_{l\alpha}(\bm{r})\Gamma_\alpha] \, \hat{\psi}^{\phantom\dagger}_l(\bm{r})
\end{equation}
where $\hat{u}_{l\alpha}(\bm{r}) = \mathcal{D}\sum_{\bm{q}}[\hat{a}^{\phantom\dagger}_{l\alpha}(\bm{q}) + \hat{a}^\dagger_{l\alpha}(-\bm{q})]$, $\hat{\psi}_l$ is a spinor in spin ($s$), valley ($\tau$) and sublattice ($\sigma$) space, and $\mathcal{F}$ and $\mathcal{D}$ collect various phonon parameters. The coupling matrices are given by $\Gamma_{1} = \tau_x\sigma_x$ and $\Gamma_{2} = \tau_y\sigma_x$. The coupling strength is 
$g = A\frac{\mathcal{F}^2\mathcal{D}^2}{\hbar\omega},$
where $A$ is the area of the system. We take the coupling strength to be around $g=70$\,meVnm\textsuperscript{2}~\cite{Wu2018}, though estimates can vary widely in the literature~\cite{Basko2008}.

Using a Schrieffer–Wolff transformation, we obtain an effective phonon-mediated electron-electron interaction given by~\cite{chatterjee_symmetry_2020}
\begin{align}\label{eq:EPC}
\hat{H}^{\text{inter}}_{\text{ph}}=&-
\frac{2g}{A}\sum_{\bp\bk\bk'\bG}\sum_{\substack{l\tau s s'\\abcd}}f_{l,\tau ab}(\bk,\bp + \bG)f^*_{l,\tau dc}(\bk',\bp + \bG) \nonumber \\
\times &c^\dagger_{\bk, \tau s a}c^\dagger_{\bk'+\bp, \bar{\tau}s' c}c^{\phantom\dagger}_{\bk', \tau s' d}c^{\phantom\dagger}_{\bk+\bp, \bar{\tau}s b},
\end{align}
where the momenta $\bk, \bk', \bp$ are summed over the first moir\'e Brillouin zone (mBZ), and $\bG$ is a moir\'e reciprocal lattice vectors. $c^\dagger_{\bm{k},\tau s a}$ is an electron creation operator for mBZ momentum $\bm{k}$, valley $\tau$, spin $s$, and moir\'e band $a$. The matrix elements $f$ are defined as
\begin{equation}
 f_{l,\tau aa'}(\bk,\bp)=\braket{u_{\tau a}(\bk)|\sigma_x P_l |u_{\bar{\tau} a'}(\bk+\bp)},
\end{equation}
where $P_l$ is the projection operator to layer $l$, with 
$\bar \tau$ meaning the opposite valley from $\tau$. In SM Sec.~\ref{App:comparing_methods}, we discuss results obtained using a different method of treating the electron-phonon coupling, which was utilized in Ref.~\cite{Kwan2024}. In this alternative `direct distortion' formalism, Eq.~\ref{eq:H_EPC} is treated directly in mean-field within a variational manifold consisting of products of a single Slater determinant and a coherent state of phonons.

\subsection{Intervalley Coulomb Scattering}\label{subsec:IVS}

\begin{figure}[h]
    \centering
    \includegraphics[width=\linewidth]{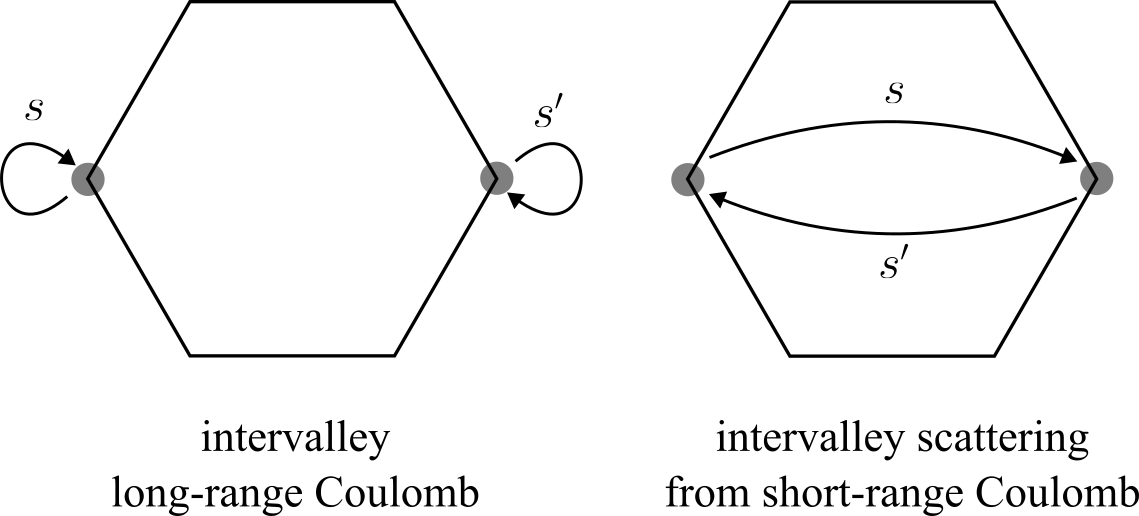}
    \caption{Long-range and short-range Coulomb interactions. The hexagons represent the Brillouin zone of graphene. While the long-range interaction also contains terms that couple the two valleys, each electron is scattered into the valley from which it originates. The short-range piece, on the other hand, exchanges electrons between the valleys, such that the net $\text{U}(1)_v$ valley charge is unchanged but the interacting electrons have swapped their valley indices. Thus, while both terms are $U(1)_v$-symmetric, the former enjoys an enhanced $[\text{SU}(2)_K\times \text{SU}(2)_{K'}]_s$ symmetry of independent spin rotations in each valley, whereas the latter only exhibits the physical $\text{SU}(2)_s$ symmetry.}
    \label{fig:scattering_illust}
\end{figure}

In Section~\ref{subsec:Coulomb}, we have ignored density operators $\rho_{\bm{q}}$ with momentum transfer $\bm{q}$ that scatters an electron to another valley, as illustrated in Fig.~\ref{fig:scattering_illust}. To include the intervalley Coulomb scattering process, we can consider
\begin{equation}
    \rho_{\bm{q} + \tau\Delta\bm{K}} = \sum_{\bm{k} \in \text{BZ}, ab\tau s} \braket{u_{\bm{k}+\bm{q},\tau a}|u_{\bm{k},\bar{\tau}b}}c^\dagger_{\bm{k} + \bm{q},\tau sa} c^{\phantom\dagger}_{\bm{k},\bar{\tau}sb},
\end{equation}
where $\Delta\bm{K} = \bm{k}_{+\gamma} - \bm{k}_{-\gamma}\approx \bm{K}_+ - \bm{K}_-$ is the momentum difference between the two valleys (strictly speaking, it is measured from the $\gamma$-point of moir\'e Brillouin zone in each valley, but one can approximate it as measured from the original graphene Dirac points), and $\bm{q}$ is now some small momentum on the moir\'e scale. We can then approximate $V(\bm{q} + \tau\Delta\bm{K}) \approx V(| \bm{K}_+ - \bm{K}_-|) \equiv V_{\text{inter}}$, and the Hamiltonian is given by
\begin{align}\label{eq:IVscatter}
\hat{H}^{\text{inter}}_{\text{e-e}}=&
\frac{V_{\text{inter}}}{2A}\sum_{\bp\bk\bk'\bG}\sum_{\substack{\tau s s'\\abcd}}f'_{\tau ab}(\bk,\bp + \bG)f'^*_{\tau dc}(\bk',\bp + \bG) \nonumber \\
\times &c^\dagger_{\bk, \tau s a}c^\dagger_{\bk'+\bp, \bar{\tau}s' c}c^{\phantom\dagger}_{\bk', \tau s' d}c^{\phantom\dagger}_{\bk+\bp, \bar{\tau}s b}
\end{align}
with matrix elements
\begin{equation}
 f'_{\tau aa'}(\bk,\bp)=\braket{u_{\tau a}(\bk)|u_{\bar{\tau} a'}(\bk+\bp)}.
\end{equation}
Assuming the dual gate-screened Coulomb interaction is applicable at all momentum scales with the same dielectric constant $\epsilon_r = 10$, one finds $V_{\text{inter}}\simeq 50$ meVnm$^2$. However, it may be more appropriate to use the dielectric constant of the substrate hBN of $\epsilon_r = 4.4$~\cite{hunt_direct_2017} instead, as screening from remote bands is expected to be insignificant at the microscopic graphene scale~\cite{alicea_graphene_2006}. This gives $V_{\text{inter}} \simeq 120$~meVnm$^2$.

This simple treatment above involves some approximations, as we now summarize. First, we have ignored the actual form of wavefunctions on graphene scale, including the spatial offset of the two sublattices. This can be relevant in certain cases, e.g.~when studying symmetry breaking in the valley-sublattice locked zeroth Landau level of monolayer graphene~\cite{alicea_graphene_2006},  but is not expected to be of qualitative significance in the present setting. Second, at the lattice scale the separation between the layers becomes non-negligible. Ref.~\cite{chatterjee_symmetry_2020} finds that the interlayer separation can sometimes \textit{increase} the effective intervalley scattering, though this effect is dependent on the particular many-body wavefunction in consideration. Many of the results in this work only depend on the qualitative features of $\hat{H}^{\text{inter}}_{\text{e-e}}$, which are impervious to such modeling details. When the actual strength of intervalley Coulomb scattering is significant, we treat $V_{\text{inter}}$ as a tunable parameter, whose true value is presumably not too far off from our rough estimate of $V_{\text{inter}} \simeq 120$~meVnm$^2$.

\subsection{Choice of Approach}
The interacting BM model and its various modifications can be numerically studied using Hartree-Fock \cite{Kang2021,Zhang2020,Bultinck2020,Liu2021b,Xie2020,Xie2021,Cea2020,Zhang_2022,Lian2021,Liu2021c,Kwan2021,Xie2021c,Faulstich2023,Kwan_2022,Cea2019,Guinea2018,Xie2023,Hejazi2021,kwan2024texturedexcitoninsulators,Kwan2021dw,Wang2025cti}, density-matrix renormalization group \cite{Soejima2020,Kang2020,Faulstich2023}, Monte Carlo \cite{Hofmann2022,zhang2021momentum,Zhang2023} or exact diagonalization \cite{Potasz_2021,Xie2021b,Repellin2020}. Sign-problem-free Monte Carlo is limited to a specific filling (charge neutrality), and exact diagonalization and density-matrix renormalization group are limited to small system sizes or a restricted number of bands. This leaves Hartree-Fock as the numerical technique that is most suited for sweeping the entire phase diagram for reasonable system sizes. Hartree-Fock can account for a number of experimental observations such as the correlated insulating states at integer filling of the moiré unit cell \cite{Cao2018,Cao:2018aa,Cao2021a,Das:2021aa,Liu2021i,Park:2021aa,Rozen:2021aa,Saito_2020,Saito:2021aa,Uri:2020aa,Yankowitz2019,Zondiner:2020aa,Lu2019ins,Pierce:2021aa,Polshyn:2019aa,Serlin2020ins,sharpe2019ins,Wu:2021aa,Stepanov_2020,Stefanov2021} and the quantized anomalous Hall effect at $\nu=+3$ in the presence of an aligned hBN substrate \cite{Serlin2020ins,sharpe2019ins}. It also qualitatively captures nearly all aspects of the IKS state at integer~\cite{Kwan2021,Wang2024} and non-integer fillings~\cite{Wagner2022}, and works reasonably well even quantitatively for a subset of these (e.g., it provides quite a good estimate for the IKS modulation vector $\boldsymbol{q}$). In this work we will use Hartree-Fock to examine the phase diagram and low energy excitations, while including the effects of the various beyond-BM perturbations outlined above.

It is often desirable to obtain band structures with higher momentum-space resolution than one can numerically compute with self-consistent Hartree-Fock on reasonable system sizes. When such band structures are necessary, we perform a ``Hartree-Fock interpolation", as explained in App.~\ref{sec:interpolation}, which essentially uses the ground state of a sparse grid to extract the mean-field parameters that are then used as a `background' to generate the band structure on a finer grid.

\subsubsection{IKS within Hartree-Fock}

The IKS is a time-reversal symmetric state characterized by intervalley coherence (IVC) at a finite wavevector $\bm{q}$. Since the order occurs at a single wavevector, we adopt an efficient approach for the HF calculations, where we work in a valley-dependent boosted reference frame with the momenta of single particle states in one valley boosted by $\bm{q}_{\text{IKS}}$ relative to the other valley. We define the boosted momenta $\tilde{\bm{k}}=\bm{k}+\tau\bm{q}_{\text{IKS}}/2$, where $\tau=\pm1$ is the valley index and is $\vec{q}_{\text{IKS}}$ is the IVC spiral wavevector. In this boosted frame, the IKS can be represented with a $\tilde{\mathbf{k}}$-diagonal Hartree-Fock projector, significantly speeding up the computation. We perform HF calculations for different possible wavevectors $\bm{q}_{\text{IKS}}$ and pick the solution with the lowest energy solution as the ground state. We project into the two central bands per spin and valley sector.

\section{Lifting spin-valley degeneracy in the IKS Manifold}\label{sec:spinvalley}

As a first application of the extended microscopic theory, we use it to establish the spin structure of the IKS states. To motivate the problem, recall that Refs.~\cite{Kwan2021, Wagner2022} performed HF calculations of TBG at integer and non-integer fillings with moderate strain, and found that IKS order is present at all nonzero integer fillings and a large range of non-integer fillings. While mapping out important features of the phase diagram in broad strokes, these previous works did not fully address the spin structure of these IKS states. This is because the interacting Hamiltonians considered there only included the dominant terms in the interaction, namely the long-range Coulomb interaction, leading to an enlarged $\text{SU}(2)_K\times \text{SU}(2)_{K'}$ symmetry, i.e. independent spin rotations within each valley. In a more realistic treatment, inter-valley processes such as intervalley Coulomb scattering and phonons are present, reducing the $\text{SU}(2)_K\times \text{SU}(2)_{K'}$ symmetry to a global spin rotation symmetry, i.e.  $\text{SU}(2)_s$. While these terms are smaller than the dominant intra-valley Coulomb scattering, they are crucial in determining the spin structure of the IKS states.

In the following, we  treat the cases of $|\nu| = 2$ and $|\nu| = 3$ separately. The $\text{SU}(2)_K\times \text{SU}(2)_{K'}$-symmetric degenerate manifold of $|\nu| = 3$ IKS states can be parametrized by the choice of spin polarization in each valley. Our main task is therefore  to determine their relative orientation i.e. whether the spins in the different valleys are  ferromagnetically or  anti-ferromagnetically aligned. 
In contrast, $|\nu| = 2$ IKS consists of two copies of spinless IKS states and has no net spin polarization in each valley. Instead, the key physics is the relative $\text{U}(1)_v$ phase between the two copies, which will be crucial in determining the microscopic Kekul\'e pattern.

\subsection{$\nu=\pm3$}\label{subsec:nu3}

The most primitive picture of the IKS state is as an intervalley coherent state of spinless electrons with a single occupied (or unoccupied) band. Introducing the physical spin, at $|\nu| = 3$ and for a Hamiltonian that preserves $\text{SU}(2)_K\times \text{SU}(2)_{K'}$-symmetry, the ground state manifold can be generated by the independent rotation of spins in the two valleys. Restricting to the central bands, the density matrix at $\nu = -3$ can be parameterized as
\begin{equation}P(\bm{k}) = \ket{\psi_{\bm{k}}}\bra{\psi_{\bm{k}}},\end{equation}
where the single occupied  HF band is
\begin{equation}\label{eq:psi_k}
\ket{\psi_{\bm{k}}} = \alpha_{\bm{k}} \ket{K, {\bm{k}}_+} \otimes \ket{\hat{\bm{n}}_+} + \beta_{\bm{k}} e^{i\phi}\ket{K', {\bm{k}}_-} \otimes \ket{\hat{\bm{n}}_-},
\end{equation}
where ${\bm{k}}_\pm = \bm{k}\pm \frac{\bm{q}_{\text{IKS}}}{2}$, for some IVC angle $\phi$, spin direction vectors $\ket{\hat{\bm{n}}_\pm}$ and coefficients $|\alpha_{\bm{k}}|^2 + |\beta_{\bm{k}}|^2 = 1$. Here, $\ket{K, \bm{k}}$ ($\ket{K', \bm{k}}$) is some Bloch state in valley $K$ ($K'$) with moir\'e momentum $\bm{k}$. Physically, the state has a Kekul\'e charge density of $\braket{\rho_{\bm{q} \sim \bm{K}_+ - \bm{K}_-}} \propto \braket{\hat{\bm{n}}_-|\hat{\bm{n}}_+} \propto \cos(\theta/2)$, where $\theta$ is the angle between $\hat{\bm{n}}_+$ and $\hat{\bm{n}}_-$, and a Kekul\'e spin density of $\braket{\bm{S}_{\bm{q} \sim \bm{K}_+ - \bm{K}_-}} \propto \braket{\hat{\bm{n}}_-|\bm{s}|\hat{\bm{n}}_+}$. At $\nu = +3$, the state can be parameterized by $P(\bm{k}) = I -  \ket{\psi_{\bm{k}}}\bra{\psi_{\bm{k}}}$, where $I$ is the identity matrix (corresponding to fully filled central bands) and $\ket{\psi_{\bm{k}}}$ is the wavefunction of the unoccupied HF band, which can also be put into the form of Eq.~\ref{eq:psi_k}.

Intervalley processes, such as the coupling of electrons to intervalley phonons and intervalley Coulomb scattering, lift the degeneracy of the ground state manifold and select out a particular relative orientation of $\hat{\bm{n}}_+$ and $\hat{\bm{n}}_-$ (of course,  $\text{SU}(2)_s$ symmetry precludes fixing the global spin axis.)  The electron-phonon coupling (Eq.~\ref{eq:EPC}) can be schematically represented as (suppressing band and momentum indices)
\begin{equation}
H^{\text{inter}}_{\text{ph}} \sim -gc^\dagger_{\tau s}c^\dagger_{\bar{\tau}s'}c^{\phantom\dagger}_{\tau s'}c^{\phantom\dagger}_{\bar{\tau}s}.
\end{equation}
Performing a mean-field decoupling, we find that the Fock term yields 
\begin{equation}
    E^{F}_{\text{ph}} \sim +g\sum_{s s'\tau}\braket{c^\dagger_{\tau s}c^{\phantom\dagger}_{\tau s'}}\braket{c^\dagger_{\bar{\tau}s'}c^{\phantom\dagger}_{\bar{\tau}s}}  \sim +g \hat{\bm{n}}_+ \cdot \hat{\bm{n}}_-,
\end{equation}
which is the anti-ferromagnetic Hund's coupling usually referred to in the literature (we have dropped some constant terms). However, we note that as the state has finite intervalley coherence, we also obtain a non-zero contribution from the Hartree term of the decoupling, given by
\begin{equation}
    E^{H}_{\text{ph}} \sim - g\sum_{ss'\tau}\braket{c^\dagger_{\tau s}c^{\phantom\dagger}_{\bar{\tau}s}}\braket{c^\dagger_{\bar{\tau}s'}c^{\phantom\dagger}_{\tau s'}} \sim -g \hat{\bm{n}}_+ \cdot \hat{\bm{n}}_-,
\end{equation}
which is ferromagnetic (again, we have dropped some constant terms). This is similar to the intuition that a Kekul\'e charge pattern lowers the ground state energy due to its coupling to physical lattice distortion, as discussed in Ref.~\cite{Kwan2024}. Numerically (see SM Sec.~\ref{app:additional_numerical_integer} for the range of parameters used), we find that the antiferromagnetic `Fock' coupling dominates, thereby favouring $\hat{\bm{n}}_+ = -\hat{\bm{n}}_-$. We note that the result, as obtained here from the Schrieffer–Wolff transformation, is qualitatively different from that using the `direct distortion' approach~\cite{Kwan2024}, as discussed in SM Sec.~\ref{App:comparing_methods}.

Intervalley Coulomb scattering (cf. Eq.~\ref{eq:IVscatter}) has a very similar form to that of the electron-phonon coupling, but with the opposite sign. It similarly exhibits competition between Hartree and Fock terms; numerically we find that it is effectively ferromagnetic, i.e.~favoring $\hat{\bm{n}}_+ = \hat{\bm{n}}_-$. We note that Ref.~\cite{chatterjee_symmetry_2020} also found that electron-phonon coupling gives rise to an anti-ferromagnetic intervalley Hund's coupling, and intervalley Coulomb scattering gives rise to a ferromagnetic intervalley Hund's coupling. However, their work considered non-IVC states where the Hartree contribution, and hence its competition with the Fock term, is absent. The role of IVC is thus to effectively suppress the absolute strength of the two couplings relative to the Fock-only value.

The antiferromagnetic ground state at $\nu = \pm 3$ is invariant under $\tau_zs_z$ (here, for definiteness, we have chosen $\braket{\tau_z\bm{s}}$ to point in the $z$-direction) so that the intervalley coherence occurs between opposite spin sectors. This induces a spin Kekul\'e pattern rather than a charge Kekul\'e pattern. However, $\tau_zs_z$ is not a true symmetry of the Hamiltonian, but rather a symmetry of the ground state due to the mean-field approximation inherent in HF. Beyond mean-field, we do not expect $\tau_zs_z$ to be preserved, so that the charge Kekul\'e pattern would be generally non-vanishing.

In realistic TBG, both  phonons and intervalley Coulomb scattering are of course present. Since the two effects favor competing ground states, we observe a sharp phase transition as we tune the relative strength of the two perturbations. At typical values of $g/V_{\text{inter}} \sim 0.6$, we find phonons to be dominant (see SM Sec.~\ref{subsec:competition}). This is consistent with electron spin resonance measurements which find the intervalley coupling to be antiferromagnetic~\cite{morissette_dirac_2023}.

\subsection{$\nu=\pm2$}\label{subsec:nu2}
The $|\nu| = 2$ IKS can be constructed from two copies of $|\nu| = 3$ IKS states, as we now show. For definiteness, we focus on the $\nu = -2$ case; $\nu = +2$ can be easily obtained by particle-hole conjugation.

Let us consider some $\nu = -3$ IKS given by (Eq.~\ref{eq:psi_k})
\begin{equation}
\ket{\psi_{\bm{k}}} = \alpha_{\bm{k}} \ket{K, {\bm{k}}_+} \otimes \ket{\hat{\bm{n}}_+} + \beta_{\bm{k}} e^{i\phi}\ket{K', \bm{k}_-} \otimes \ket{\hat{\bm{n}}_-},
\end{equation}
where as before we have $\bm{k}_\pm = \bm{k}\pm \frac{\bm{q}_{\rm IKS}}{2}$. On physical grounds, we expect that $\alpha_{\bm{k}}$, $\beta_{\bm{k}}$ and $\bm{q}_{\rm IKS}$
are fixed by the dominant energy scales such as single-particle dispersion under strain and long-range Coulomb interaction (we can choose some definite gauge and absorb the phase degree of freedom into the factor  $e^{i\phi}$), so that $\hat{\bm{n}}_+$,  $\hat{\bm{n}}_-$ and $\phi$ parameterize the degenerate manifold at $\text{SU}(2)_K\times \text{SU}(2)_{K'}$-level. The selection of preferred ground states from this degenerate manifold under $\text{SU}(2)_K\times \text{SU}(2)_{K'}$-breaking perturbation is the topic of the preceding section.

To construct $\nu = -2$ IKS, we introduce another copy of IKS (with the same modulation vector $\bm{q}_{\rm IKS}$), given by
\begin{equation}
\ket{\psi'_{\bm{k}}} = \alpha_{\bm{k}} \ket{K, {\bm{k}}_+} \otimes \ket{\hat{\bm{n}}'_+} + \beta_{\bm{k}} e^{i\phi'}\ket{K', {\bm{k}}_-} \otimes \ket{\hat{\bm{n}}'_-},
\end{equation}
and we demand that the two copies are orthogonal, i.e. $\braket{\psi'_{\bm{k}}|\psi_{\bm{k}}} = 0$ for every $\bm{k}$, such that the density matrix of $\nu = -2$ IKS state is given by
\begin{equation}
    P(\bm{k}) = \ket{\psi_{\bm{k}}}\bra{\psi_{\bm{k}}} + \ket{\psi'_{\bm{k}}}\bra{\psi'_{\bm{k}}}.
\end{equation}
Now, since 
\begin{equation}
\braket{\psi'_{\bm{k}}|\psi_{\bm{k}}} =  |\alpha_{\bm{k}}|^2\braket{\hat{\bm{n}}'_+|\hat{\bm{n}}_+} + |\beta_{\bm{k}}|^2e^{i(\phi - \phi')}\braket{\hat{\bm{n}}'_-|\hat{\bm{n}}_-},    
\end{equation}
and $|\alpha_{\bm{k}}|^2$ and $|\beta_{\bm{k}}|^2$ are non-constant functions of $\bm{k}$ (since the valley polarization $|\alpha_{\bm{k}}|^2 - |\beta_{\bm{k}}|^2$ varies in the Brillouin zone~\cite{Kwan2021} for topologically mandated reasons~\cite{kwan2024texturedexcitoninsulators}), to ensure that $\braket{\psi'_{\bm{k}}|\psi_{\bm{k}}}$ vanishes at every $\bm{k}$ in the BZ, we must have $\hat{\bm{n}}'_\pm = - \hat{\bm{n}}_\pm$.  This allows us to write
\begin{widetext}
\begin{align}
P(\bm{k}) &=|\alpha_{\bm{k}}|^2 \ket{K,{\bm{k}_+}}\bra{K,{\bm{k}_+}} \otimes (\ket{\hat{\bm{n}}_+}\bra{\hat{\bm{n}}_+} + \ket{-\hat{\bm{n}}_+}\bra{-\hat{\bm{n}}_+}) \nonumber \\  & +  \alpha^{\phantom*}_{\bm{k}}\beta_{\bm{k}}^*\ket{K,{\bm{k}_+}}\bra{K',{\bm{k}_-}} \otimes (e^{-i\phi}\ket{\hat{\bm{n}}_+}\bra{\hat{\bm{n}}_-} + e^{-i\phi'}\ket{-\hat{\bm{n}}_+}\bra{-\hat{\bm{n}}_-}) \nonumber \\ & +  \alpha_{\bm{k}}^*\beta^{\phantom*}_{\bm{k}}\ket{K',{\bm{k}_-}}\bra{K,{\bm{k}_+}} \otimes (e^{i\phi}\ket{\hat{\bm{n}}_-}\bra{\hat{\bm{n}}_+} + e^{i\phi'}\ket{-\hat{\bm{n}}_-}\bra{-\hat{\bm{n}}_+}) \nonumber \\ & + |\beta_{\bm{k}}|^2 \ket{K',{\bm{k}_-}}\bra{K',{\bm{k}_-}} \otimes (\ket{\hat{\bm{n}}_-}\bra{\hat{\bm{n}}_-} + \ket{-\hat{\bm{n}}_-}\bra{-\hat{\bm{n}}_-}). 
\end{align}
\end{widetext}

Using $\ket{\hat{\bm{n}}_+}\bra{\hat{\bm{n}}_+} + \ket{-\hat{\bm{n}}_+}\bra{-\hat{\bm{n}}_+} = I_2$  (the 2-dimensional identity matrix), and defining $V \equiv e^{-i\phi}\ket{\hat{\bm{n}}_+}\bra{\hat{\bm{n}}_-} + e^{-i\phi'}\ket{-\hat{\bm{n}}_+}\bra{-\hat{\bm{n}}_-}$, we can write the density matrix more succinctly as
\begin{equation}\label{eq:PV}
P(\bm{k}) =    \begin{pmatrix}
        \tilde{P}_{++}(\bm{k}) \otimes I_2 & \tilde{P}_{+-}(\bm{k}) \otimes V \\
            \tilde{P}_{-+}(\bm{k}) \otimes V^\dagger & \tilde{P}_{--}(\bm{k}) \otimes I_2 \\
    \end{pmatrix}
\end{equation}
where we have defined $\tilde{P}_{++}(\bm{k}) = |\alpha_{\bm{k}}|^2 \ket{K,\tilde{\bm{k}}}\bra{K,\tilde{\bm{k}}}$ and similarly for the other entries.

It is straightforward to show that $VV^\dagger = I_2$, i.e. $V \in \text{U}(2)$. Importantly, this means that $V$ can be parameterized by 4 real parameters, for example, through $V = e^{i\delta}e^{i\gamma  (\hat{\bm{n} } \cdot \bm{s})}$, for some angles $\delta, \gamma$ and some unit vector $\hat{\bm{n}}$. In considering two copies of $\nu = -3$ IKS, we have introduced a total of 6 real parameters ($\hat{\bm{n}}_\pm$, $\phi$ and $\phi'$). As such, this formulation is actually redundant (ultimately, this is because there is no unique way to split $P(\bm{k})$ into $\ket{\psi_{\bm{k}}}\bra{\psi_{\bm{k}}} + \ket{\psi'_{\bm{k}}}\bra{\psi'_{\bm{k}}}$).

It is also straightforward to demonstrate that the manifold of states parameterized by unitary matrix $V$ with fixed $\tilde{P}_{++}(\bm{k})$ \emph{etc.} are indeed related by transformations in $\text{SU}(2)_K\times \text{SU}(2)_{K'}$. Choosing $V = I_2$, we have the spin singlet state, which preserves global spin rotations $s_x, s_y$ and $s_z$ symmetries (i.e. $\text{SU}(2)_s$) but breaks symmetry of independent spin rotations in each valley generated by $\{\tau_zs_x,\tau_zs_y,\tau_zs_z\}$, and the valley U(1) symmetry generated by $\tau_z$. This state can be written as 
\begin{equation}\label{eq:singlet}
P_{\text{singlet}}(\bm{k}) = \tilde{P}(\bm{k}) \otimes I_2.
\end{equation}
Consider the unitary transformation $U  = e^{i\frac{\delta}{2}\tau_z} e^{i\frac{\gamma}{2}\tau_z (\hat{\bm{n} }\cdot \bm{s})}$, where $\hat{\bm{n}}$ is some unit vector in spin space. We have
\begin{align}\label{eq:U}
 & UP_{\text{singlet}}(\bm{k})U^\dagger = \nonumber \\ & \begin{pmatrix}
    \tilde{P}_{++}(\bm{k}) \otimes I & \tilde{P}_{+-}(\bm{k}) \otimes e^{i\delta}e^{i\gamma  (\hat{\bm{n} }\cdot \bm{s})}\\
 \tilde{P}_{-+}(\bm{k})\otimes e^{-i\delta}e^{-i\gamma  (\hat{\bm{n} }\cdot \bm{s})} & \tilde{P}_{--}(\bm{k}) \otimes I
\end{pmatrix},
\end{align}
which is exactly the manifold of states defined by Eq.~\ref{eq:PV} with $V = e^{i\delta}e^{i\gamma  (\hat{\bm{n} }\cdot \bm{s})}$. This shows that $V$ is a parameterization of the manifold of degenerate $\nu = -2$ IKS states under $\text{SU}(2)_K\times \text{SU}(2)_{K'}$ symmetry, and we will sometimes write $P(\bm{k};V)$ to make this explicit, where it is understood that $[\tilde{P}(\bm{k})]_{\tau \tau'}$ is fixed by the dominant $\text{SU}(2)_K\times \text{SU}(2)_{K'}$-preserving terms in Hamiltonian. We will also use the notations $\delta$, $\gamma$ and $\hat{\bm{n}}$ to refer to the particular parametrization of $V$ given by $V = e^{i\delta}e^{i\gamma  (\hat{\bm{n} }\cdot \bm{s})}$. Physically, the state consists of one copy of IKS in each of the sector with spin parallel or anti-parallel to $\hat{\bm{n}}$, with a relative $\text{U}(1)_v$ phase difference of $2\gamma$ between the two spin sectors. This should  be contrasted with the situation for $|\nu| = 3$, where we considered intervalley coherence between basis states with different spin directions (Eq.~\ref{eq:psi_k}). This difference stems from the additional redundancy entailed in describing $|\nu| = 2$ IKS as two copies of $|\nu| = 3$ IKS states. 

While the states $P(\bm{k};V)$ are all energetically degenerate for a Hamiltonian that preserves $\text{SU}(2)_K\times \text{SU}(2)_{K'}$, they correspond to different spin and charge patterns on the graphene scale. To be precise, the Kekul\'e charge density $\braket{\rho_{\bm{q} \sim \bm{K}_+ - \bm{K}_-}} \propto \text{tr}(V) \propto \cos\gamma$, while the Kekul\'e spin density $\braket{\bm{S}_{\bm{q} \sim \bm{K}_+ - \bm{K}_-}} \propto \text{tr}(\bm{s}V) \propto \hat{\bm{n}}\sin\gamma$.

Coupling to intervalley phonons, as described in Sec.~\ref{subsec:EPC_model}, breaks the symmetry of independent spin rotations in each valley and (partially) lifts the ground state degeneracy. Since the valley diagonal part of $P(\bm{k})$ is actually fixed, the Fock term that arises from decoupling the electron-phonon coupling [cf. Eq.~\ref{eq:EPC}] expression is  independent of $V$. As such, the only non-trivial energetic contribution comes from the Hartree piece, schematically given by
\begin{equation}E_{\text{EPC}} \sim - \sum_{ss'\tau}\braket{c^\dagger_{\tau s}c^{\phantom{\dagger}}_{\bar{\tau}s}}\braket{c^\dagger_{\bar{\tau}s'}c^{\phantom{\dagger}}_{\tau s'}}, \end{equation}
which implies
\begin{equation}
E_{\text{EPC}} \sim -g |\text{tr}(V)|^2
\end{equation}
Since $\text{tr}(V) = 2e^{i\delta}\cos\gamma$, we conclude that the preferred ground state is a spin singlet state corresponding to $V = e^{i\delta}$ for $\delta \in [0, 2\pi)$ (and $\gamma=0$). While there will be sub-leading modifications to $\tilde{P}(\bm{k})$ for finite electron-phonon coupling, the parameterization Eq.~\ref{eq:PV} remains valid. We have numerically confirmed our results with self-consistent HF calculations (see SM Sec.~\ref{app:additional_numerical_integer} for the range of parameters used). We also comment that the spin singlet IKS state preserves the physical (spinful) time-reversal symmetry $\mathcal{T}_s = s_y\tau_x \mathcal{K}$.

On the graphene scale, the spin singlet state corresponds to the Kekul\'e interference pattern from the two spin sectors being locked in phase, giving rise to maximal Kekul\'e charge density pattern. This is consistent with the notion that Kekul\'e charge pattern lowers the ground state energy due to its coupling to physical lattice distortion~\cite{Kwan2024}.

The intervalley Coulomb scattering takes a similar form to the phonon-induced interaction but with the opposite sign; in other words, we have
\begin{equation}
    E^{\text{inter}}_{\text{e-e}} \sim +V_{\text{inter}}|\text{tr}(V)|^2,
\end{equation}
which immediately implies that the energetically favored ground state in this case is given by $V = e^{i\delta}e^{i\frac{\pi}{2}(\hat{\bm{n}}\cdot \bm{s})}$ with $\text{tr}(V) = 0$, for some arbitrary angle $\delta$ and some arbitrary unit vector $\hat{\bm{n}}$. This again is verified in our HF calculations (see SM Sec.~\ref{app:additional_numerical_integer} for the range of parameters used), and describes one copy of IKS state in each spin sector (parallel and anti-parallel to  $\hat{\bm{n}}$) with $\pi$ relative phase difference in the IVC angle. On the graphene scale, this corresponds to vanishing charge density Kekul\'e pattern but maximal spin Kekul\'e pattern. A minimal charge density Kekul\'e pattern is favored since it reduces the electrostatic Hartree penalty at wavevectors corresponding to the graphene Dirac momentum.

At this point, it is appropriate to refer back to the results at $|\nu| = 3$ and clarify the main differences. First, as already noted, while the IKS at $|\nu| = 3$ has intervalley coherence between states with different spins, we can always decompose a $|\nu| = 2$ IKS state as two copies of spin collinear IKS states, each of which exhibits intervalley coherence within a single spin sector. [Note that putting two copies of spin-valley-locked $\nu = -3$ IKS together results in a $\nu = -2$ IKS that admits a spin collinear decomposition along an appropriately chosen axis.] This also means that the picture of ``ferromagnetic" and ``antiferromagnetic" IKS relevant to $|\nu| = 3$ is no longer appropriate for $|\nu|=2$, as the spin polarization in each valley always vanishes. Instead, the key parameter is $2\gamma$, the difference in IVC phase in the two spin sectors. Second, for $|\nu| = 3$ and for the intervalley processes considered here, both the Hartree and Fock contributions are relevant, with the latter found to dominate numerically. For $|\nu| = 2$, the Fock contribution is constant, and therefore only the Hartree contribution is non-trivial. As such, the energetic considerations at $|\nu| = 2$ and $|\nu| = 3$ are very different.

As in the case of $|\nu| = 3$, we find electron-phonon coupling to dominate over intervalley Coulomb scattering in a realistic system at $|\nu| = 2$ (see SM Sec.~\ref{app:additional_numerical_integer}). This leads to a spin singlet state for $|\nu|= 2$ that maximizes the charge density Kekul\'e pattern, consistent with the observation of such patterns in STM~\cite{Nuckolls_2023, Kim_2023}. Alternatively, since estimates of $g$ and $V_{\text{inter}}$ vary, one can use the experimental observation of charge density Kekul\'e pattern at $|\nu| = 2$ as a means to bound the ratio $g/V_{\text{inter}}$, which can be then used to infer the ground states at other fillings such as $|\nu| = 3$. [However, there is a small caveat to this: at $|\nu| = 2$ the  phonon-dominated regime is $g/V_{\text{inter}} > 0.094$, but at $|\nu| = 3$ the threshold is $g/V_{\text{inter}} > 0.12$, as discussed in SM Sec.~\ref{subsec:competition}.]

\subsection{Applied Fields}
\begin{table*}
\begin{tabular}{| c|c|c|c|c|} 
 \hline
 Perturbation& Symmetries of Hamiltonian & Ground state manifold\textsuperscript{\textdagger}& Broken generators&Nambu-Goldstone modes\\ 
 \hline
 Electron-Phonon Coupling & $\tau_z$, $s_x$, $s_y$, $s_z$ & $\hat{\bm{n}}_+ = - \hat{\bm{n}}_-$& $\tau_z,\bm{s}_\perp$\textsuperscript{\textdaggerdbl}& 3 linear \\
  \hline
 Intervalley Coulomb Scattering & $\tau_z$, $s_x$, $s_y$, $s_z$ & $\hat{\bm{n}}_+ = \hat{\bm{n}}_-$& $\tau_z$, $\bm{s}_\perp$\textsuperscript{\textdaggerdbl}& 1 linear, 1 quadratic  \\
 \hline
 Zeeman $s_z$ & $\tau_z$, $s_z$, $\tau_z s_z$ & $\hat{\bm{n}}_+ = \hat{\bm{n}}_- = \hat{\bm{z}}$& $\tau_z$& 1 linear\\
 \hline
 Ising SOC $\tau_z s_z$ & $\tau_z$, $s_z$, $\tau_z s_z$ &  $\hat{\bm{n}}_+ = -\hat{\bm{n}}_- = \hat{\bm{z}}$& $\tau_z$& 1 linear\\
 \hline
\end{tabular}
\caption{Summary of effects of perturbations at $\nu=\pm 3$. Note that without any perturbations, the IKS would have 1 linear and 2 quadratic Nambu-Goldstone modes. \textdagger: $\hat{\bm{n}}_+,\hat{\bm{n}}_-$ indicate the direction of spin polarization in valleys $K,K'$ respectively. The ground state manifold also includes an additional U(1) coordinate parameterizing the in-plane valley angle. \textdaggerdbl: $\bm{s}_\perp$ refers to the two spin operators perpendicular to $\hat{\bm{n}}_\pm$.}\label{tab:summary_nu3}
\end{table*}

\begin{table*}
\begin{tabular}{|c|c|c|c|c|} 
 \hline
 Perturbation& Symmetries of Hamiltonian & Ground state manifold\textsuperscript{\textdagger}& Broken generators&Nambu-Goldstone modes\\ 
 \hline
 Electron-Phonon Coupling& $\tau_z$, $s_x$, $s_y$, $s_z$ & $e^{i\delta}$& $\tau_z$ & 1 linear \\
  \hline
 Intervalley Coulomb Scattering& $\tau_z$, $s_x$, $s_y$, $s_z$ & $e^{i\delta} e^{i\frac{\pi}{2}(\hat{\bm{n}} \cdot \bm{s})}$& $\tau_z$, $\bm{s}_\perp$\textsuperscript{\textdaggerdbl}& 3 linear  \\
 \hline
 Zeeman $s_z$ & $\tau_z$, $s_z$, $\tau_z s_z$ & $e^{i\delta}e^{i\frac{\pi}{2}(\hat{\bm{n}}_{xy} \cdot \bm{s})}$\textsuperscript{\textdaggerdbl\textdaggerdbl, \S}& $\tau_z$, $s_z$ & 2 linear
  \\
 \hline
 Ising SOC $\tau_z s_z$ & $\tau_z$, $s_z$, $\tau_z s_z$ &  $e^{i\delta}e^{i\gamma s_z}$\textsuperscript{\S}& $\tau_z$, $\tau_zs_z$& 2 linear
  \\
 \hline
\end{tabular}
\caption{Summary of effects of perturbations at $\nu=\pm 2$. Note that without any perturbations, the IKS would have 4 linear Nambu-Goldstone modes. \textdagger: The ground state is given by $P(\bm{k};V)$ according to Eq.~\ref{eq:PV}, where $V$ is the entry listed in the table. \S: The expressions only hold in the limit of infinitesimal perturbation. \textdaggerdbl: $\bm{s}_\perp$ refers to the two spin operators perpendicular to $\hat{\bm{n}}$.  \textdaggerdbl\textdaggerdbl: $\hat{\bm{n}}_{xy}$ refers to a unit vector in the $xy$-plane.}\label{tab:summary_nu2}
\end{table*}

So far, we have considered how the intervalley processes from various interaction terms fix the spin structure of the IKS state. With this in hand, we can now consider the response of resulting state(s) to applied fields or other extrinsic single-particle terms enter the  Hamiltonian. Two such terms are of particular interest: a Zeeman splitting  $\frac{1}{2}\Delta_{s_z}s_z$ (without spin-orbit coupling, the Zeeman axis is arbitrary; for definiteness we take it to be along $s_z$), and (b) an Ising-like spin orbit coupling (SOC) term $\frac{1}{2}\Delta_{\tau_zs_z}\tau_zs_z$. Zeeman splitting can be generated by applying a magnetic field. However, as we are not considering the orbital effects of magnetic field here, the discussion is most relevant to the case of in-plane magnetic field in twisted symmetric trilayer graphene (TSTG); although closely related to TBG, unlike TBG~\cite{kwan_twisted_2020} for symmetry reasons this material  only has a negligible orbital coupling to in-plane magnetic fields~\cite{lake_reentrant_2021,qin_-plane_2021}. SOC, on the other hand, can be induced by proximity-coupling to a transition-metal dichalcogenide, where $\Delta_{\tau_zs_z}$ typically reaches a few meV~\cite{sun_determining_2023}.

\begin{figure}
    \centering
    \includegraphics[width=\linewidth]{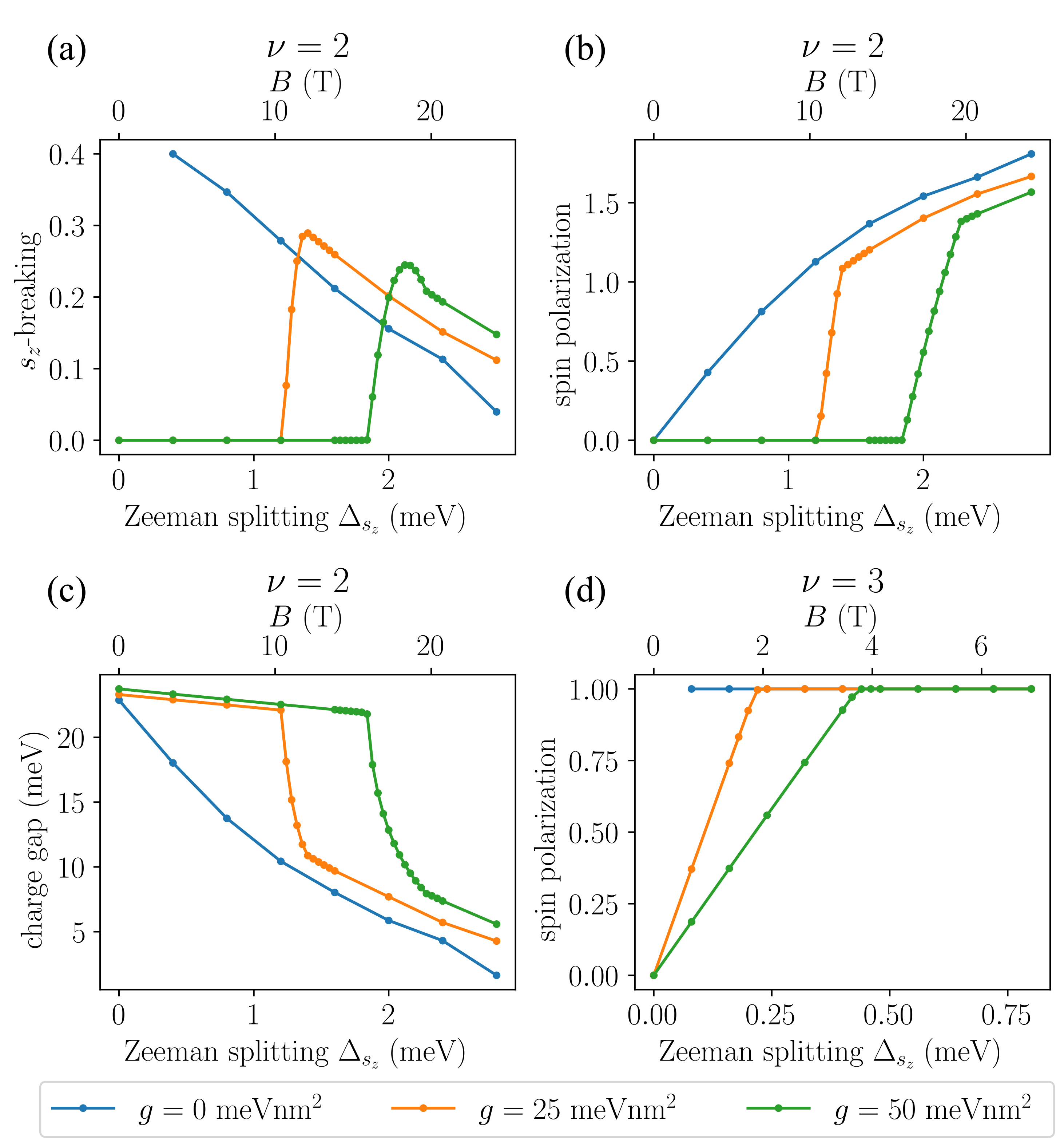}
    \caption{(a) Degree of breaking of the U(1) symmetry generated by $s_z$, defined as $\norm{P_{\uparrow\downarrow}}^2/(N_1N_2)$ for a $N_1 \times N_2$ system (the vertical bars denote Frobenius norm), (b) spin polarization, and (c) charge gap as a function of Zeeman splitting for different electron-phonon coupling strengths at $\nu = 2$. (d) Spin polarization as a function of Zeeman splitting at $\nu = 3$. Results are computed using $10 \times 10$ HF with $\epsilon = 0.3\%$, $w_{\text{AA}} = 80$~meV, and $\theta = 1.05^\circ$. The IKS vector $\bm{q}_{\text{IKS}} = 0.5\bm{G}_1$ for all of the states shown.}
    \label{fig:spinpol_g}
\end{figure}

For $|\nu| = 3$, the effects of both types of applied field are straightforward to understand: the Zeeman splitting aligns the spins in both valleys along the field, whereas the Ising SOC anti-aligns them. As discussed in Sec.~\ref{subsec:nu3}, in the absence of applied fields we expect the ground state of realistic systems to be a valley anti-ferromagnet due to electron-phonon coupling, which dominates over the intervalley Coulomb scattering. Upon applying a Zeeman field, the magnetization axes of the spins will progressively cant towards the external field axis until they are parallel to it, at which point the magnetization saturates (Fig.~\ref{fig:spinpol_g}d). On the other hand, both Ising SOC and electron-phonon coupling favor valley anti-ferromagnet, so the addition of Ising SOC to a realistic system only serves to select out a particular direction for the valley anti-ferromagnetic order. These effects are listed in Table~\ref{tab:summary_nu3}.

\begin{figure}
    \centering
    \includegraphics[width=0.9\linewidth]{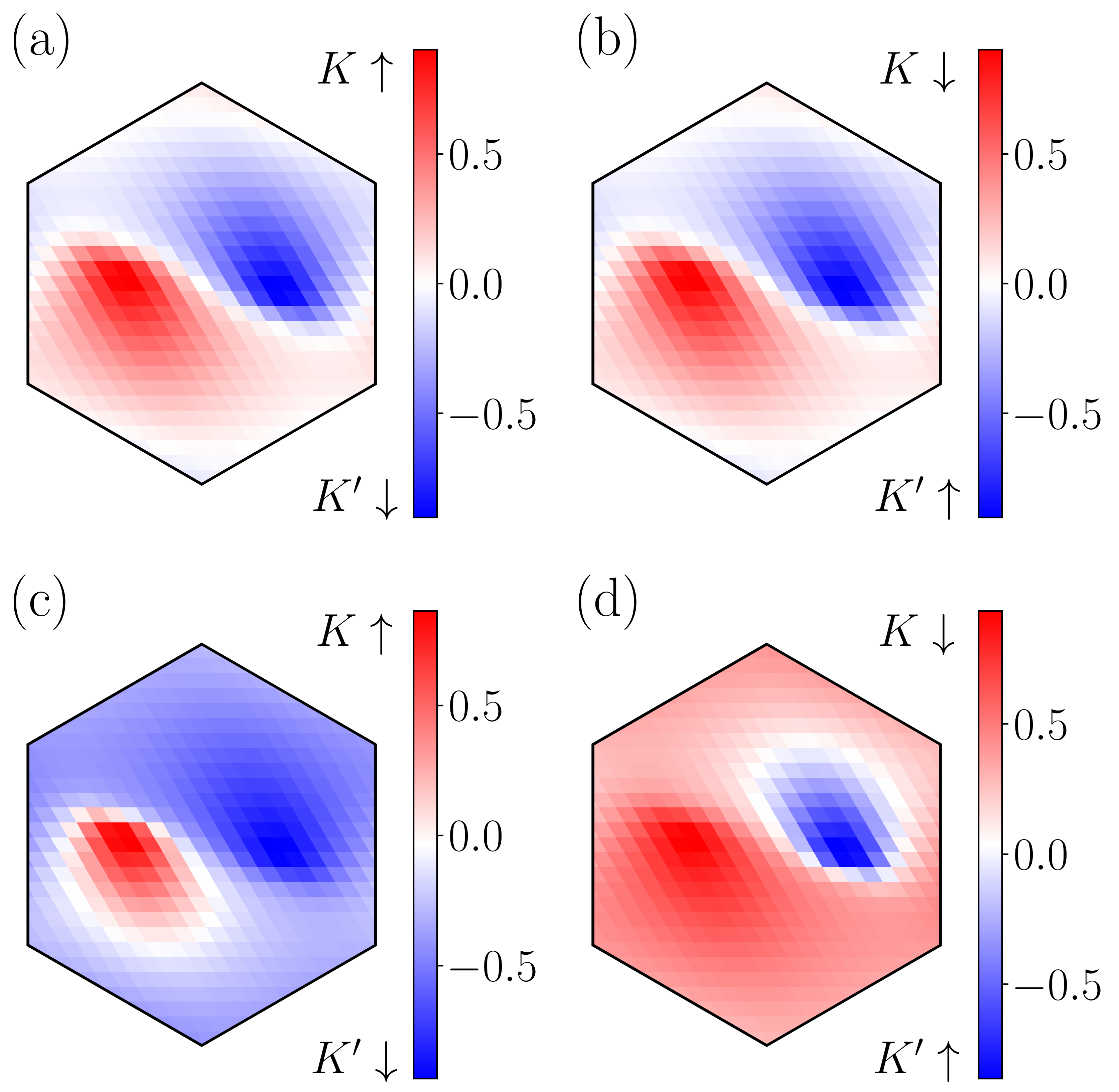}
    \caption{(a),(b) Valley/spin polarization of unoccupied bands of the IKS state at $\nu = 2$ without Zeeman splitting, showing $s_z\tau_z = \pm 1$ sectors respectively. We have chosen the state that would be selected out by infinitesimal Zeeman splitting from the degenerate manifold. (c),(d) 
    Same as (a),(b) except with Zeeman splitting $\Delta_{s_z} = 0.8$\,meV. The system develops a net spin polarization but no net valley polarization. Results are computed using $20 \times 20$ HF with $\epsilon = 0.3\%$, $w_{\text{AA}} = 80$~meV, $\theta = 1.05^\circ$, including 2 bands per spin and valley sector, and without non-local tunneling. The IKS vector is given by $\bm{q}_{\text{IKS}} = 0.5\bm{G}_1$.}
    \label{fig:zeeman}
\end{figure}

More careful analysis is required to understand the effects of of applied fields on the IKS ground state at $|\nu| = 2$. Consider for the moment switching off the $\text{SU}(2)_K\times \text{SU}(2)_{K'}$-breaking interaction terms. From the perspective of the simplest $|\nu| = 2$ spin singlet IKS, the expectation is that a Zeeman term $\propto s_z$ will just shrink the gap and leave the density matrix unchanged. However, we find that the actual ground state exhibits a spontaneous breaking of $\tau_z$ (since it has IVC) and $s_z$, but the combined symmetry $\tau_z s_z$ is preserved, i.e. separate charge conservation in the $\tau_zs_z = +1$  ($K\uparrow $ and $ K'\downarrow$) and $\tau_zs_z = -1$  ($K\downarrow $ and $ K'\uparrow$) sectors. Application of $\Delta_{s_z}$ smoothly deforms the bands in each sector, as shown in Fig.~\ref{fig:zeeman}, leading to a net spin polarization $\langle s_z\rangle$ that grows linearly with applied field (see SM Sec.~\ref{app:additional_numerical_integer}).  This is favorable compared to a state that preserves $s_z$ symmetry, which can only develop spin polarization by moving electrons across the charge gap. The paramagnetic response can be traced back to the existence of gapless spin-valley Goldstone modes at $\Delta_{s_z}=0$, as discussed below in Sec.~\ref{sec:collective}. In the zero field limit, the preferred ground state can be written as $P(\bm{k};V)$ where $V = e^{i\delta}e^{i\frac{\pi}{2}(\hat{\bm{n}}_{xy} \cdot \bm{s})}$ for some unit vector $\hat{\bm{n}}_{xy}$ in the $xy$-plane. However, we emphasize that all such $P$ have vanishing spin polarization, so the ground state at finite $\Delta_{s_z}$ does not admit such a parameterization. For the case of applied Ising SOC, $\tau_z$ and $\tau_z s_z$ are spontaneously broken while $s_z$ is preserved, such that $\braket{\tau_z s_z}$ increases linearly with applied $\Delta_{\tau_zs_z}$. The results are summarized in Table.~\ref{tab:summary_nu2}.

Now we consider a more realistic system with electron-phonon coupling and under Zeeman splitting (intervalley Coulomb scattering is left out as its effects are assumed to be dominated by electron-phonon coupling). Referring to Table~\ref{tab:summary_nu2}, we note that electron-phonon coupling favors ground states that preserve all global spin rotation symmetries, but Zeeman splitting favors states that spontaneously break $s_z$ symmetry. With a finite electron-phonon coupling strength, we find that $s_z$ symmetry remains unbroken up to a finite $\Delta_{s_z}$ (Fig.~\ref{fig:spinpol_g}a). As noted previously, if $s_z$ is preserved, no magnetization is expected to develop unless the Zeeman splitting is large enough to overcome the charge gap of the order of 10\,meV. In reality, beyond some finite threshold well below the charge gap, the Zeeman perturbation leads to the spontaneous breaking of $s_z$ symmetry and development of spin polarization (Fig.~\ref{fig:spinpol_g}b).

The non-trivial dependence of spin polarization on Zeeman field also leads to non-trivial behavior of other properties, such as the charge gap, as shown in Fig.~\ref{fig:spinpol_g}c. For finite electron-phonon coupling, the charge gap first decreases linearly with externally applied field, corresponding to the intrinsic $g_s$-factor of electronic spins. The spontaneous breaking of $s_z$ symmetry and development of spin polarization beyond some finite threshold leads to a significant distortion of the band structure and rapid decrease of the charge gap. Similar `kinks' in the charge gap as a function of external field have been experimentally observed~\cite{yu_spin_2023, zhang_heavy_2025}. However, one should be cautious in interpreting this similarity, since for both out-of-plane direction as in the experiments and in-plane direction, magnetic field also couples to orbital degrees of freedom in TBG~\cite{kwan_twisted_2020}. 

\section{Phonons at non-integer filling}\label{sec:nonint}

\begin{figure}
    \centering
    \includegraphics[width=\linewidth]{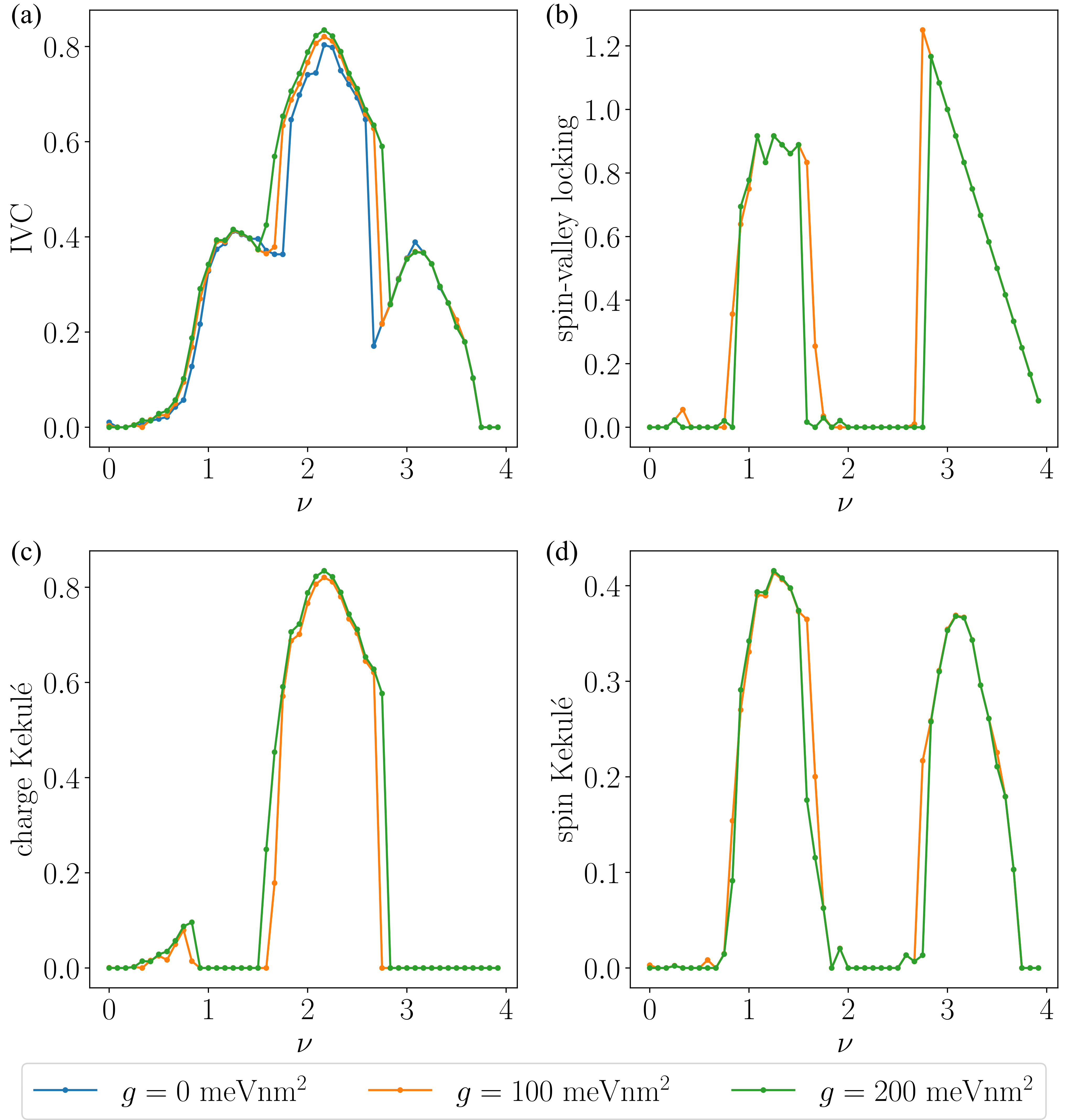}
     \caption{(a) IVC order parameter, (b) spin-valley locking, strength of (c) charge, and (d) spin Kekul\'e patterns (see the text for the definitions used) of the Hartree-Fock ground states for different values of the electron-phonon coupling constant $g$. The system size used is $12\times12$ and the magnitude of strain is $0.5\%$. We have omitted $g = 0$ for (b,c,d), since $\text{SU}(2)_K\times \text{SU}(2)_{K'}$ symmetry implies that there are no definite values for the plotted quantities.}
    \label{fig:ivc_nonint_04}
\end{figure}

In the previous sections, we have studied the effects of electron-phonon coupling on IKS states at integer fillings $\nu = \pm 2$ and $\nu = \pm 3$. Here, we study the effects at non-integer fillings, focusing on how the IVC strength is influenced by the electron-phonon coupling $g$. In Fig.~\ref{fig:ivc_nonint_04}a, we show the IVC strength, defined as $2\norm{P_{+-}}^2/(N_1N_2)$ for an $N_1 \times N_2$ system ($\norm{\cdot}$ denotes the Frobenius norm), in the filling interval $0 < \nu < 4$ (the results for negative $\nu$ are similar as the model has approximate particle-hole symmetry). Regardless of electron-phonon coupling, we observe peaks in IVC at integer fillings $\nu =2,3$ and around $\nu = 1$, with the peak at $\nu = 2$ about twice as high as those at or around $\nu = 1, 3$. This is consistent with the notation that $|\nu| = 2$ IKS states consist of two copies of $|\nu| = 3$ IKS states, as elaborated in Sec.~\ref{sec:spinvalley}. We note that the ground state at $\nu = 1$ does not have a finite charge gap~\cite{Kwan2021}.

In order to capture the effect of phonons, we introduce an indicator of the strength of the charge Kekul\'e pattern given by $\norm{\text{tr}_sP_{+-}}^2/(N_1N_2)$ ($\text{tr}_s$ is the trace over the spin indices), and that of spin Kekul\'e pattern, given by $\sum_i\norm{\text{tr}_s(s_iP_{+-})}^2/(N_1N_2)$. With the inclusion of electron-phonon coupling, the IVC around $\nu = 2$ is strengthened. This is expected, as coupling to zone-corner phonons favors a Kekul\'e charge density of the IKS~\cite{Kwan2024}, as discussed for $|\nu|=2$ in Sec.~\ref{subsec:nu2}, and this extends to non-integer fillings around $\nu = 2$, as shown in Fig.~\ref{fig:ivc_nonint_04}c\footnote{Strictly speaking, the measure we have introduced does not distinguish between charge density and charge current order. Nevertheless, since the states at and near $\nu = 2$ preserves the physical time-reversal symmetry, the Kekul\'e pattern must be of charge density character.}. As discussed in Sec.~\ref{subsec:nu3}, due to the antiferromagnetic nature of electron-phonon coupling, we found the IKS state at $|\nu|=3$ to be spin-valley locked, leading to no charge density Kekul\'e pattern at the mean-field level (but we expect non-vanishing charge density order beyond mean field, as explained in Sec.~\ref{subsec:nu3}). Instead, we find non-vanishing spin Kekul\'e pattern. As shown in Fig.~\ref{fig:ivc_nonint_04}b,d, non-integer fillings near $\nu = 3$ also have significant spin-valley locking (defined as $|\text{tr}(\bm{s}\tau_zP)|/(N_1N_2)$ for an $N_1 \times N_2$ system) and exhibit a spin Kekul\'e pattern. As shown in Fig.~\ref{fig:ivc_nonint_04}b, spin-valley locking is also present around $\nu = 1$. Since electron-phonon coupling does not directly couple to the spin-valley locked IVC, we do not observe a similarly marked increase of the IVC strength near $\nu = 1,3$. We note a first order transition at about $\nu_c \approx 2.6$ (the value shifts closer to 3 with larger electron-phonon coupling) that separates non-integer filling states with same symmetry as $\nu = 2$ and states with the same symmetry as $\nu = 3$. This is a relevant threshold that will feature in later discussions. A similar threshold with $\nu_c \approx -2.6$ exists on the hole-doped side.

In magic-angle twisted symmetric trilayer graphene (TSTG), a material closely related to TBG, the Kekul\'e charge density pattern as detected in STM has a maximal intensity near $|\nu| = 2$~\cite{kimResolvingIntervalleyGaps2025} (though some samples show peaks closer to $\nu = -2.5$ and particle-hole asymmetry~\cite{Kim_2023}), which is broadly consistent with our calculation in TBG showing maximal IVC at $|\nu| = 2$. No secondary peak in the intensity of charge density Kekul\'e pattern is observed near $|\nu| = 1,3$, which is also consistent with our calculation, since near those fillings we expect predominantly spin Kekul\'e patterns, which are invisible to STM measurements. However, we do not expect the charge density Kekul\'e pattern to vanish completely, as discussed in Sec.~\ref{sec:spinvalley}. Indeed, in Ref.~\cite{Nuckolls_2023}, for strained TBG, the Kekul\'e density pattern is observed in the entire range of $2 \leq \nu \leq 3$.

In SM Sec.~\ref{App:comparing_methods}, we perform similar calculations using the `direct distortion' method of incorporating the phonons. The observation of strengthened IVC near $\nu = 2$ with increasing $g$ is unchanged, though details can differ. We also present data with a smaller strain magnitude in SM Sec.~\ref{app:additional_numerical_integer}.

\section{Collective modes}\label{sec:collective}

The $\text{SU}(2)_K\times \text{SU}(2)_{K'}$ breaking perturbations described in Sec.~\ref{sec:spinvalley} select out states from the degenerate manifold, which impacts the ground state charge and spin density patterns on the graphene scale. In this section, we discuss how the lifting of degeneracies affects the collective modes about these states, with a particular focus on the Nambu-Goldstone modes (NGM) arising from spontaneously broken continuous symmetries.

To calculate the number of NGMs, we use the method discussed in Ref.~\cite{watanabeannualrev}, outlined below. For a set  of linearly independent spontaneously broken generators $\hat{Q}_i$ for $i = 1, ... ,n_{\text{BG}}$, one defines a real antisymmetric matrix $\rho$ such that
\begin{equation}\label{eq:rho}
    \rho_{ij} \propto -i\braket{[\hat{Q}_i, \hat{Q}_j]},
\end{equation}
where the expectation value is taken with respect to the symmetry-broken ground state. The number of NGBs is given by
\begin{equation}n_{\text{NGM}} = n_{\text{BG}} - \frac{1}{2} \text{rank} \rho.\end{equation}
The NGMs can be further divided into Type-A and Type-B, whose counting satisfies
\begin{equation}
n_A + 2n_B = n_{\text{BG}}.
\end{equation}
Barring systems with fine-tuned parameters, Type-A NGMs have linear dispersion and Type-B NGMs have quadratic dispersion.

Before including the perturbations, we first consider the NGMs of IKS states in the Hamiltonian with full $\text{SU}(2)_K\times \text{SU}(2)_{K'}$ symmetry. While the broken symmetry generators depend on which broken symmetry state from the degenerate manifold is considered, they yield equivalent excitation spectra. As such, we may analyze the most convenient choice. 

At $|\nu| = 2$, we pick the spin singlet ground state which spontaneously breaks $\tau_z$, as well as $\tau_z s_i$ for $i = x, y, z$. It can be straightforwardly shown that the matrix $\rho$ in Eq.~\ref{eq:rho} is identically zero. Therefore, there are four linear NGMs, equal to the number of broken generators.  
When electron-phonon coupling is taken into account, the spin singlet ground state is selected.  $\tau_z s_i$ are now explicitly broken and therefore are no longer broken generators. The only remaining broken generator is $\tau_z$, and this leads to one linear NGM, with small gaps for the other three former NGMs. The effects of various other perturbations and their effects on the number of NGMs are summarized in Table~\ref{tab:summary_nu2}.

At $|\nu|=3$, in the absence of perturbations, we pick the fully ferromagnetic IKS corresponding to $\hat{\bm{n}}_\pm=\hat{z}$ in Eq.~\ref{eq:psi_k}, which breaks the independent generators $\tau_z,\tau_zs_x,\tau_zs_y,s_x,s_y$. Since the state has non-zero $\langle s_z\rangle$, it can be shown that there are 1 linear and 2 quadratic NGMs. Table~\ref{tab:summary_nu3} summarizes the effects of the perturbations on the NGM counting for $|\nu| = 3$.

\begin{figure}
    \centering
    \includegraphics[width=\linewidth]{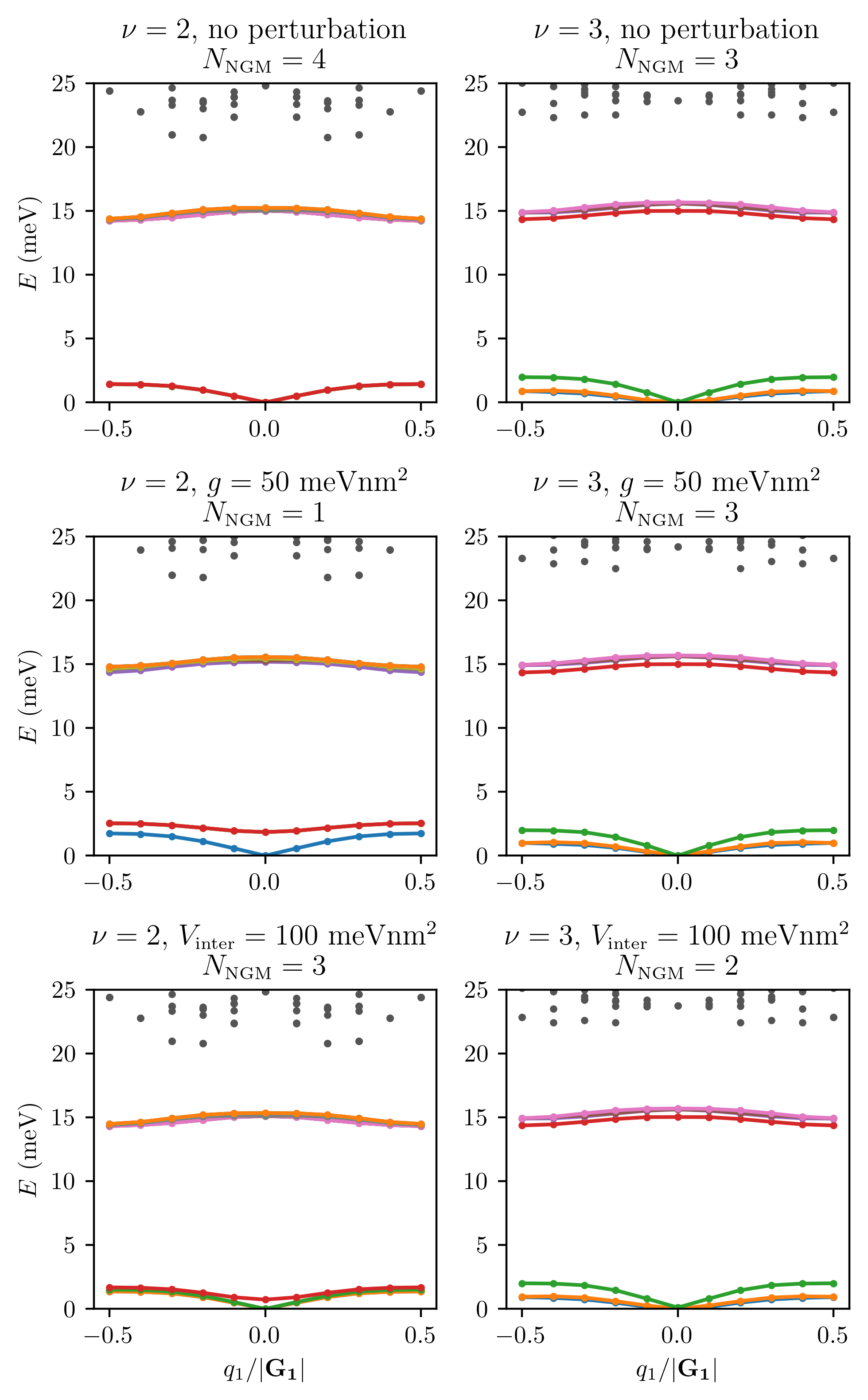}
    \caption{Neutral excitation spectra at $\nu = 2$ and $\nu = 3$ for IKS states with no perturbation (top row), with finite electron-phonon coupling (middle row), or with finite intervalley Coulomb scattering (bottom row). Collective excitations are shown with solid lines and particle-hole excitations are shown with gray dots. The plots show the spectra along the $\bm{G}_1$ direction in the mBZ with $q_2 = 0$ (additional data is presented in SM Sec.~\ref{app:additional_numerical_integer}). The results are obtained from $10 \times 10$ TDHF with $0.3\%$ strain, including 2 bands per spin/valley, and not including the effects of non-local tunneling. The ground state IKS has IVC at $\bm{q}_{\text{IKS}}= 0.5\bm{G}_1$. The numbers of gapless modes agree with theoretical expectations. }\label{fig:collective}
\end{figure}

We further numerically compute the neutral excitation spectra of these states using time-dependent Hartree-Fock (TDHF) theory, as explained in the App.~\ref{app:TDHF}. The results of the calculations for $\nu=2,3$ with or without electron-phonon coupling and intervalley scattering are shown in Fig.~\ref{fig:collective}. The total numbers of gapless modes agree with predictions. In SM Sec.~\ref{app:int_collective_modes}, we show results from larger-scale calculations to confirm the counting of linear and quadratic modes separately, as well as results under external fields.
In App.~\ref{app:NLSM_nu2}, we construct a non-linear sigma model (NLSM) to better understand the NGMs at $|\nu| = 2$. In particular, we show that the velocity of the valley U(1) mode is proportional to the square root of the $\bm{q}_{\text{IKS}}$-stiffness of the IKS state. 

Besides the NGMs, we observe an additional set of gapped collective modes at about $E \approx 15$~meV for both $\nu = 2$ and $\nu = 3$, below the particle-hole continuum. There are 8 such modes at $\nu = 2$, and 4 such modes at $\nu = 3$. Including the NGMs (and their gapped versions under perturbations), there are in total 12 collective modes at $\nu = 2$ and 7 collective modes at $\nu = 3$. This is consistent with the counting of $16 - \nu^2$ soft modes from the broken U(8) symmetry of the 8 central TBG bands~\cite{khalafSoftModesMagic2020}. We note that, even without the perturbations discussed in this section, the BM Hamiltonian already explicitly breaks the U(8) symmetry. For 0.3\% strain, the Brillouin-zone averaged energy difference of the central bands is about 15~meV, very close to the energy of this additional set of collective excitations\footnote{For systems without strain, the dominant U(8)-breaking term is the difference between inter-Chern and intra-Chern coupling instead of the band dispersion~\cite{khalafSoftModesMagic2020}.}. Finally, we remark that, due to the presence of remote bands and beyond mean-field effects, the actual charge gap is likely to be much smaller than the $\sim$20~meV shown in Fig.~\ref{fig:collective}. As such, this additional set of collective modes may overlap with particle-hole continuum in a real system.

\begin{figure}[t]
    \centering
    \includegraphics[width=\linewidth]{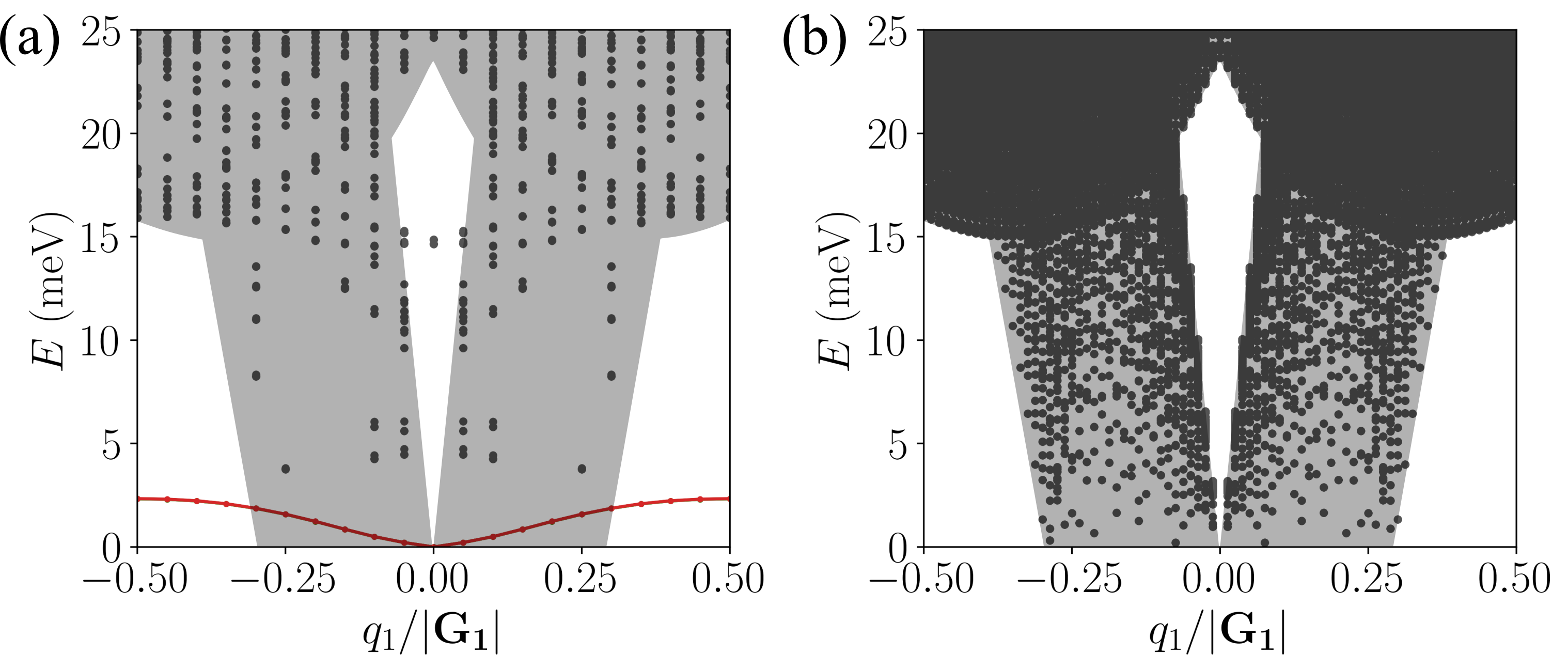}
    \caption{(a) Neutral excitation spectra at $\nu = 2.16$ along the $\bm{G}_1$ direction in the mBZ with $q_2=0$. The results are obtained from $20 \times 20$ TDHF, with $0.3\%$ strain, 2 bands per spin/valley and without including non-local tunneling or electron-phonon coupling. The solid lines denote low-energy collective excitations, and gray dots denote other excitations. The shaded region is the particle-hole continuum, as extracted from (b). We observe Nambu-Goldstone modes that are qualitatively the same as those for $\nu = 2$. (b) The particle-hole excitation spectrum extracted from the $80 \times 80$ interpolated Hartree-Fock band structure at the same parameters. The shaded region represents the particle-hole continuum extracted from the data.}
    \label{fig:coll_nonint}
\end{figure}

Fig.~\ref{fig:coll_nonint}a shows the neutral excitation spectrum at $\nu = 2.16$ in the $\text{SU}(2)_K\times \text{SU}(2)_{K'}$ limit. The ground state is a doped version of the $\nu = 2$ IKS with the same set of broken symmetries. This leads to a low-energy collective mode spectrum that is essentially the same as that at $\nu = 2$. However, in contrast with the case of integer fillings, here the particle-hole excitations are not gapped, though the finite grid size in momentum space leads to an apparent `gap' for the particle-hole excitations. To better study these particle-hole excitations, we compute the particle-hole excitation spectrum from an interpolated HF band structure and map out the boundaries of particle-hole continuum, as shown in Fig.~\ref{fig:coll_nonint}b. While the small-momentum part of the low-energy collective modes overlaps with the particle-hole continuum, they may remain relatively undamped since the latter are intra-band excitations that do not carry significant flavor character and hence couple weakly to NGMs. The U(1) valley mode has been invoked to explain anomalously long relaxation times in optical pump probe experiments in the vicinity of $|\nu|=2$~\cite{xie_long-lived_2024}.

An in-plane magnetic field in TBG appears to quench the isosopin ordering, as quantified by the reduction of electronic entropy~\cite{Saito:2021aa,Rozen:2021aa}, in a broad range of fillings, related to the so-called ``isospin Pomeranchuk effect''. As discussed in previous sections, IKS states at and around $\nu = 3$ have finite spin polarizations in each of the two valleys, which are placed into  a spin-valley locked configuration since the electron-phonon coupling favours an anti-ferromagnetic alignment. This is also true near $\nu = 1$, as seen in Fig.~\ref{fig:ivc_nonint_04}. An in-plane magnetic field aligns the spin moments along the field, thereby quenching isospin fluctuations and potentially explaining the experimental findings near $\nu=1$. For fillings near $\nu = 2$, a more careful analysis is required, as we now sketch. In the SU(2)$_K\times$SU(2)$_{K'}$ limit, the IKS state at $|\nu| = 2$ has 4 NGMs (see Tab.~\ref{tab:summary_nu2}). With electron-phonon coupling, the NGM arising from valley U(1) symmetry-breaking remains gapless, while the 3 NGMs due to valley-spin fluctuations develop a small gap. However, since the experimental temperature (about 10 K in Ref.~\cite{Rozen:2021aa}) is similar to the electron-phonon coupling strength, we expect that these fluctuations remain active and contribute to the entropy of the system. With applied Zeeman field, two of these spin-valley fluctuations are further quenched\footnote{The neutral modes related to the $\tau_z$ and $\tau_zs_z$ generators are unaffected by the Zeeman field.}, leading to a reduction in entropy. Note that strictly speaking, there are no Goldstone modes at $T > 0$ due to the absence of continuous symmetry breaking at finite temperature. However, the Goldstone mode count should still  quantify the  number of soft fluctuations expected to contribute electronic entropy. The arguments above apply equally to non-integer fillings around $|\nu| = 2$ up to the transition at $\nu_c$, since for these fillings the doped IKS retains the same symmetries as at $|\nu|=2$.

\section{Phenomenological Implications for superconductivity}\label{sec:sc}
We now turn to an assessment of (doped) IKS states as possible parent states for superconductivity in TBG in the filling range $-3 < \nu < -2$. We note that TSTG under zero or small perpendicular displacement field exhibits very similar physics to TBG, so the discussion is applicable there as well. As microscopic calculations of superconductivity are challenging, we consider the phenomenological aspects of the experimentally observed superconductors, and how they may relate to the IKS. This pursuit is in part motivated by a very recent report of a ``double gap" in the STM spectrum~\cite{kimResolvingIntervalleyGaps2025} for TSTG, where the ``outer gap" is attributed to the intervalley coherence order and the ``inner gap" is attributed to superconductivity, suggesting the coexistence of the two orders. The coexistence of two energy gaps is also observed in Ref.~\cite{park_simultaneous_2025}.

Previous work~\cite{zhou2024doubledomeunconventionalsuperconductivitytwisted}, involving four of the present authors, examined superconductivity in TSTG, focusing on the double-dome feature at finite displacement field. The discussion in this section adds to the previous work in two ways. First, by including $\text{SU}(2)_K\times \text{SU}(2)_{K'}$-breaking terms, such as intervalley phonons and intervalley Coulomb scattering, we have now gained a better understanding of the spin structure of the non-superconducting ground state of TBG. From this, we can make more informed deductions about the spin structure of the superconducting state that emerges from the normal state. Second, the scope of the following discussion will be somewhat different from the previous work. At zero or low displacement field of TSTG and within the range $-3 \lesssim \nu < -2$, there is a critical filling $\nu_c$ that separates region of weaker superconductivity ($\nu < \nu_c$) and stronger superconductivity ($\nu > \nu_c$), and the focus was on the stronger side. Nevertheless, the superconductivity does not completely vanish on the weaker side, and we will include it in our discussion below.

\begin{figure}[h!]
    \centering
    \includegraphics[width=\linewidth]{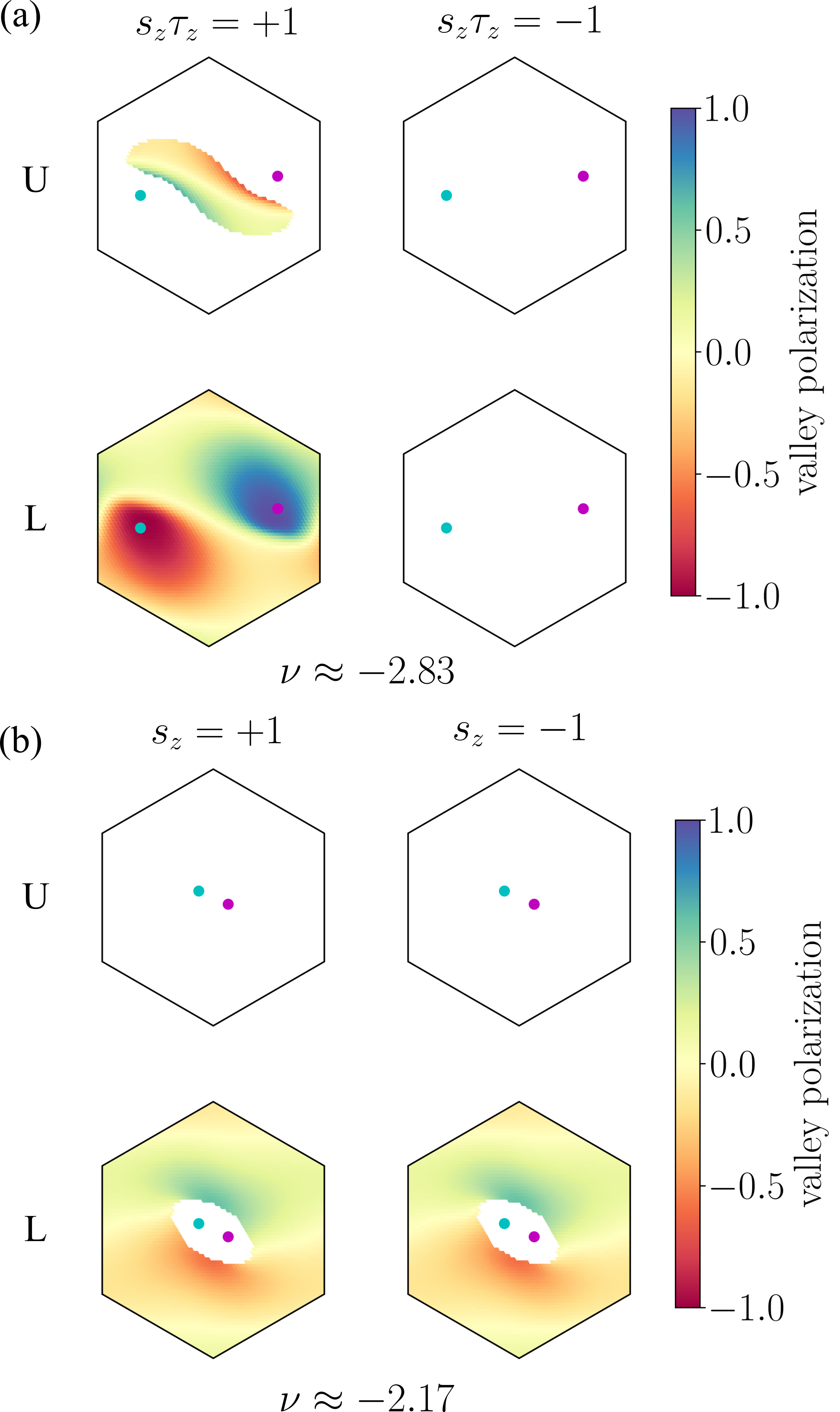}
    \caption{Fermi surfaces of the doped IKS states for (a) $\nu < \nu_c$ and (b) $\nu > \nu_c$. For (a), the state is spin-valley locked and we show the electron occupation for each $s_z\tau_z$ sector (for definiteness we have assumed $\braket{\bm{s}\tau_z}$ is along the $z$-direction). For (b), the state is spin singlet and we show the electron occupation for the two identical spin sectors. We label the lowest band in each charge sector as L and the second lowest band as U. The color plots show valley polarization of occupied orbitals, whereas white spaces indicate unoccupied orbitals. Cyan and magenta dots indicate the $\gamma$-points of valley $K$ and $K'$ respectively (due to finite $\bm{q}_{\text{IKS}}$, the two $\gamma$-points do not align). We performed $12 \times 12$ Hartree-Fock with $0.3\%$ strain, $g = 70$~meVnm\textsuperscript{2} and $V_{\text{inter}}= 100$~meVnm\textsuperscript{2}. The data are then interpolated to $60 \times 60$ with HF interpolation scheme.}
    \label{fig:fermi_surface}
\end{figure}

Within the filling range of $-3 < \nu < -2$, mean field calculations show a first order transition signaled by an abrupt jump in the intervalley coherence and spin-valley locking order parameter,\footnote{The Hamiltonian used in this calculation is approximately particle-hole symmetric.} at a critical density $\nu_c$ as shown in Fig.~\ref{fig:ivc_nonint_04}. Our  numerical calculations find $\nu_c \approx-2.6$ in the limit of vanishing electron-phonon coupling (although it can depend on the magnitude of strain and values of other parameters used), and it evolves towards  $-3$ with increasing electron-phonon coupling. The critical density $\nu_c$ separates two distinct behaviors. For $\nu_c < \nu < -2$, the state is a doped version of IKS at $\nu = -2$, with unbroken global spin rotation symmetry SU(2)$_s$ (assuming non-zero electron-phonon coupling). At $-3 < \nu < \nu_c$, the state is a doped version of IKS at $\nu = -3$, with finite spin-valley locking. The symmetry properties of the ground states are also reflected in the symmetries of the Fermi surfaces. In Fig.~\ref{fig:fermi_surface}, we show typical Fermi surfaces for both regimes. For $-3 < \nu < \nu_c$, there is a single spin-valley locked Fermi surface (though we find that the topology of the Fermi surface can change from closed to open across the filling range, as discussed in SM Sec.~\ref{subsec:fermi_surface_extra}), while for $\nu_c < \nu < -2$, there are two identical Fermi surfaces in the two spin sectors.[We will assume that the superconducting order parameters only involve these partially filled Hartree-Fock bands, as the energy scale of the superconducting order is much smaller than the Hartree-Fock energy scale. We provide a more formal justification in SM Sec.~\ref{app:HFB} within the Hartree-Fock-Bogoliubov framework.] We believe that both regimes can potentially produce superconductivity, albeit with possibly different features. There is experimental evidence that supports such a multi-regime picture in TSTG, including the transition from a U-shaped to a V-shaped tunneling gap in Ref.~\cite{kim_evidence_2022}. However, we flag that this remains a matter of ongoing debate, as Ref.~\cite{oh_evidence_2021} only found V-shaped superconducting gaps in TBG, and similarly  Ref.~\cite{park_simultaneous_2025, kimResolvingIntervalleyGaps2025} found only V-shaped superconducting gaps in TSTG.

The authors of Ref.~\cite{lake_pairing_2022} argued for superconductivity originating from a (non-IVC) spin-valley locked parent state in the filling range $-3 < \nu < -2$. The IKS at $-3 < \nu < \nu_c$ similarly has a finite spin-valley locking, though the Fermi surface is contained in a single IKS band with intervalley coherence between opposite spin sectors, as shown in Fig.~\ref{fig:fermi_surface}a, whereas the spin-valley locked state considered in  Ref.~\cite{lake_pairing_2022}  has two bands at the Fermi level related by a simultaneous flip of spin and valley index. Nevertheless, as detailed in the App.~\ref{app:SC_OP_IVC}, we find that the pairing symmetry of a superconductor evolving out of such an IKS parent state shows a mixture of spin singlet and spin triplet characteristics, similar to that considered in Ref.~\cite{lake_pairing_2022}, and much of their discussion applies here as well. In particular, such a state remains consistent with the experimentally observed Pauli limit violation~\cite{cao_pauli-limit_2021, kim_evidence_2022, liu_isospin_2022}, enhancement of superconductivity by spin-orbit coupling (SOC)~\cite{arora_superconductivity_2020}, and quenching of isospin ordering by in-plane magnetic field~\cite{Saito:2021aa,Rozen:2021aa}.

Nevertheless, we caution that the experimental observations mentioned above cannot fully rule out superconductivity from an SU(2)$_s$-symmetric parent state. Violation of the Pauli limit is possible with spin-triplet pairing or strong electron correlations, and the enhancement of superconductivity with SOC does not rule out the co-existence of a different superconducting mechanism in a system without externally applied SOC. Furthermore, as we have argued in Sec.~\ref{sec:collective}, the quenching of isospin order by in-plane magnetic field is also consistent with the doped IKS state at $\nu_c < \nu < -2$, which is SU(2)$_s$ symmetric.

While the pairing symmetry in the $-3 < \nu < \nu_c$ doping range would contain a specific mixture of spin singlet and spin triplet characteristics due to the broken SU(2)$_s$-symmetry, in the $\nu_c < \nu < -2$ doping regime, the pairing could either be spin singlet or triplet as SU(2)$_s$ is preserved in the parent state (see App.~\ref{app:SC_OP_IVC} for more discussion). For $-2.5 < \nu < -2.2$, Ref.~\cite{kim_evidence_2022} observed evidence for a gapped superconductor in TSTG. This may naturally suggest a spin singlet $s$-wave pairing symmetry, though the authors proposed that a superconductor with a nodal order parameter can remain gapped in the strong-coupling regime. On the other hand, Ref.~\cite{oh_evidence_2021} reported signatures of nodal superconductivity in the entire filling range $-3 < \nu < -2$ in TBG. We also note the Pauli-limit violation observed in Ref.~\cite{cao_pauli-limit_2021} for TSTG, suggesting a spin-triplet pairing mechanism instead. However, as we have already remarked, the Pauli limit is only valid for the weakly coupled regime, which may not be appropriate near $\nu = -2$.

In Ref.~\cite{zhou2024doubledomeunconventionalsuperconductivitytwisted}, without fixing the detailed spin structure of IKS states, some of us proposed a modified time-reversal symmetry for the doping regime of $\nu_c < \nu < -2$ and a related generalized Anderson's theorem that protects the superconductivity from the effect of defects at the moir\'e scale defects, but not those at the microscopic graphene scale. From the present work, we see that electron-phonon coupling  selects out the spin-singlet normal ground state and enforces the time-reversal symmetry for $\nu_c < \nu < -2$. As such, $s$-wave superconductors that form from such a parent state would satisfy the standard version of Anderson's theorem and hence be robust against  any form of quenched disorder that preserves time reversal symmetry.

In the above discussion, $\nu_c$ divides the filling range $-3 < \nu < -2$ into two regimes with the same isospin symmetries as the correlated insulators at $\nu = -3$ and $\nu = -2$ respectively. Experimentally, the correlated insulators on the hole-doped side are less robust than those on the electron side, and the above discussion is expected to be most applicable to samples that exhibit insulating behaviors at both $\nu = -2$ and $\nu = -3$~\cite{Cao2018, Lu2019ins, Stepanov_2020, Cao2021a, Saito:2021aa}. In a parallel ongoing work~\cite{ongoing}, we investigate systems that do not form insulators at $\nu = -3$ or $\nu = -2$ by properly modeling the particle-hole asymmetry in TBG.

In TSTG, a perpendicular displacement field provides another experimental tuning knob. Ref.~\cite{zhou2024doubledomeunconventionalsuperconductivitytwisted} experimentally observed that the robustness of the superconductor is displacement field dependent, and proposed a transition between different spin-valley coupled states driven by displacement field within the filling range $-3 < \nu < \nu_c$. Theoretically, the distinction between the two regimes is linked to the flavor sector entered by  electrons doped into the $\nu = -3$ state. At low displacement field, electrons enter into the spin-valley sector that the state is polarized into, similar to TBG, while at higher displacement field, electrons enter into the opposite spin-valley sector. This leads to Fermi surfaces that are contained in different quasi-particle bands for the two cases. Experimentally, the state at higher displacement field is found to host more robust superconductivity.

\section{Conclusions}\label{sec:conclusion}

As we flagged at the outset, this work was prompted by the observation of Kekul\'e orders in STM spectroscopy on samples of MA-TBG/TSTG with both generic and ultralow strain~\cite{Nuckolls_2023,Kim_2023, kimResolvingIntervalleyGaps2025}. The former case, wherein the order is modulated incommensurately on the moir\'e scale, served as an important validation of the theoretical prediction of IKS order~\cite{Kwan2021,Wagner2022} and underscored the importance of strain to the global phase diagram. The latter case, where the order is apparently commensurate and at $\bm{q}=0$, prompted a revisiting of the microscopic model of TBG, and the resulting proposal that incorporating electron-phonon interactions was important to a complete phenomenology of the zero-strain limit~\cite{Blason2022, Kwan2024,Shi_2025}.

In the present work, we have built upon and substantially extended these results: to wit, we have demonstrated that the inclusion of additional perturbations is also central to establishing the {\it spin structure} of the IKS state in strained samples, and furthermore, that a careful analysis of the resulting charge Kekul\'e patterns provides important clues about the relative strength of various perturbations. Specifically, we incorporate coupling to intervalley phonons, intervalley Coulomb scattering, and possibly external fields. The importance of these perturbations is highlighted by the fact that they break the $\text{SU}(2)_K\times \text{SU}(2)_{K'}$ symmetry (i.e. independent rotations of spin in the two valleys) enjoyed by the strained, interacting BM model (which conventionally includes only the long-range piece of the Coulomb interaction).  This motivates a picture of an energy scale hierarchy wherein the dominant terms --- strain and long-range interactions --- select out a manifold of degenerate IKS states related by $\text{SU}(2)_K\times \text{SU}(2)_{K'}$, whose degeneracy is (partially) lifted by the smaller residual terms.

The degeneracy-lifting leads to a range of competing IKS states distinguished most clearly by their spin structures, but with direct implication on the types of graphene-scale density-wave patterns. Notably, we find that electron-phonon coupling and intervalley Coulomb scattering have the opposite effects. At $|\nu| = 2$, phonons favor a spin-singlet ground state, which is consistent with the STM observation of charge density Kekul\'e pattern (the dominance of phonons is also consistent with estimates of the coupling strengths). Within this regime, we find that the IKS state at $|\nu| = 3$ is spin-valley locked. The analysis of ground state symmetries at integer fillings naturally carries over to nearby non-integer fillings, where we find that strength of charge Kekul\'e pattern is peaked around $|\nu| = 2$, in agreement with STM measurements, especially on the electron-doped side~\cite{Kim_2023, kimResolvingIntervalleyGaps2025}. The evolution of gaps under applied fields is also distinct for the various broken-symmetry states.

Besides ground state properties, we also study the collective excitation spectra of the IKS states under various perturbations. We confirm that the numbers and types of Nambu-Goldstone modes computed numerically from TDHF  agree with theoretical expectations based on symmetry analysis. In addition, as the Zeeman splitting gaps out some collective modes, we expect electronic entropy to be quenched by in-plane magnetic field even when there is no net spin polarization nor spin-valley locking at zero field, as it is around $|\nu| = 2$. This is consistent with experimental observations~\cite{Saito:2021aa,Rozen:2021aa}.

The discussion of the ground state spin structure also impacts the analysis of possible superconducting pairing order parameters for the doped IKS states. The normal ground states in filling range of $-3 < \nu < -2$, where superconductivity is most commonly observed, are divided into two regimes around $\nu_c\approx -2.6$, with $-3 < \nu < \nu_c$ being spin-valley locked and $\nu_c < \nu < -2$ being SU(2)$_s$-symmetric. In the former regime,  the broken spin rotation symmetry of the parent state precludes classifying the superconducting order parameter in terms of spin singlets or triplets. In the latter regime, the superconducting order parameter could still be either singlet or triplet, although the determination of which of these persists is beyond the scope of this work. Nevertheless, as the normal ground state preserves time-reversal symmetry, if the superconductor is $s$-wave, it would satisfy Anderson's theorem \cite{ANDERSON195926} and be robust to the inclusion of nonmagnetic impurities.

Our work thus builds a more complete picture of the phase diagram of realistic TBG, as well as a more nuanced view of the IKS order itself. Crucially, it establishes that with minimal assumptions about the relative strength of various couplings --- themselves bolstered by numerical estimates and STM data --- additional aspects of the phenomenology can also be captured by an IKS normal state. In future, it would be appealing to understand the extent to which alternative pictures of the IKS states~\cite{herzogarbeitman2025kekulespiralorderstrained,ledwith_nonlocal_2025} are amenable to a similar analysis, and if the insights thus gleaned can provide further clues into the enigma of superconductivity in magic-angle moir\'e graphene.

\acknowledgements
We are grateful for support from a Leverhulme Trust International Professorship (Grant Number LIP-202-014, Z.W.), a  Swiss National Science Foundation (SNSF)  Ambizione Grant  (Number PZ00P2-216183, G.W.), a postdoctoral research fellowship at the Princeton Center for Theoretical Science (Y.H.K.), the European Research Council under the European Union Horizon 2020 Research and Innovation Programme  (Grant Agreement No. 101076597-SIESS, N.B.), the UK Engineering and Physical Sciences Research Council EPSRC (Grant EP/X030881/1, S.H.S.), and the  UKRI Horizon Europe Guarantee (Grant No. EP/Z002419/1, S.A.P.).

\begin{appendix}

\section{Time-Dependent Hartree-Fock}\label{app:TDHF}

We summarize the time-dependent Hartree-Fock (TDHF) equations used to compute the neutral collective mode spectrum. Consider the interacting Hamiltonian expressed in a self-consistent Hartree-Fock (HF) basis
\begin{equation}
    \hat{H}=\sum_{ab}T_{ab}c^\dagger_ac^{\phantom\dagger}_b+\frac{1}{2}\sum_{abcd}V_{ab,cd}c^\dagger_bc^\dagger_ac^{\phantom\dagger}_cc^{\phantom\dagger}_d.
\end{equation}
The HF eigenvalue of orbital $a$ is $E_a$. Let $i,j,k,\ldots$ and $m,n,o,\ldots$ refer to occupied and unoccupied HF orbitals respectively.  Define the $A$- and $B$-matrices as~\cite{ring_nuclear_2004}
\begin{align}\label{eq:ABmatrices}
    A_{mi,nj}&=(E_m-E_i)\delta_{ij}\delta_{mn}+V_{mj,in}-V_{mj,ni}\\
    B_{mi,nj}&=V_{mn,ij}-V_{mn,ji}.
\end{align}
$A$ is Hermitian and $B$ is symmetric. 

The Hermitian stability matrix $\mathcal{H}$ is
\begin{equation}\label{eq:stability}
    \mathcal{H}=
    \begin{pmatrix}
    A & B \\ B^* & A^*
    \end{pmatrix}.
\end{equation}
If the HF state corresponds to a local variational minimum, then $\mathcal{H}$ is positive definite~\cite{thouless1960stability}.

The $\mathcal{L}$-matrix is 
\begin{equation}\label{eq:L}
    \mathcal{L}=\sigma_Z\mathcal{H}=
    \begin{pmatrix}
    A & B \\ -B^* & -A^*
    \end{pmatrix}
\end{equation}
with corresponding eigenvectors $w^\nu=(X^\nu,Y^\nu)$, symplectic norm $\Gamma^{\nu\nu'}=(X^\nu)^\dagger X^{\nu'}-(Y^\nu)^\dagger Y^{\nu'}$ and eigenvalues $\Omega^\nu$. Note the ``particle-hole symmetry'' $\mathcal{L}=-\sigma_X\mathcal{L}^*\sigma_X$, such that $({Y^\nu}^*,{X^\nu}^*)$ is also an eigenvector of $\mathcal{L}$ and  has eigenvalue $-{\Omega^\nu}^*$. If $\mathcal{H}$ is positive definite, then we can normalize such that $\Gamma^{\nu\nu'}=\text{sgn}(\Omega^\nu)\delta_{\nu\nu'}$.

Due to the translation symmetry of the HF ground state, $\mathcal{L}$ block-decomposes into different momentum transfer sectors. More specifically, $ A_{mi,nj}$ is only non-zero for $\bm{k}_m - \bm{k}_i = \bm{k}_n - \bm{k}_j$, and $B_{mi,nj}$ is only non-zero for $\bm{k}_m - \bm{k}_i = \bm{k}_j - \bm{k}_n$. As such, we write $A(\bm{q}) \equiv (A_{mi,nj})_{\bm{k}_n - \bm{k}_j = \bm{k}_m - \bm{k}_i = \bm{q}}$ and $B(\bm{q}) \equiv (B_{mi,nj})_{\bm{k}_n - \bm{k}_j = -\bm{k}_m + \bm{k}_i = \bm{q}}$ to denote such non-vanishing blocks of $A$- and $B$-matrices. For $\bm{q}$  being a time-reversal invariant momentum (TRIM), we define
\begin{equation}
    \mathcal{L}_{\bm{q}\in\text{TRIM}}= \begin{bmatrix}
        A(\bm{q}) & B(\bm{q})\\
        -B^*(\bm{q}) & -A^*(\bm{q})
    \end{bmatrix}.
\end{equation}
We can straightforwardly diagonalize the above to obtain collective mode energies $\omega(\bm{q}) =  \text{sgn}(X^\dagger X-Y^\dagger Y)\Omega(\bm{q})$ and their wavefunctions. Each pair of modes related by the aforementioned ``particle-hole symmetry" is only included once.

On the other hand, for a non-TRIM momentum $\bm{q}$, we find the following structure
\begin{equation}
    \mathcal{L}_{\bm{q}\notin\text{TRIM}}= \begin{bmatrix}
        A(\bm{q}) & 0 & 0 & B(\bm{q})\\
        0 & A(-\bm{q}) & B^T(\bm{q}) & 0\\
        0 & -B^*(\bm{q}) & -A^*(\bm{q}) & 0 \\
        -B^\dagger(\bm{q}) & 0 & 0 & -A^*(-\bm{q})
    \end{bmatrix}
\end{equation}
where the matrix index is ordered as $X(\bm{q}),X(-\bm{q}),Y(-\bm{q}),Y(\bm{q})$, and we have used the fact that the full $A$-matrix is Hermitian and the full $B$-matrix is symmetric. The advantage of writing $\mathcal{L}_{\bm{q}\notin\text{TRIM}}$ this way is that we can study just the outer block, since it is related to the inner block simply as
\begin{equation}
    \mathcal{L}_{\bm{q},\text{outer}}=\begin{bmatrix}
        A(\bm{q}) & B(\bm{q}) \\ -B^\dagger(\bm{q}) & -A^*(-\bm{q})
    \end{bmatrix}=-\sigma_X\mathcal{L}^*_{\bm{q},\text{inner}}\sigma_X.
\end{equation}
Since $\mathcal{L}_{\bm{q},\text{outer}}$ (and $\mathcal{L}_{\bm{q}\notin\text{TRIM}}$ more generally) mixes particle-hole labels from the $\bm{q}$ and $-\bm{q}$ sectors, we make the assignment such that 
\begin{itemize}
    \item If $\text{sgn}(\omega)>0$ and $\text{sgn}(X^\dagger X-Y^\dagger Y)>0$, we assign $\omega$ to $+\bm{q}$
    \item If $\text{sgn}(\omega)<0$ and $\text{sgn}(X^\dagger X-Y^\dagger Y)<0$, we assign $-\omega$ to $-\bm{q}$
    \item If $\text{sgn}(\omega)>0$ and $\text{sgn}(X^\dagger X-Y^\dagger Y)<0$, we assign $-\omega$ to $-\bm{q}$
    \item If $\text{sgn}(\omega)<0$ and $\text{sgn}(X^\dagger X-Y^\dagger Y)>0$, we assign $\omega$ to $+\bm{q}$
\end{itemize}

\section{Hartree-Fock interpolation}\label{sec:interpolation}

In order to capture details of the Hartree-Fock (HF) band structure beyond the resolution of self-consistent HF scheme, we use a physically-motivated interpolation scheme, based on the method employed in Ref.~\cite{Xie2021c}. For convenience, we absorb $\bm{q}_{\text{IKS}}$ into the definition of crystal momentum such that the single-particle density matrix is diagonal in momentum, i.e. $ \braket{c^\dagger_{\bm{k}'\beta}c^{\phantom\dagger}_{\bm{k}\alpha}} = P_{\alpha\beta}(\bm{k})\delta_{\bm{k}\bm{k}'}$, where $\alpha, \beta$ labels all other degrees of freedom other than momentum. Suppose we have found the variational ground state of the system with $N_1 \times N_2$ self-consistent HF. To find the interpolated band structure, we perform a one-shot Hartree-Fock using the mean-field Hamiltonian generated by the $N_1 \times N_2$ variational ground state. In the following, we write down the expressions explicitly for the Hamiltonian with only long-range Coulomb interaction, i.e. $ \hat{H} = \hat{H}_{\text{BM}} + \hat{H}_{\text{int}}$. The inclusion of the perturbations $\hat{H}'$ is straightforward. 

We let the ground state of $N_1 \times N_2$ system be $P_0$. For concreteness, we consider the interpolated band structure at some momentum $\bm{k}$, which needs not to be on the original $N_1 \times N_2$ grid. We can write down the mean-field Hamiltonian at $\bm{k}$
\begin{equation}
    h_{\text{MF}}(\bm{k}) = h_{\text{BM}}(\bm{k}) + h_{\text{HF}}[P_0 - P_{\text{ref}}](\bm{k})
\end{equation}
where $P_{\text{ref}}$ is the reference Hamiltonian in the subtraction scheme. We can decompose the HF Hamiltonian into Hartree and Fock parts, i.e. $ h_{\text{HF}} =  h_{\text{H}} +  h_{\text{F}}$, with ($h_{\text{HF}}(\bm{k}) \equiv [h_{\text{HF}}(\bm{k})]_{\alpha, \beta}c^\dagger_{\bm{k},\alpha}c^{\phantom\dagger}_{\bm{k},\beta}$)
\begin{widetext}
  \begin{align}
    [h_{\text{H}}[P](\bm{k})]_{\tau sa, \tau's'b} &= \frac{\delta_{\tau \tau'}\delta_{ss'}}{A}\sum_{\bm{G}}\sum_{\bm{k}',\tau''s''cd}V(\bm{G})\braket{u_{\bm{k}+\bm{G},\tau a}|u_{\bm{k},\tau b}}\braket{u_{\bm{k}'-\bm{G},\tau''c}|u_{\bm{k}',\tau''d}}P_{\tau''s''d,\tau''s''c}(\bm{k}')
\\
    [h_{\text{F}}[P](\bm{k})]_{\tau sa, \tau's'b} &= -\frac{1}{A}\sum_{\bm{G}}\sum_{\bm{k}',\tau's'cd}V(\bm{k}'-\bm{k}+\bm{G})\braket{u_{\bm{k}+\bm{G},\tau a}|u_{\bm{k}',\tau d}}\braket{u_{\bm{k}'-\bm{G},\tau'c}|u_{\bm{k},\tau'b}}P_{\tau sd,\tau's'c'}(\bm{k}')
\end{align}
\end{widetext}
Here, $A$ is the area of the $N_1 \times N_2$ system, $\bm{k}'$ is summed over the discrete $N_1 \times N_2$ grid, but $\bm{k}$ is some arbitrary momentum not restricted to the grid. This allows map out the band structure of $\bm{k}$ over some denser grid $N'_1 \times N'_2$, In practice, we usually choose $N'_1$ and $N'_2$ to be multiples of $N_1$ and $N_2$ respectively.

\section{Landau theory for phase transition within $\nu = \pm 2$ IKS}

Following the same arguments as Sec.~\ref{subsec:nu2}, we construct $\nu = -2$ (or $\nu = 2$ after particle-hole conjugation) from two copies of $\nu = -3$ IKS state, labeled `band A' and `band B' respectively. Similar to the discussion in the main text, band $A$ is restricted to the spin-valley sector $\ket{K, \bm{s}_K}$ and $\ket{K, \bm{s}_K'}$, while band $B$ is restricted to the spin-valley sector $\ket{K, -\bm{s}_K}$ and $\ket{K, -\bm{s}_K'}$. In anticipation of the effects of external field, however, we will not always require the two bands to be identical up to spin and valley U(1)$_v$-angle. The density matrix due to band $A$ is 

\begin{equation}
P^A =    \begin{pmatrix}
        \tilde{P}^A_{++} \otimes \ket{\bm{s}_K}\bra{\bm{s}_K} & \tilde{P}^A_{+-} \otimes \ket{\bm{s}_K}\bra{\bm{s}_K'} \\
            \tilde{P}^A_{-+} \otimes \ket{\bm{s}_K'}\bra{\bm{s}_K} & \tilde{P}^A_{--} \otimes \ket{\bm{s}_K'}\bra{\bm{s}_K'} \\
    \end{pmatrix}
\end{equation}
(where we use tilde to denote spinless density matrices) and that due to the band $B$ is 
\begin{equation}
P^B =    \begin{pmatrix}
        \tilde{P}^B_{++} \otimes \ket{-\bm{s}_K}\bra{-\bm{s}_K} & \tilde{P}^B_{+-} \otimes \ket{-\bm{s}_K}\bra{-\bm{s}_K'} \\
            \tilde{P}^B_{-+} \otimes \ket{-\bm{s}_K'}\bra{-\bm{s}_K} & \tilde{P}^B_{--} \otimes \ket{-\bm{s}_K'}\bra{-\bm{s}_K'} \\
    \end{pmatrix}
\end{equation}
The total density matrix is simply the sum, i.e.
\begin{equation}
    P = P^A + P^B
\end{equation}

We would like to investigate the competition between electron-phonon coupling (EPC) and Zeeman splitting. We will assume that both EPC and Zeeman splitting are small, such that we \textit{do not consider the cross terms in the energy expression}. That is, we will deduce the energy contribution from EPC based on the case without Zeeman splitting. In such a case, we can take $\tilde{P}^A$ and $\tilde{P}^B$ to be the same up to some $\text{U}(1)$ phase, i.e. we may write  $\tilde{P}^A_{++} =  \tilde{P}^B_{++} \equiv \tilde{P}_{++}$, $\tilde{P}^A_{+-}e^{-i\phi_A} =  \tilde{P}^B_{+-}e^{-i\phi_B} \equiv \tilde{P}_{+-}$. We have now reduced our construction to the same form as that in the main text, i.e.
\begin{equation}\label{eq:P_V}
P =    \begin{pmatrix}
        \tilde{P}_{++} \otimes I & \tilde{P}_{+-} \otimes V \\
            \tilde{P}_{-+} \otimes V^\dagger & \tilde{P}_{--} \otimes I \\
    \end{pmatrix}
\end{equation}
for some \begin{equation}V = e^{i\phi_A}\ket{\bm{s}_K}\bra{\bm{s}_K'} + e^{i\phi_B}\ket{-\bm{s}_K}\bra{-\bm{s}_K'}.\end{equation}
According to the main text, we have
\begin{equation}
\mathcal{E}_{\text{EPC}} = -\bar{g}|\text{tr}(V)|^2
\end{equation}
where $\mathcal{E}$ is energy density and $\bar{g}$ is some coupling constant proportional to $g$ used in the main text. From
\begin{equation}|\text{tr}(V)| = |e^{i\phi_A}\braket{\bm{s}_K'|\bm{s}^{\phantom\prime}_K} +  e^{i\phi_B}\braket{-\bm{s}_K'|-\bm{s}_K^{\phantom\prime}}|\end{equation}
we have
\begin{equation}|\text{tr}(V)| \leq |\braket{\bm{s}_K'|\bm{s}^{\phantom\prime}_K}| + |\braket{-\bm{s}_K'|-\bm{s}^{\phantom\prime}_K}|\end{equation}where the inequality can always be saturated with appropriate choice of $\phi_A$ and $\phi_B$. Since there is no other penalty for choosing the phases to saturate the inequality (later we will see that the Zeeman coupling term depends on $\bm{s}_K$ and $\bm{s}_K'$, but not on $\phi_A$ and $\phi_B$), we will always choose the phases as such. Note that $|\braket{\bm{s}_K'|\bm{s}^{\phantom\prime}_K}| = |\cos(\frac{\alpha}{2})|$, where $\alpha$ is the angle between $\bm{s}_K$ and $\bm{s}_K'$, and we have
\begin{equation}
\mathcal{E}_{\text{EPC}} = -2\bar{g}(\cos\alpha + 1).
\end{equation}
Now, if we apply some external field, we will deform the density matrices such that
\begin{equation}\tilde{P}^{A/B} = \tilde{P} \pm  \epsilon \delta \tilde{P}\end{equation}which describes the deformation of the valley-polarized `lobes' in momentum space~\cite{Kwan2021}. The spin polarization is given by
$\braket{\bm{S}} = \epsilon (\bm{s}^{\phantom\prime}_K - \bm{s}_K') $, and $|\bm{s}^{\phantom\prime}_K - \bm{s}_K'| = 2|\sin(\alpha/2)|$. It is clear that it is always preferable to align $\bm{s}_K - \bm{s}_K'$ with the external field, such that the Zeeman energy is given by
\begin{equation}
\mathcal{E}_Z = -2\epsilon\bar{\Delta}\sin(\alpha/2)
\end{equation}
where $\bar{\Delta}$ is proportional to Zeeman splitting. Without perturbation, $\epsilon = 0$ corresponds to the ground state. As such, we introduce \begin{equation}\mathcal{E}_{\text{dist}} = \frac{1}{2}\kappa \epsilon^2\end{equation} to penalize distortion. We will assume the distortion is small such that only quadratic term is retained. Minimizing $\mathcal{E}_Z + \mathcal{E}_{\text{dist}}$ over $\epsilon$, we find that $\epsilon = 2\bar{\Delta}\sin(\alpha/2)/\kappa$, and the total energy density is 
\begin{equation}\mathcal{E}_\text{tot} = -2\bar{g}\left(\cos\alpha + 1) + \frac{\bar{\Delta}^2}{\kappa} (\cos\alpha - 1)\right).\end{equation}
Dropping $\alpha$ independent constants, we have
\begin{equation}
\mathcal{E}_{\text{tot}} = \left(-2\bar{g} + \frac{\bar{\Delta}^2}{\kappa}\right)\cos \alpha
\end{equation}
leading to two possible grounds states with $\alpha = 0$ (EPC-dominated) or $\alpha = \pi$ (Zeeman-dominated). The transition between the two regimes is first order.

In numerical calculations, as shown in Fig.~\ref{fig:spinpol_g} of the main text, the transition is close to but not strictly first order. In the beginning of the section, we postulated that the system consists of two occupied mean-field bands with orthogonal spin-valley polarization. In the sliver between the two phases, the picture no longer holds. [Numerically, a direct gap opens up at where band A and B used to cross, indicating hybridization.] We also note that the above analysis is performed in the limit where both electron-phonon coupling and Zeeman splitting are infinitesimal, while in actual numerical simulation both perturbations are finite.

\section{Nonlinear sigma model for IKS at $\nu = \pm 2$}\label{app:NLSM_nu2}
We consider the case with $\text{SU}(2)_K\times \text{SU}(2)_{K'}$ symmetry, and to focus on the gapless modes, we only consider fluctuation in the soft directions. We suppress sublattice and layer indices (one can take them as implicitly summed over).

We consider the order parameter given by $\tilde{n}_{ss'}(\bm{r}) = \braket{\psi^\dagger_{+s}(\bm{r})\psi^{\phantom\dagger}_{-s'}(\bm{r})}$, subject to the constraint $\tilde{n}^\dagger \tilde{n} = |n_0|^2 I$ for some fixed scalar $n_0$ which is taken to be real and positive. For simplicity, we rescale the order parameter as $n = \tilde{n}/n_0 $ such that $n^\dagger n = I$. The system has the symmetry $n \rightarrow UnV^\dagger$ for $2 \times 2$ unitary matrices $U$ and $V$.  This allows us to write the Lagrangian up to quadratic order as
\begin{equation}
\mathcal{L} = \frac{1}{2} \text{tr}\left(\partial_tn^\dagger\partial_tn \right) - \frac{1}{2}\kappa_{ij}\text{tr}\left(\partial_in^\dagger\partial_jn \right)
\end{equation}
where matrix multiplication over spin indices and summation over spatial directions are implied. 

Considering fluctuations around the uniform state $n = I$, we have 
\begin{equation}    
n \approx \left(1 - \frac{1}{2}\sum_\mu \pi^2_\mu\right)I + i \sum_\mu \sigma_\mu \pi_\mu,
\end{equation}
for small quantities $\pi_\mu$ with temporal and spatial dependence and $\sigma_0 \equiv I$. The linearized Lagrangian is given by 
\begin{equation}
    \mathcal{L} = \sum_\mu \left[ (\partial_t \pi_\mu)^2 - \kappa_{ij} (\partial_i \pi_\mu)(\partial_j\pi_\mu)\right],
\end{equation}producing 4 modes with identical dispersion given by $\omega = \sqrt{\kappa_{ij}q_iq_j}$. Electron-phonon coupling introduces an additional energy term $-g|\text{tr}(n)|^2$. As $|\text{tr}(n)| \approx 2(1 - \sum_{i = 1,2,3} \pi_i^2)$, it gaps out 3 out of the 4 modes, as expected.

\begin{figure}[h]
    \centering
    \includegraphics[width=0.8\linewidth]{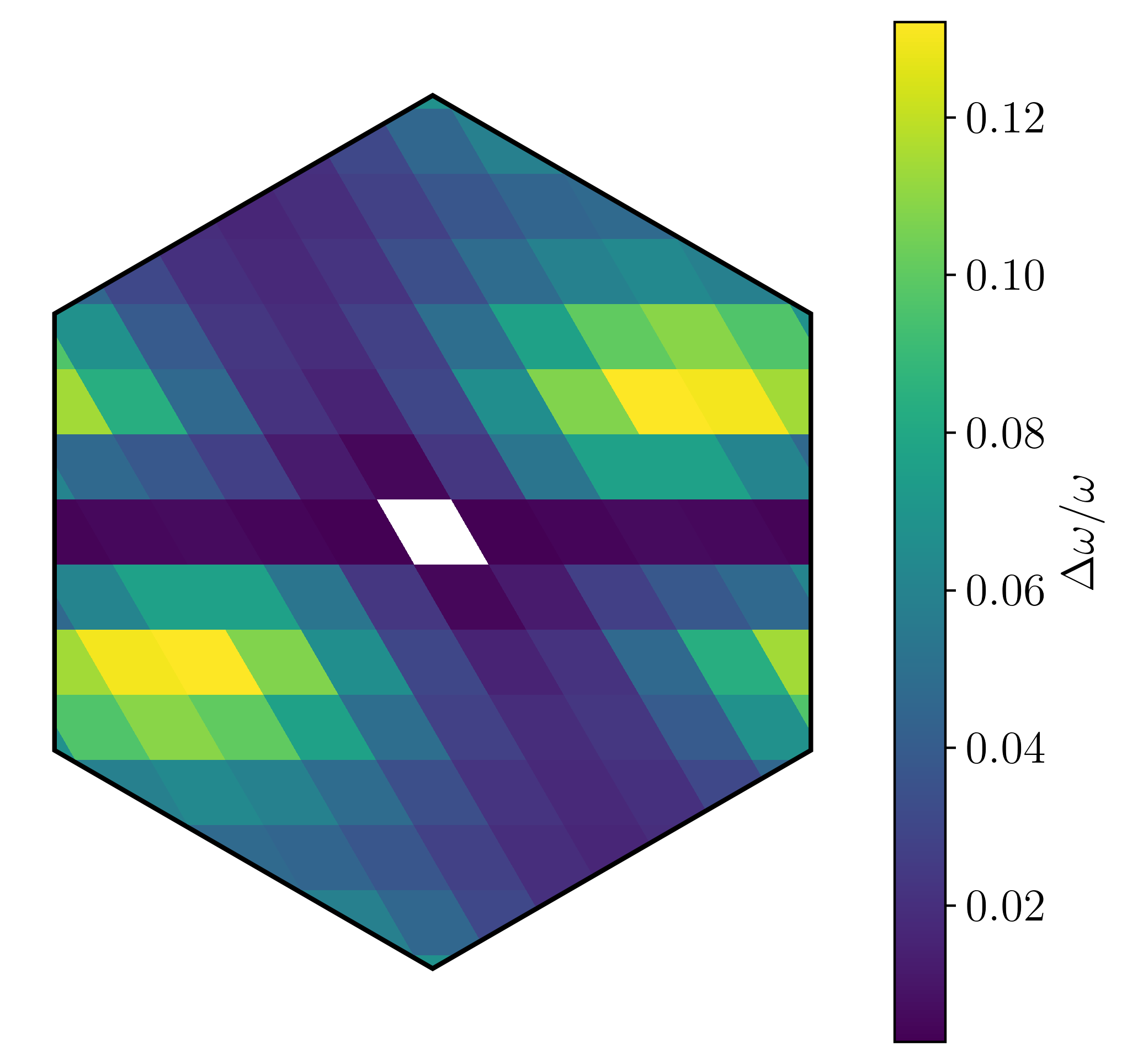}
    \caption{Splitting of the low energy excitations of $\nu = 2$ IKS from $10 \times 10$ HF at $\epsilon = 0.3\%$. $\Delta \omega/\omega \equiv (\omega_4 - \omega_1)/\omega_1$ is the fractional splitting of the excitation modes ($\omega_i$ is the $i$-th lowest energy mode at a particular $\bm{q}$). The results are for $\text{SU}(2)_K\times \text{SU}(2)_{K'}$ symmetric Hamiltonian.}
    \label{fig:excitation_splitting}
\end{figure}

Numerically, with $10 \times 10$ TDHF at $\epsilon = 0.3\%$, we find 3 modes with exactly the same velocities (up to machine accuracy) and 1 mode with velocity that differs by $< 0.5\%$, where the difference is due to finite grid size limiting the numerical accuracy for the velocity at $\bm{q} = \bm{0}$. At finite momenta the lowest 4 excitation modes split into 3 degenerate modes and 1 non-degenerate mode, with their splitting shown in Fig.~\ref{fig:excitation_splitting}.

\section{Superconducting order parameter from IVC Fermi surfaces}\label{app:SC_OP_IVC}

In this section, we discuss the symmetry of the superconducting order parameters that can form from IKS parent states. For convenience, we absorb $\bm{q}_{\text{IKS}}$ into the definition of momentum.

\subsection{Spin polarized IKS}

The following discussion is relevant to IKS at $-3 < \nu < \nu_c$ but with ferromagnetic intervalley Hund's coupling. Consider a spin polarized IVC system. Within the non-trivial spin sector (we suppress the spin index for brevity), we have a Fermi surface where the corresponding band (which we call band 1) is described by the creation operator
\begin{equation}c^\dagger_1(\bm{k}) = \frac{1}{\sqrt{2}}(e^{-i\varphi_{\bm{k}}/2}c^\dagger_K(\bm{k}) + e^{i\varphi_{\bm{k}}/2}c^\dagger_{K'}(\bm{k}))\end{equation}
where $\varphi_{\bm{k}}$ is some function of $\bm{k}$. Suppose the system has some spinless time-reversal symmetry that acts as
\begin{equation}
\mathcal{T}c^\dagger_\tau(\bm{k})\mathcal{T}^{-1} = c^\dagger_{\bar{\tau}}(-\bm{k}).
\end{equation}
We also assume that the IVC state does not spontaneously break this time-reversal symmetry, such that
\begin{equation}\mathcal{T}c^\dagger_1(\bm{k})\mathcal{T}^{-1} = c^\dagger_1(-\bm{k}).\end{equation}
This immediately implies
\begin{equation}\varphi_{-\bm{k}} = \varphi_{\bm{k}}.\end{equation}
We can also define the orthogonal IVC band to be
\begin{equation}c^\dagger_2(\bm{k}) = \frac{1}{\sqrt{2}}(e^{-i\varphi_{\bm{k}}/2}c^\dagger_K(\bm{k}) - e^{i\varphi_{\bm{k}}/2}c^\dagger_{K'}(\bm{k})).\end{equation}
We can invert the relation to find
\begin{equation}\begin{pmatrix}
    c^\dagger_{K}(\bm{k}) \\ c^\dagger_{K'}(\bm{k}) 
\end{pmatrix} = \frac{1}{\sqrt{2}}\begin{pmatrix}
    e^{i\varphi_{\bm{k}}/2} & e^{i\varphi_{\bm{k}}/2} \\
    e^{-i\varphi_{\bm{k}}/2} & -e^{-i\varphi_{\bm{k}}/2}
\end{pmatrix}\begin{pmatrix}
    c^\dagger_1(\bm{k}) \\ c^\dagger_2(\bm{k})
\end{pmatrix}.\end{equation}
Suppose IVC band 1 undergoes superconducting pairing with \begin{equation}\Delta_1(\bm{k}) = \braket{c^\dagger_1(\bm{k})c^\dagger_1(-\bm{k})} = - \Delta_1(-\bm{k}).\end{equation}
In the valley basis, we define the SC order parameter as
\begin{equation}[\Delta_{\bm{k}}]_{\tau\tau'} = \braket{c^\dagger_\tau(\bm{k})c^\dagger_{\tau'}(-\bm{k})}.\end{equation}
We find 
\begin{equation}\Delta_{\bm{k}} = \frac{1}{2}\begin{pmatrix}
    e^{i\varphi_{\bm{k}}} & 1 \\
    1 & e^{-i\varphi_{\bm{k}}}
\end{pmatrix}\Delta_1(\bm{k})\end{equation}
where we have used $\varphi_{-\bm{k}} = \varphi_{\bm{k}}$ to simplify the expression. Note that $\Delta_{\bm{k}} = - \Delta_{-\bm{k}}^T$, as expected from fermionic statistics.

Borrowing standard notation for triplet superconductivity and treating the valley index as a pseudo-spin index, we can write
\begin{equation}\Delta_{\bm{k}} = i(\bm{d}_v(\bm{k})\cdot \bm{\tau})\tau_y\end{equation} (where the subscript $v$ is meant to emphasize that the vector is in valley space instead of physical spin space) with
\begin{equation}
\bm{d}_v(\bm{k}) = \frac{1}{2}\Delta_1(\bm{k})\begin{pmatrix}
    -i\sin\varphi_{\bm{k}} \\ -i\cos \varphi_{\bm{k}} \\ 1
\end{pmatrix}
\end{equation}
satisfying $\bm{d}_v(\bm{k}) = - \bm{d}_v(-\bm{k})$. 

We note that this is ``non-unitary", in the sense that
\begin{equation}\bm{q}_v \equiv i\bm{d} _v\times \bm{d}_v^* = \frac{1}{2}|\Delta_1(\bm{k})|^2 \begin{pmatrix}
    \cos \varphi_{\bm{k}} \\ - \sin \varphi_{\bm{k}} \\ 0
\end{pmatrix} \end{equation}
is non-zero. Being non-unitary indicates a preferred direction in the valley space. This is expected, since the parent state breaks the valley U(1) symmetry.

If we were to express the order parameter in the full basis $(c_{K,\uparrow}^\dagger,c_{K,\downarrow}^\dagger,c_{K',\uparrow}^\dagger,c_{K',\downarrow}^\dagger)$, we have

\begin{equation}\Delta_{\bm{k}} = \frac{1}{2}\begin{pmatrix}
    e^{i\varphi_{\bm{k}}} & 0 & 1 & 0 \\
    0 & 0 & 0& 0 \\
    1 & 0 & e^{-i\varphi_{\bm{k}}} & 0 \\
    0 & 0 & 0& 0
\end{pmatrix}\Delta_1(\bm{k}).\end{equation}
or, in $\bm{d}$-vector formalism, 
\begin{equation}\Delta_{\bm{k}} = \left(\frac{1 + s_z}{2}\right)i(\bm{d}_v(\bm{k})\cdot \bm{\tau})\tau_y\end{equation}
In this discussion, we find the pairing is purely ``triplet" in valley index. This is a result of assuming maximum IVC (i.e. no valley polarization at any $\bm{k}$-point). Consider a more general parameterization
\begin{equation}c^\dagger_1(\bm{k}) = \alpha_{\bm{k}} c^\dagger_K(\bm{k}) + \beta_{\bm{k}} c^\dagger_{K'}(\bm{k})\end{equation}
with $|\alpha_{\bm{k}}|^2 + |\beta_{\bm{k}}|^2 = 1$. Time-reversal symmetry now demands 
\begin{equation}\alpha_{-\bm{k}} = \beta^*_{\bm{k}}, \quad \beta_{-\bm{k}} = \alpha^*_{\bm{k}}\end{equation}
and we find, in the basis of $(c^\dagger_{K\uparrow}, c^\dagger_{K'\downarrow})$,
\begin{equation}
\Delta_{\bm{k}} = \begin{pmatrix}
    \alpha_{\bm{k}}^*\beta_{\bm{k}} & |\alpha_{\bm{k}}|^2 \\
    |\beta_{\bm{k}}|^2 & \alpha_{\bm{k}}\beta_{\bm{k}}^*
\end{pmatrix} \Delta_1(\bm{k}).
\end{equation}
In general, if $|\alpha_{\bm{k}}|^2 \neq |\beta_{\bm{k}}|^2$, the order parameter has a mixture of ``singlet" and ``triplet" characters.

\subsection{Spin-valley locked IKS}

The following discussion is relevant to IKS at $-3 < \nu < \nu_c$ with the expected anti-ferromagnetic intervalley Hund's coupling. Consider a spin-valley locked IVC system, which we can assume to be non-trivial only in the $s_z\tau_z = + 1$ sector. We have an IVC band described by 
\begin{equation}c^\dagger_1(\bm{k}) = \frac{1}{\sqrt{2}}(e^{-i\varphi_{\bm{k}}/2}c^\dagger_{K\uparrow}(\bm{k}) + e^{i\varphi_{\bm{k}}/2}c^\dagger_{K'\downarrow}(\bm{k})).\end{equation}
We will consider a time-reversal-symmetry-like operator that acts as
\begin{align}
    \mathcal{T'}c^\dagger_{K \uparrow}(\bm{k})\mathcal{T'}^{-1} = c^\dagger_{K'\downarrow}(-\bm{k}) \\
    \mathcal{T'}c^\dagger_{K' \downarrow}(\bm{k})\mathcal{T'}^{-1} = c^\dagger_{K\uparrow}(-\bm{k}).
\end{align}
We note that this operator is not the physical spinful time-reversal operator, as $\mathcal{T'}^2 = + 1$, but this is the relevant operator for the case of anti-ferromagnetic IKS.

The discussion in the preceding section follows similarly, by simply replacing $c^\dagger_{K'\uparrow}$ with $c^\dagger_{K'\downarrow}$. In the basis $(c_{K,\uparrow}^\dagger,c_{K,\downarrow}^\dagger,c_{K',\uparrow}^\dagger,c_{K',\downarrow}^\dagger)$, the superconducting order parameter is given by
\begin{equation}\Delta_{\bm{k}} = \frac{1}{2}\begin{pmatrix}
    e^{i\varphi_{\bm{k}}} & 0 & 0 &1 \\
    0 & 0 & 0 & 0\\
    0 & 0 & 0 & 0\\
    1 &0& 0& e^{-i\varphi_{\bm{k}}}
\end{pmatrix}\Delta_1(\bm{k})\end{equation}
where $\Delta_1(\bm{k})= \braket{c^\dagger_1(\bm{k})c^\dagger_1(-\bm{k})}$. We can write this as in the $\bm{d}$-formalism \textit{for the physical spin} as
\begin{widetext}
    \begin{equation}\Delta_{\bm{k}} = \frac{1}{4}\Delta_1(\bm{k})\{[(-i\sin\varphi_{\bm{k}} \tau_0 - \cos \varphi_{\bm{k}} \tau_z)s_x + (-i \cos \varphi_{\bm{k}}\tau_0 + \sin \varphi_{\bm{k}} \tau_z)s_y + \tau_x s_z](is_y) - \tau_y s_y \}.\end{equation}
\end{widetext}
That is, \begin{equation}\Delta_{\bm{k}} = i(\tilde{\bm{d}}(\bm{k}) \cdot \bm{s} + \tilde{d}_0 s_0)s_y\end{equation}
with
\begin{equation}
\tilde{\bm{d}}(\bm{k}) =  \frac{1}{4}\Delta_1(\bm{k})\begin{pmatrix}
    -i\sin\varphi_{\bm{k}} \tau_0 - \cos \varphi_{\bm{k}} \tau_z \\  -i \cos \varphi_{\bm{k}}\tau_0 + \sin \varphi_{\bm{k}} \tau_z \\ \tau_x
\end{pmatrix}
\end{equation}
and
\begin{equation}\tilde{d}_0 =   \frac{i}{4}\Delta_1(\bm{k}) \tau_y.\end{equation}
In general, the pairing is a mixture of spin singlet and spin triplet. 
\subsection{Spin singlet IKS}
For IKS at $\nu_c < \nu < -2$, we have two intervalley-coherent bands. Here we focus on the case where we have a spin singlet IKS parent state that corresponds to the preferred ground state at finite electron-phonon coupling. The relevant IVC bands where the Fermi surfaces reside in are  
\begin{equation}c^\dagger_{1\uparrow}(\bm{k}) = \frac{1}{\sqrt{2}}(e^{-i\varphi_{\bm{k}}/2}c^\dagger_{K\uparrow}(\bm{k}) + e^{i\varphi_{\bm{k}}/2}c^\dagger_{K'\uparrow}(\bm{k}))\end{equation}
\begin{equation}c^\dagger_{1\downarrow}(\bm{k}) = \frac{1}{\sqrt{2}}(e^{-i\varphi_{\bm{k}}/2}c^\dagger_{K\downarrow}(\bm{k}) + e^{i\varphi_{\bm{k}}/2}c^\dagger_{K'\downarrow}(\bm{k})).\end{equation}
Different from the single-band case, in this case the parent state does not break $\text{SU}(2)_s$ symmetry. We write the superconducting order parameter as $[\Delta_1(\bm{k})]_{ss'} = \braket{c^\dagger_{1s}(\bm{k})c^\dagger_{1s'}(-\bm{k})}$, and due to fermionic statistics, we must have $\Delta_1(\bm{k}) = - \Delta^T_1(-\bm{k})$. The order parameter could be either a spin singlet or a spin triplet, and the forms follow the usual discussion of superconducting pairing symmetry. That is, for spin singlet, we have
\begin{equation}
\Delta_1(\bm{k}) = \begin{pmatrix}
    0 & \psi(\bm{k}) \\ - \psi(\bm{k}) & 0
\end{pmatrix} = is_y\psi(\bm{k})
\end{equation}
with $\psi(-\bm{k}) = \psi(\bm{k})$. For spin triplet, we have
\begin{equation}
\Delta_1(\bm{k}) = \begin{pmatrix}
    -d_x(\bm{k}) + id_y(\bm{k}) & d_z(\bm{k}) \\ d_z(\bm{k}) & d_x(\bm{k}) + id_y(\bm{k})
\end{pmatrix} = i(\bm{d} \cdot \bm{s})s_y
\end{equation}
for some $\bm{d}(\bm{k}) = -\bm{d}(-\bm{k})$. Note that, different from the use of $\bm{d}$ notation in previous subsections, due to the unbroken $\text{SU}(2)_s$ symmetry of the parent state, the $\bm{d}$ vector is \textit{a priori} an arbitrary odd function.

We can use the procedures in the previous subsections to convert from the IVC band basis  $(c^\dagger_{1\uparrow}, c^\dagger_{1\downarrow})$ back into the single-particle valley-diagonal basis  $(c_{K,\uparrow}^\dagger,c_{K,\downarrow}^\dagger,c_{K',\uparrow}^\dagger,c_{K',\downarrow}^\dagger)$. For brevity, this is omitted here. Furthermore, the discussion of the spin structure is clearly unaffected if we assume a more general form of IVC given by
$$c^\dagger_{1s}(\bm{k}) = \alpha_{\bm{k}} c^\dagger_{Ks}(\bm{k}) + \beta_{\bm{k}} c^\dagger_{K's}(\bm{k})$$
for $s = \uparrow,\downarrow$, since $\alpha_{\bm{k}}, \beta_{\bm{k}}$ are independent of the spin index due to $\text{SU}(2)_s$ symmetry.

\end{appendix}

\bibliography{references}

\newpage
\clearpage
\onecolumngrid
	\begin{center}
		\textbf{\large --- Supplementary Material ---\\Putting a new spin on the incommensurate Kekul\'{e} spiral: spin-valley locking, fermiology, collective modes, and implications for superconductivity}\\
		\medskip
		\text{Ziwei Wang, Glenn Wagner, Yves H. Kwan, Nick Bultinck, Steven H. Simon, and S.A. Parameswaran}
	\end{center}
	
	\setcounter{equation}{0}
	\setcounter{figure}{0}
	\setcounter{table}{0}
	\setcounter{page}{1}
    \setcounter{section}{0}
    \renewcommand{\thesection}{S\arabic{section}}
	\makeatletter
	\renewcommand{\thefigure}{S\arabic{figure}}
	\renewcommand{\bibnumfmt}[1]{[S#1]}

\section{Additional numerical results}\label{app:additional_numerical_integer}

\subsection{Effects of perturbations under varying strain}
\begin{figure}[h]
    \centering
    \includegraphics[width=0.7\linewidth]{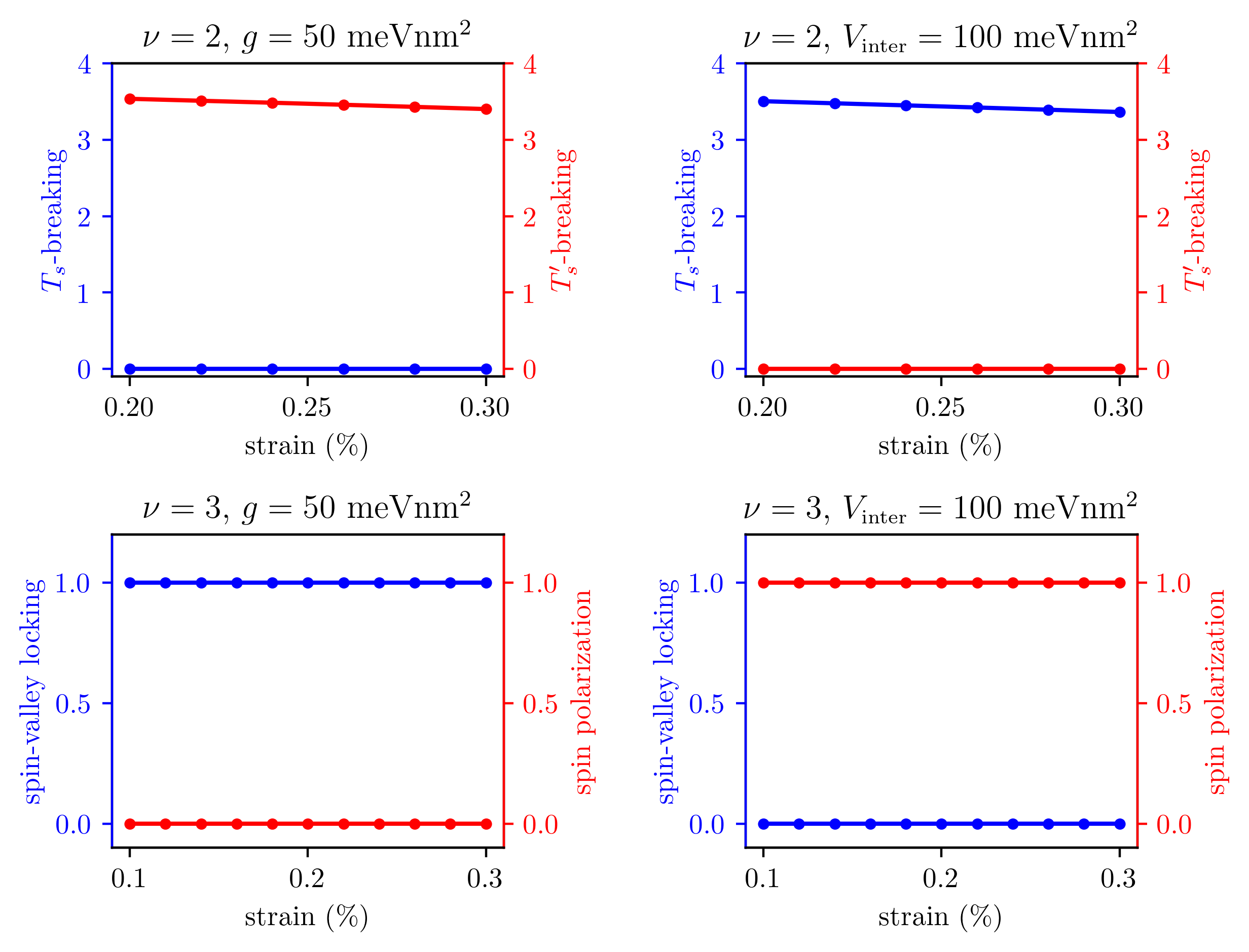}
    \caption{Properties of IKS with electron-phonon coupling and intervalley Coulomb scattering under varying magnitudes of strain. To characterize the states at $\nu = 2$, we consider two types of spinful anti-unitary time-reversal-like symmetries, given by $\mathcal{T}_s = s_y\tau_x \mathcal{K}$, and $\mathcal{T}^\prime_s = s_y \tau_y \mathcal{K}$, where $\mathcal{K}$ is complex conjugation. The spin-singlet IKS preserves $\mathcal{T}_s$ while the IKS state with $\pi$ valley U(1) phase difference in the two spin sectors preserves $\mathcal{T}^\prime_s$. To distinguish ferromagnetic and anti-ferromagnetic states at $\nu = 3$, we plot the total spin polarization and degree of spin-valley locking respectively. }
    \label{fig:varying_strain}
\end{figure}

The effects of electron-phonon coupling and intervalley Coulomb scattering at $\nu = 2,3$ at varying strengths of strain are shown in Fig.~\ref{fig:varying_strain}. We observe that the effects of these perturbations, as summarized in Tab.~\ref{tab:summary_nu3} and~\ref{tab:summary_nu2} in the main text, are insensitive to the magnitude of strain, provided it is large enough to stabilize the IKS.

\subsection{Competition between electron-phonon coupling and intervalley Coulomb scattering}\label{subsec:competition}

\begin{figure}[h]
    \centering
    \includegraphics[width=0.7\linewidth]{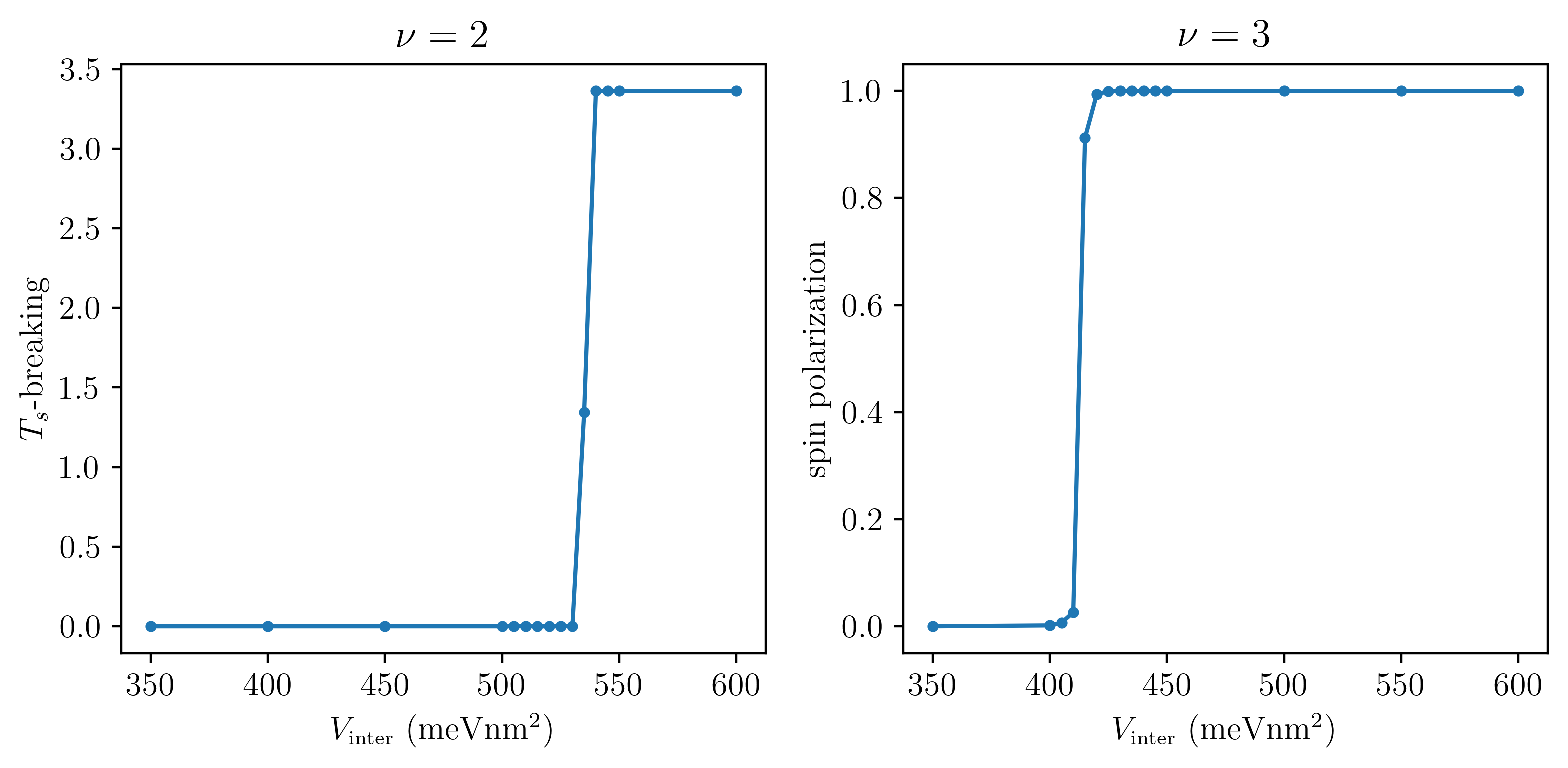}
    \caption{Properties of IKS at $\nu = 2$ and $\nu = 3$ with fixed electron-phonon coupling at $g = 50$\,meVnm\textsuperscript{2} with varying strength of intervalley Coulomb scattering $V_\text{inter}$. $\mathcal{T}_s$ refers to the spinful time-reversal symmetry, which is used as an indicator to distinguish between the ground state favored by electron-phonon coupling, which preserves $\mathcal{T}_s$, and that favored by intervalley Coulomb scattering, which breaks $\mathcal{T}_s$, at $\nu = 2$. At $\nu = 3$, the total spin polarization distinguishes the ferromagnet from the anti-ferromagnet. The data is obtained from $10 \times 10$ HF, $\epsilon = 0.3\%$, $w_{\text{AA}} = 80$~meV, and $\theta = 1.05^\circ$.}
    \label{fig:epc_inter}
\end{figure}

As discussed in Tab.~\ref{tab:summary_nu3} for $|\nu|=3$ and Tab.~\ref{tab:summary_nu2} for $|\nu|=2$ in the main text, electron-phonon coupling and intervalley Coulomb scattering favor different IKS states. Here, we numerically investigate the transition between the two regimes by varying intervalley Coulomb scattering $V_\text{inter}$ with a fixed electron-phonon coupling strength $g$, as shown in Fig.~\ref{fig:epc_inter}. At both $\nu=2$ and $\nu=3$, we observe a sharp transition between the two regimes as $V_\text{inter}$ is increased. In particular, the $\nu=2$ IKS transitions from a $\mathcal{T}_s$-symmetric spin singlet state to a a $\mathcal{T}_s'$-symmetric state, while the $\nu=3$ transitions from a ferromagnetic to an antiferromagnetic state. We notice that the transitions occur at different $V_\text{inter}$ for different fillings. This is possible because the two perturbations appear in the Hamiltonian with different form factors, and as such, their competition depends on the state in question. Nevertheless, we estimate that in a realistic system $g/V_{\text{inter}} \sim 0.6$, which lies in the phonon-dominated regimes for both $|\nu| = 2,3$.

\subsection{Paramagnetic Response of IKS}
\begin{figure}[h]
    \centering
    \includegraphics[width=0.5\linewidth]{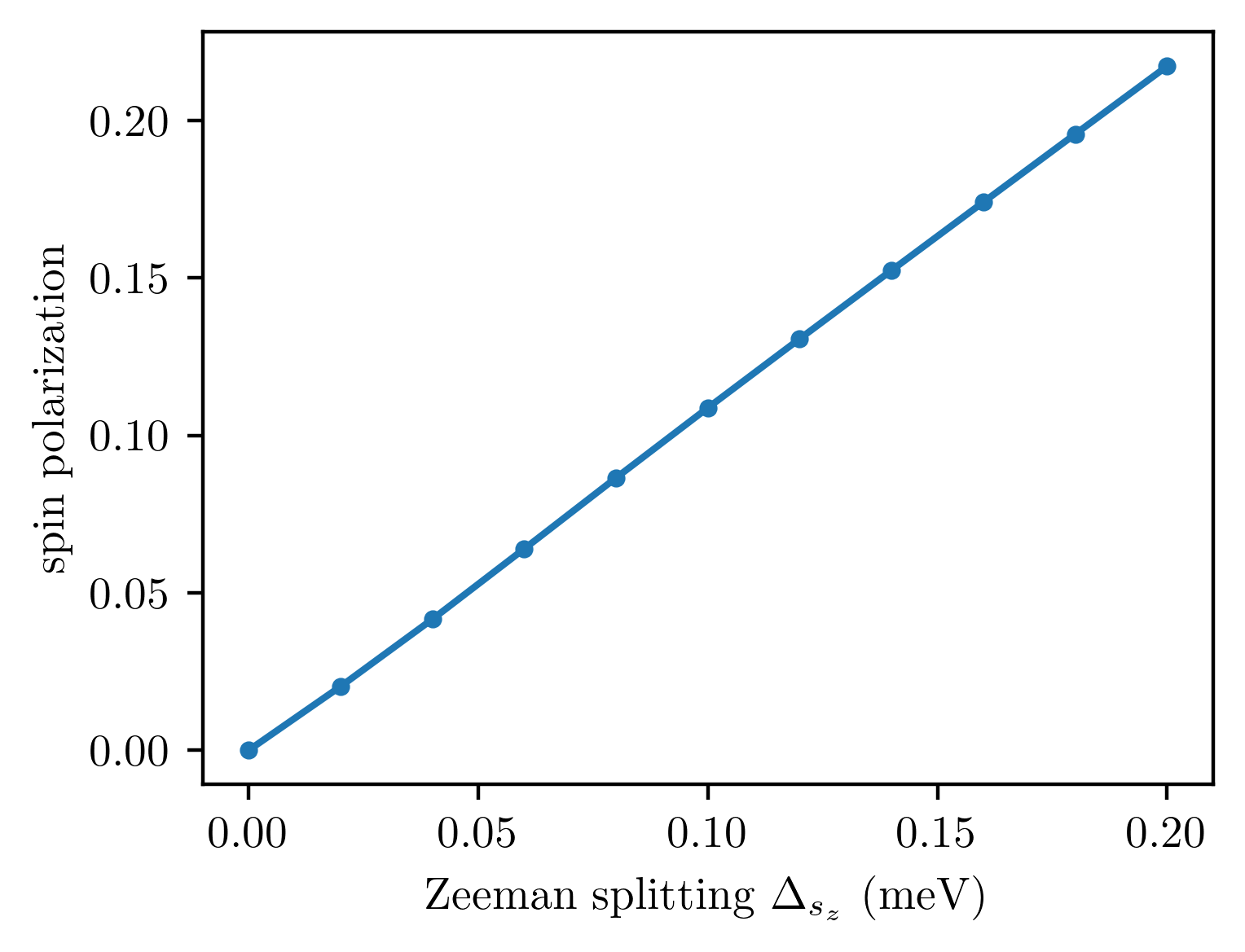}
    \caption{The spin polarization of IKS state at $\nu = 2$ under Zeeman splitting $\Delta_{s_z}$. $\epsilon = 0.3\%$, $w_{\text{AA}} = 80$~meV, $\theta = 1.05^\circ$, and the system size used is $10 \times 10$. Two bands are retained per spin and valley.}
    \label{fig:zeeman_linear}
\end{figure}

Fig.~\ref{fig:zeeman_linear} demonstrates that the IKS state at $\nu = 2$ is paramagnetic (i.e.~spin polarization increases linearly with applied field). No electron-phonon coupling or intervalley Coulomb scattering is applied.

\subsection{Additional collective modes results}\label{app:int_collective_modes}

\begin{figure}[h]
    \centering
    \includegraphics[width=0.5\linewidth]{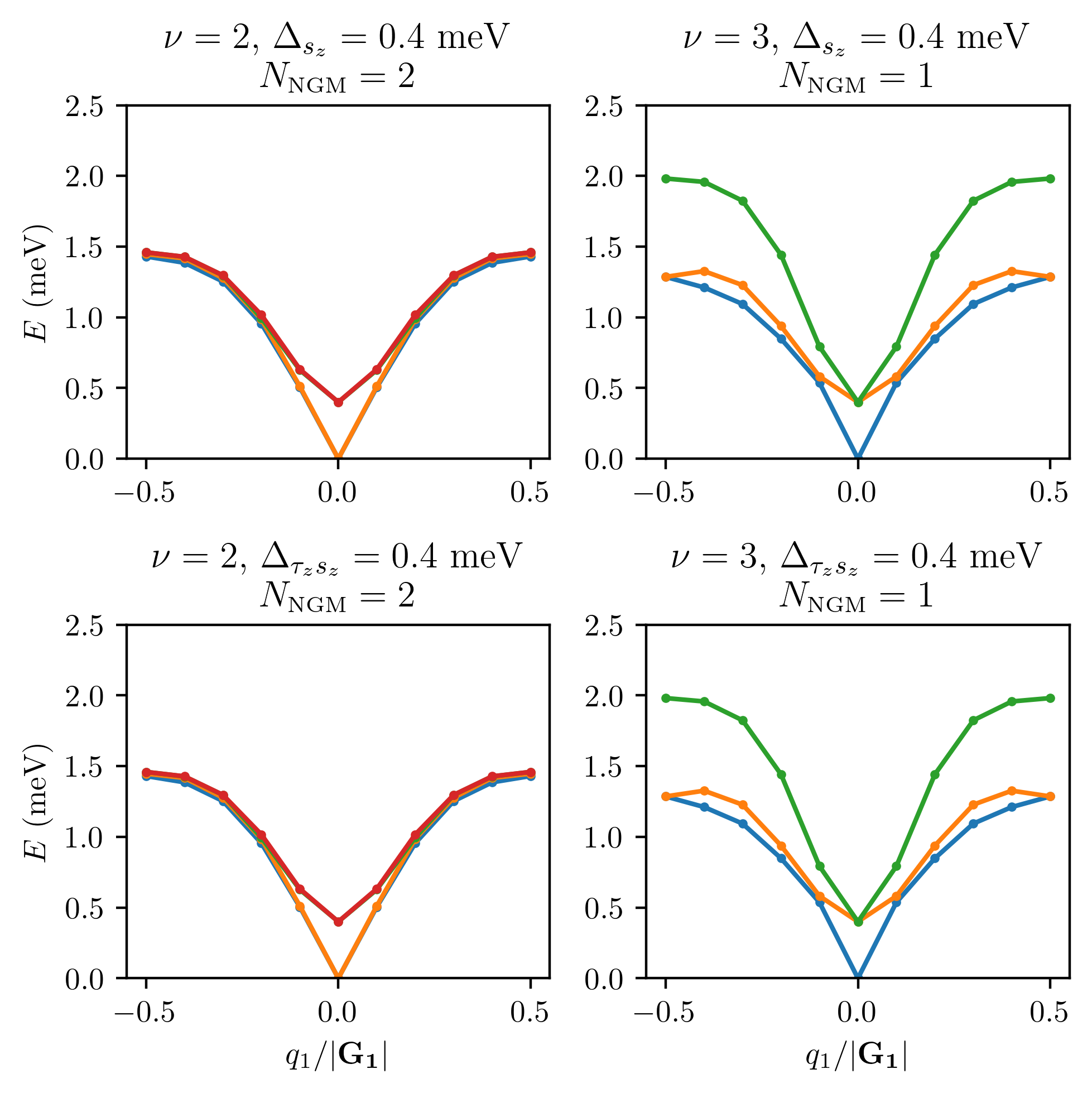}
    \caption{Neutral excitation spectra at $\nu = 2$ and $\nu = 3$ for IKS states with Zeeman splitting $\Delta_{s_z}$ or spin-orbit coupling $\Delta_{\tau_zs_z}$. The plots show the spectra along the $\bm{G}_1$ direction in the mBZ with $q_2 = 0$. The results are obtained from $10 \times 10$ TDHF with $0.3\%$ strain. The number of gapless modes agree with expectations.}
    \label{fig:coll_zeeman}
\end{figure}

In this section, we present additional numerical results on the collective modes of the IKS. In Fig.~\ref{fig:coll_zeeman}, we show excitation spectra under Zeeman splitting ($\frac{1}{2}\Delta_{s_z}s_z$) or Ising-like spin-orbit coupling ($\frac{1}{2}\Delta_{\tau_zs_z}\tau_zs_z$). The numbers of NGMs agree with expectation. We also note that the excitation spectra with Zeeman splitting is identical with that of spin-orbit coupling, when applied with the same strength. This is because spin-orbit coupling is equivalent to Zeeman splitting with a $\pi$ rotation of the spin in one of the valley.

\begin{figure}[h]
    \centering
\includegraphics[width=0.7\linewidth]{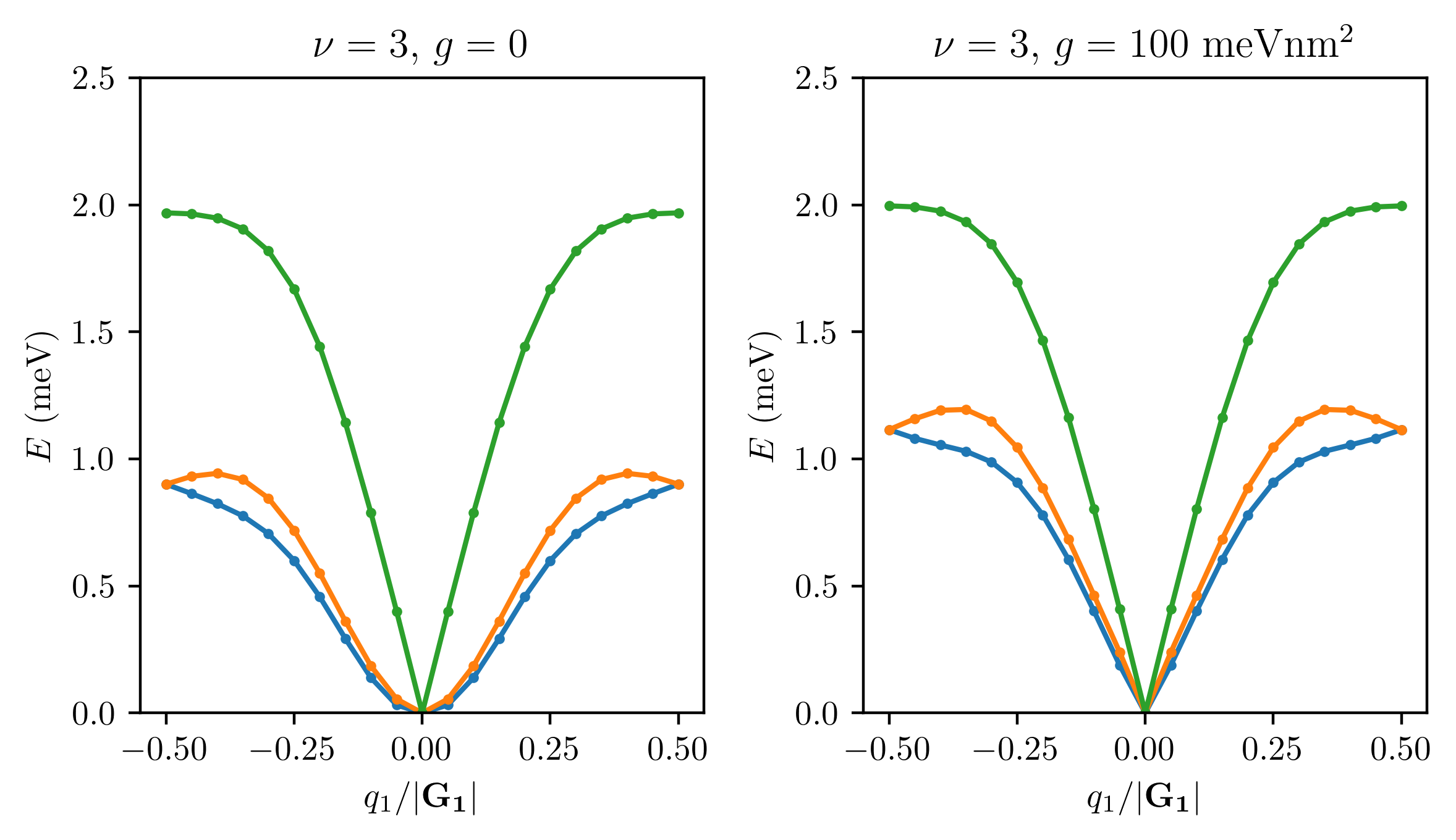}
    \caption{Low-lying neutral excitation spectra at $\nu = 3$ for IKS states without and with electron-phonon coupling. The plots show the spectra along the $\bm{G}_1$ direction in the mBZ with $q_2 = 0$. The results are obtained from $20 \times 20$ TDHF with $0.3\%$ strain. We observe 1 linear mode and 2 quadratic modes for the case without electron-phonon coupling, and 3 linear modes for the case with electron-phonon coupling, in agreement with theory (see Tab.~II of the main text).}
    \label{fig:coll_20}
\end{figure}

\begin{figure}[h]
    \centering
    \includegraphics[width=0.5\linewidth]{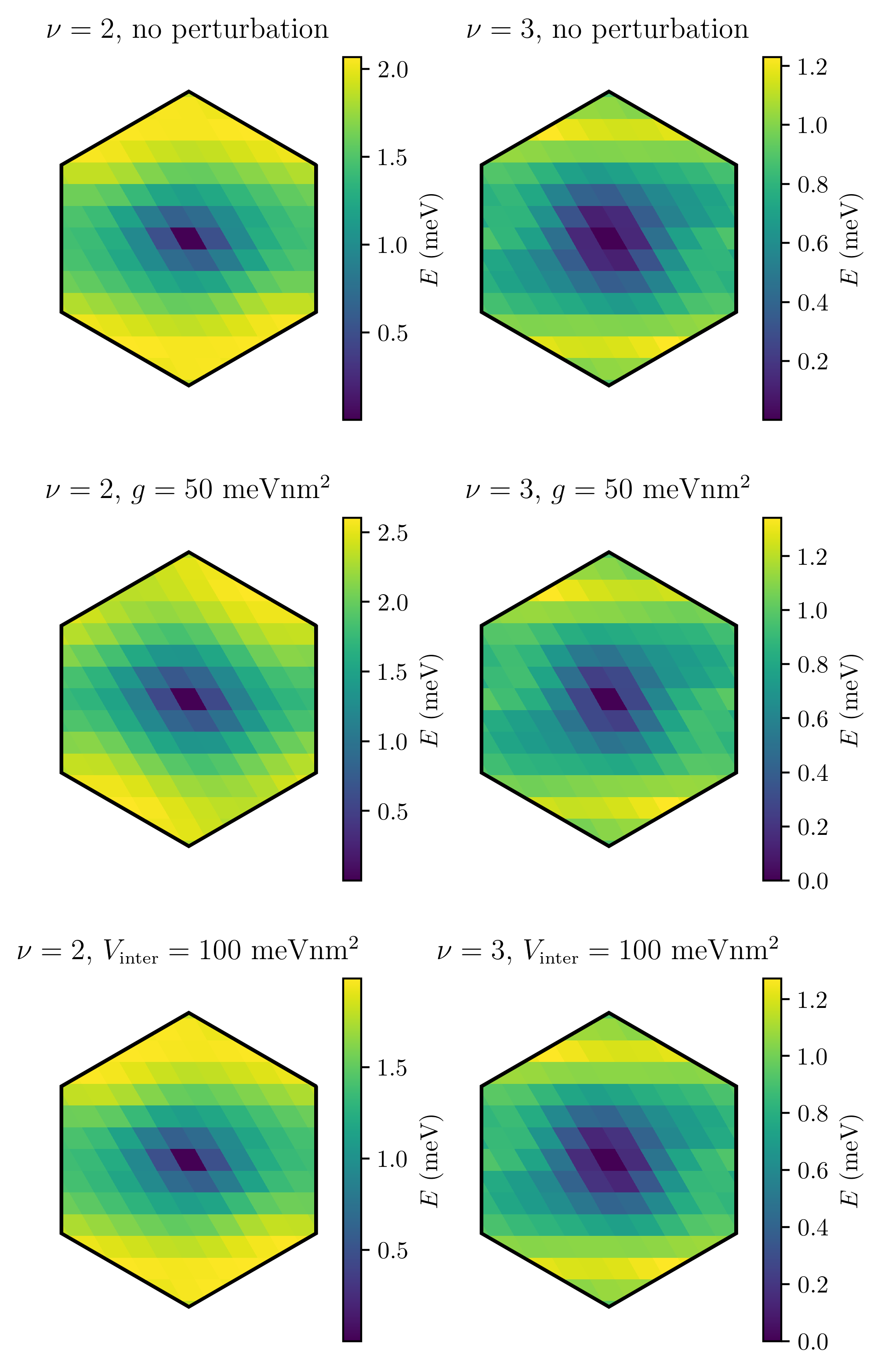}
    \caption{Lowest energy collective excitation spectra at $\nu = 2$ and $\nu = 3$ for IKS states with no perturbation (top row), with finite electron-phonon coupling (middle row), or with finite intervalley Coulomb scattering (bottom row). The results are obtained from $10 \times 10$ TDHF with $0.3\%$ strain, including 2 bands per spin/valley, and not including the effects of non-local tunneling. The ground state IKS has IVC at $\bm{q}_{\text{IKS}} = 0.5\bm{G}_1$.}
    \label{fig:coll_hex}
\end{figure}

In the main text, we verified that the numerical calculations yield the same number of NGMs as theoretical expectations. Beyond the total number, theory also predicts the number of linear and quadratic modes separately. In Fig.~\ref{fig:coll_20}, considering the case of $\nu = 3$ IKS with and without electron-phonon coupling, we confirm that the counting of linear and quadratic modes is correct.
In Fig.~\ref{fig:coll_hex}, we present the dispersion of the lowest collective excitations in the entire mBZ with the same parameters as those used in the line cuts of Fig.~\ref{fig:collective} of the main text.

\subsection{Effects of phonons at lower strain}

\begin{figure}
    \centering
    \includegraphics[width=0.5\linewidth]{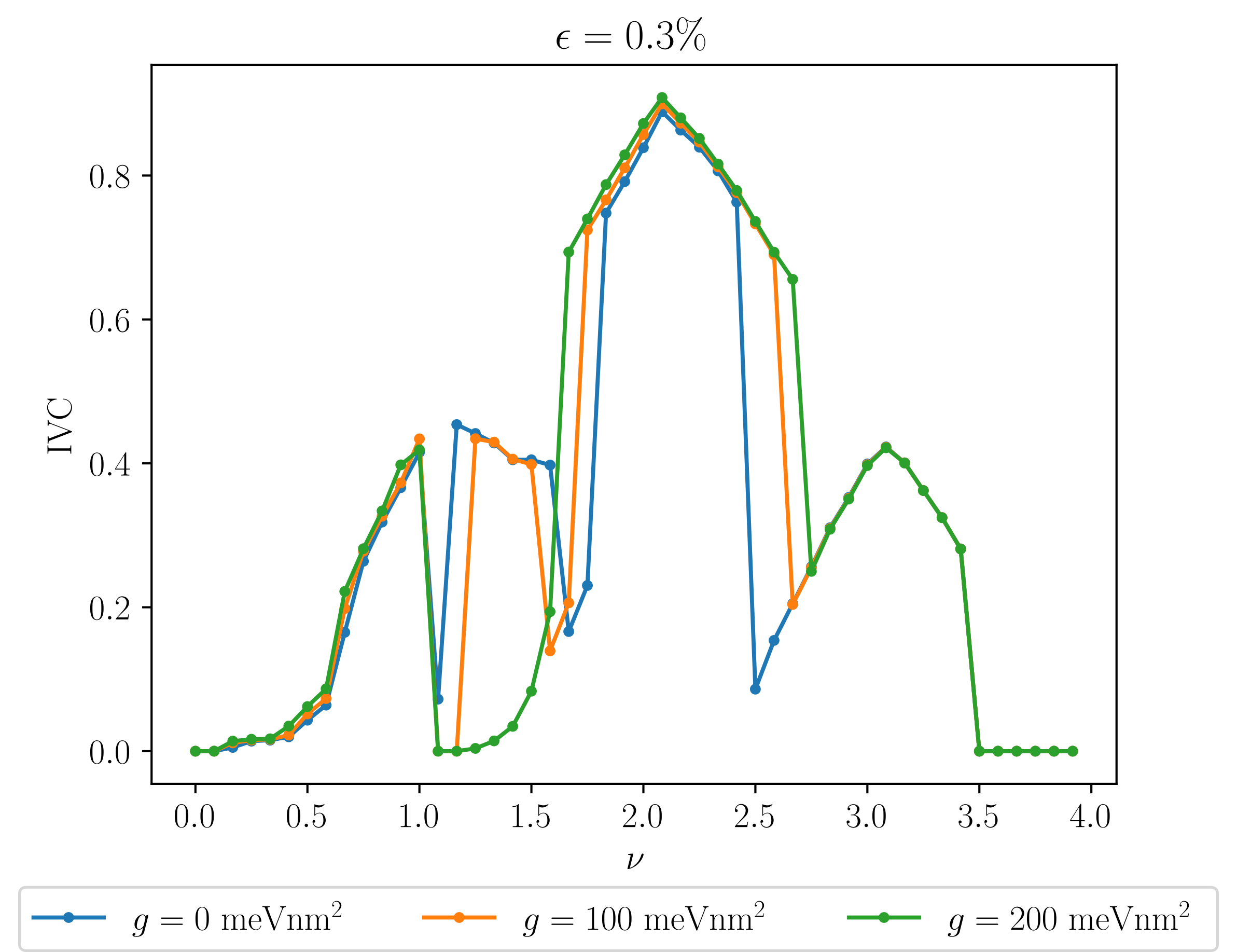}
    \caption{We compute the IVC order parameter of the $12 \times 12$ Hartree-Fock ground states for different values of the electron-phonon coupling constant $g$ at a strain of 0.3\%, smaller than that used in Fig.~\ref{fig:ivc_nonint_04} in the main text. }
    \label{fig:ivc_nonint_12_0.3}
\end{figure}

In Fig.~\ref{fig:ivc_nonint_12_0.3}, we show the IVC strength as a function of filling at 0.3\% strain. Similar to the result at larger strain (see Fig.~\ref{fig:ivc_nonint_04} in the main text), we observe an enhanced IVC near $\nu = 2$. However, at $\nu = 1 + \delta$ and $g = 200$ meVnm\textsuperscript{2} (larger than most estimates of the electron-phonon coupling strength), we observe a drastic reduction in the amount of IVC observed. Since the IVC near $\nu = 1$ does not correspond to a charge density Kekul\'e pattern, we do not expect electron-phonon coupling to strengthen the IVC there. The reduction in IVC can be understood as large electron-phonon coupling renormalizing the bands, which pushes the system out of the IKS regime (see the phase diagram of Ref.~\cite{Wagner2022}, where the presence of IKS at non-integer fillings can be quite sensitive to the strain magnitude at lower strains). At larger strain of 0.5\%, deeper in the IKS regime, this is not observed.

\begin{figure}[h]
    \centering
    \includegraphics[width=0.5\linewidth]{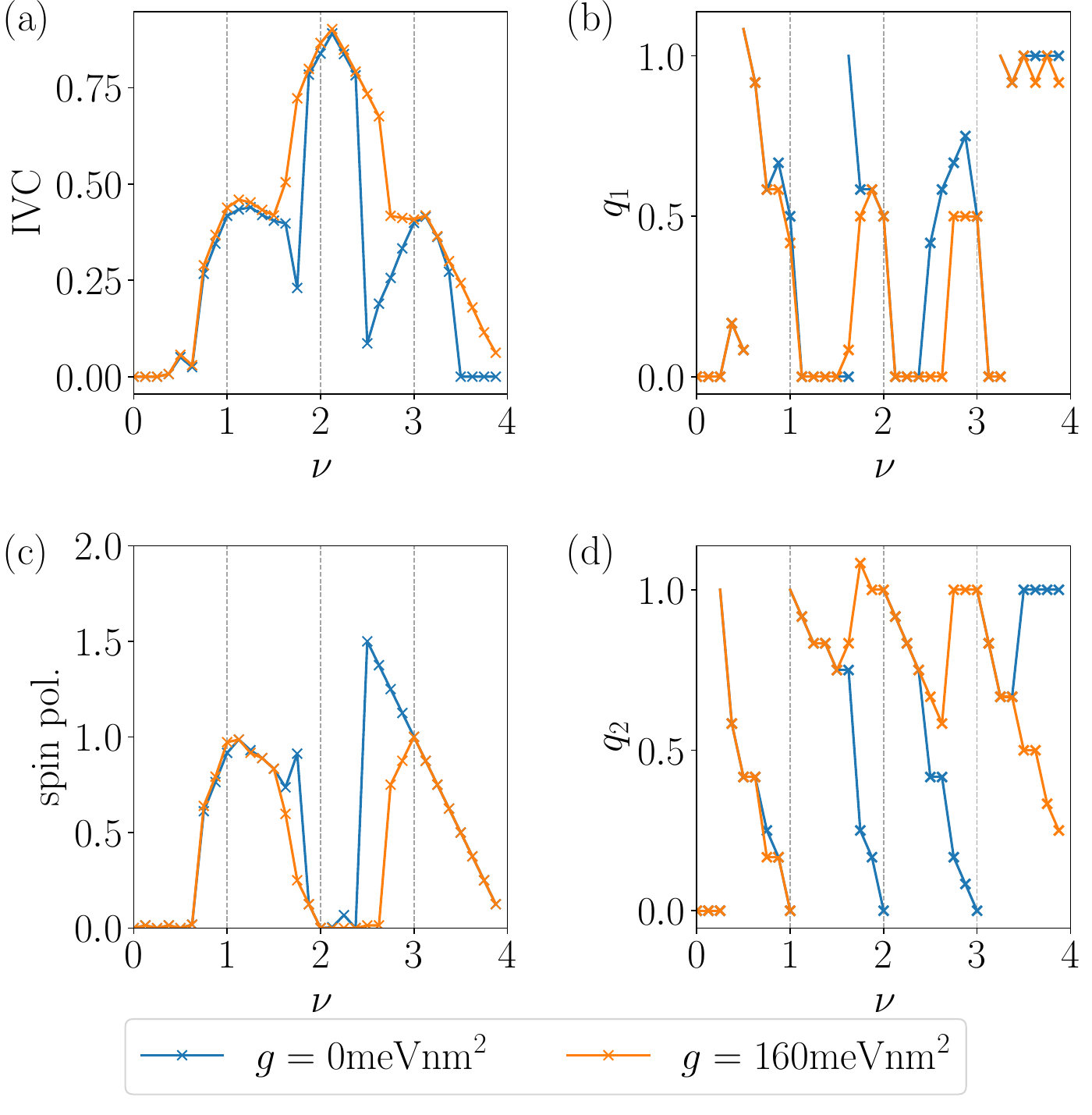}
    \caption{\textbf{Comparison of IKS with and without phonons using the direct distortion method.} We compute (a) the IVC order parameter and (c) the spin polarization of the Hartree-Fock ground state for two different values of the electron-phonon coupling constant $g$. Panels (b,d) show the components of the ideal boost vector $\mathbf{q}=q_1\bm{G}_1+q_2\bm{G}_2$ of the IKS. Phonons increase the IVC order parameter at metallic fillings, while not substantially modifying the state at integer fillings. System size $12\times12$ and strain $\epsilon=0.3\%$.}
    \label{fig:non_integer_average_scheme}
\end{figure}

\section{Additional results on Fermi surfaces}
\subsection{Topology of Fermi surfaces}\label{subsec:fermi_surface_extra}

\begin{figure}[h!]
    \centering
    \includegraphics[width=\linewidth]{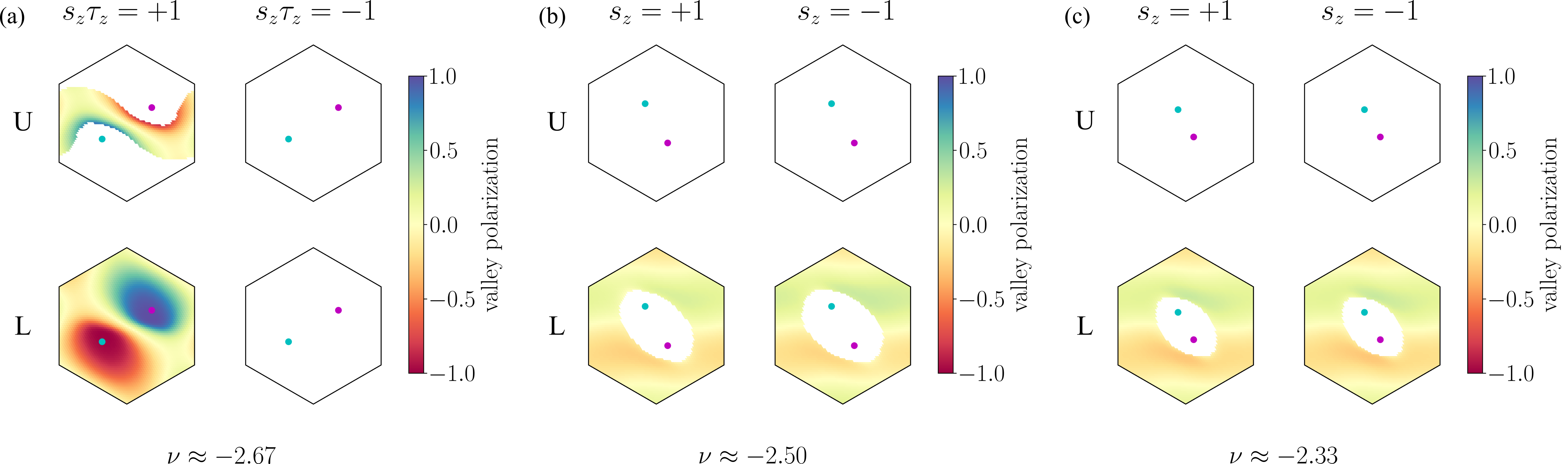}
    \caption{Fermi surface for some additional fillings. Except for filling factors, all other parameters are the same as Fig.~\ref{fig:fermi_surface} in the main text ($0.2\%$ strain, $\varphi = 0^\circ$, $g = 70$~meVnm\textsuperscript{2} and $V_{\text{inter}}= 100$~meVnm\textsuperscript{2}).}
    \label{fig:fermi_surface_extra}
\end{figure}

In Fig.~\ref{fig:fermi_surface_extra}, we show some additional plots of Fermi surfaces. While the electron pocket at $\nu \approx -2.83$ is closed (Fig.~\ref{fig:fermi_surface} of the main text), the Fermi surface at $\nu \approx -2.67$ is open, i.e. connected across the mBZ boundary. From this, we can deduce the existence of a Lifshitz transition within the $-3 < \nu < \nu_c$ range. In the $\nu_c < \nu < -2$ range, we did not observe a similar transition.

\subsection{Fermi surfaces at a smaller strain magnitude and different strain directions}

\begin{figure}[h]
    \centering
    \includegraphics[width=0.7\linewidth]{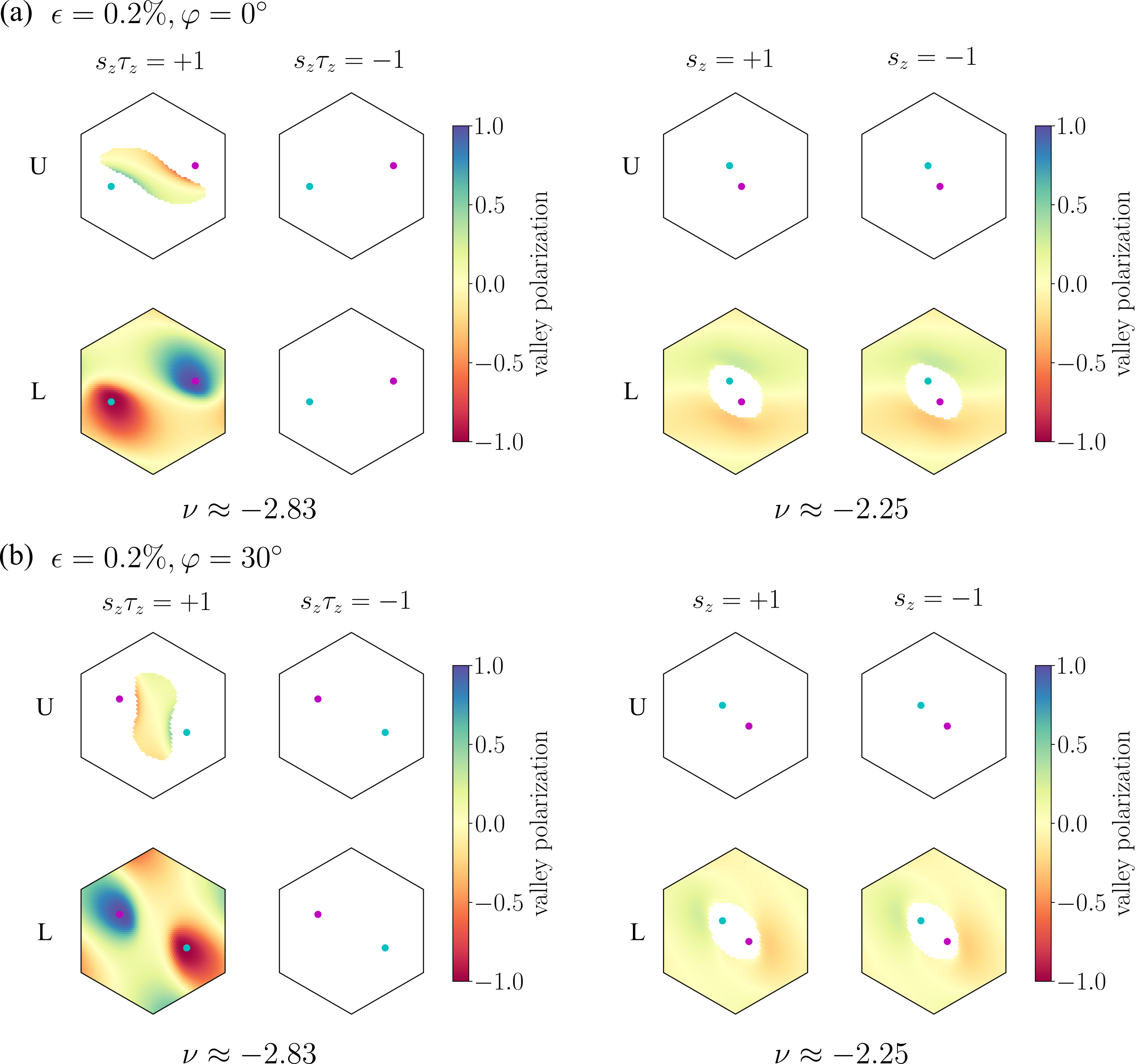}
    \caption{Fermi surfaces with $0.2\%$ strain, $g = 70$~meVnm\textsuperscript{2} and $V_{\text{inter}}= 100$~meVnm\textsuperscript{2}. (a) shows the results for strain direction $\varphi = 0^\circ$, and (b) for  $\varphi = 30^\circ$}
    \label{fig:fermi_surface_small_strain}
\end{figure}

We explore the Fermi surfaces at strain magnitude of $0.2\%$ with two different strain angles, $\varphi = 0^\circ$ and $30^\circ$, shown in Fig.~\ref{fig:fermi_surface_small_strain}. We see the symmetry and topology of Fermi surfaces are not sensitive to the magnitude (as compared to $0.3\%$ used in the main text) and direction of strain.

\section{Comparing Two Different Approaches to Electron-Phonon Coupling}
\label{App:comparing_methods}

In this section, we compare and contrast two different approaches to treating electron-phonon coupling. One method is the standard way of using Schrieffer–Wolff (SW) transformation to obtain an effective electron-electron interaction, as used in the main text. Another method is the one used in Ref.~\cite{Kwan2024}, which we will call the direct distortion method. 

With SW transformation, one usually retains only the leading order in the phonon-induced electron-electron interaction. As such, the method is only reliable in the weak coupling regime. The direct distortion method does not have this limitation. However, it assumes a product state of an electron Slater determinant and a coherent state of phonons, precluding any entanglement between the two sectors.

We note that the two methods agree at $|\nu| = 2$. As pointed out in Ref.~\cite{Kwan2024}, in general the direct distortion method favors maximizing charge Kekul\'e pattern. In this work, we have found that the SW transformation favors the spin-singlet IKS state, which maximizes the charge Kekul\'e pattern. However, the SW transformation and direct distortion method do not agree for the IKS at $|\nu| = 3$. With the SW transformation, the anti-ferromagnetic IKS state is favored, but the direct distortion method favors a ferromagnetic ground state instead, since it maximizes charge Kekul\'e pattern. 

The two methods also yield contrasting results for non-integer filling at $|\nu| = 3 - \delta$. We have computed the HF ground states at non-integer fillings with electron-phonon coupling included using direct distortion method, shown in Fig.~\ref{fig:non_integer_average_scheme}. As one dopes IKS at $|\nu| = 3$ towards charge-neutrality, IVC decreases in the absence of electron-phonon coupling, but increases with large enough electron-phonon coupling under the direct distortion method, as electron-phonon coupling modifies which band the holes enter. However, with SW transformation, the IVC strength decreases with doping towards charge neutrality in the presence of large electron-phonon couplings up to 200\,meVnm\textsuperscript{2}, as shown in Fig.~\ref{fig:ivc_nonint_04} of the main text.

We provide the following explanation of the observed agreement between the two methods at $|\nu| = 2$ and the discrepancy at $|\nu| = 3$. As derived in Ref.~\cite{Kwan2024} with the direct distortion method, the energy gain from lattice distortion has the form $E_{\text{ph}} \sim g|\text{tr}\Gamma_\alpha P |^2$, where $P$ is the single-particle density matrix of the electrons, $\Gamma_a = \tau_x\sigma_x$, $\Gamma_b = \tau_y \sigma_x$ are the intervalley coupling matrices, and $\sigma_i$ are Pauli matrices in sublattice space. The phonon-induced interaction in the SW method can be schematically written as (suppressing momentum indices and form factors)
\begin{equation}H_{\text{ph}} \sim -g\sum_{\tau,s,\sigma}c^\dagger_{\tau s \sigma}c^\dagger_{\bar{\tau}s^\prime \sigma^\prime} c_{\tau s^\prime \bar{\sigma^\prime}} c^{\phantom\dagger}_{\tau s \bar{\sigma}}.\end{equation}
The Hartree energy is given by
\begin{equation}E_H \sim -g |\sum_{s\sigma} \braket{c^\dagger_{+ s \sigma}c_{-s\bar{\sigma}}}|^2 \sim -g|\text{tr}\Gamma_\alpha P |^2\end{equation}
This shows that the direct distortion method is qualitatively equivalent to the Hartree energy of the SW transformed Hamiltonian. On the other hand, the Fock part is given by
\begin{equation}
E_F \sim g \sum_{\tau s \sigma}\braket{c^\dagger_{\tau s \sigma}c_{\tau s^\prime \bar{\sigma^\prime}}}\braket{c^\dagger_{\bar{\tau} s^\prime \sigma^\prime}c^{\phantom\dagger}_{\bar{\tau} s \bar{\sigma}}},
\end{equation}
which does not appear in the direct distortion method. Unlike the Hartree energy, the Fock energy depends on the intravalley part of the single particle density matrix. For the $|\nu| = 2$ IKS, referring to Eq.~\ref{eq:PV} of the main text, the intravalley part of the density matrix is identical for all the states in the ground state manifold. Therefore, the Fock energy is the same for all the states in the degenerate manifold, and only the Hartree energy serves to favor particular states. As the Hartree energy is qualitatively equivalent to the direct distortion method, the two methods naturally agree.

For IKS at $\nu = 3$, the situation is more more complicated. The Hartree term continues to favor charge Kekul\'e pattern (intervalley coherence between the same spin sector), thereby favoring a ferromagnetic state. But the Fock term is gives rise to an anti-ferromagnetic coupling, as explained in Sec.~\ref{subsec:nu3} of the main text. In practice, we find the $\nu = 3$ IKS to be anti-ferromagnetic with SW transformation, while the direct distortion favors the ferromagnetic state due to the maximization of charge Kekul\'e pattern.

\begin{figure}[h]
    \centering
    \includegraphics[width=0.5\linewidth]{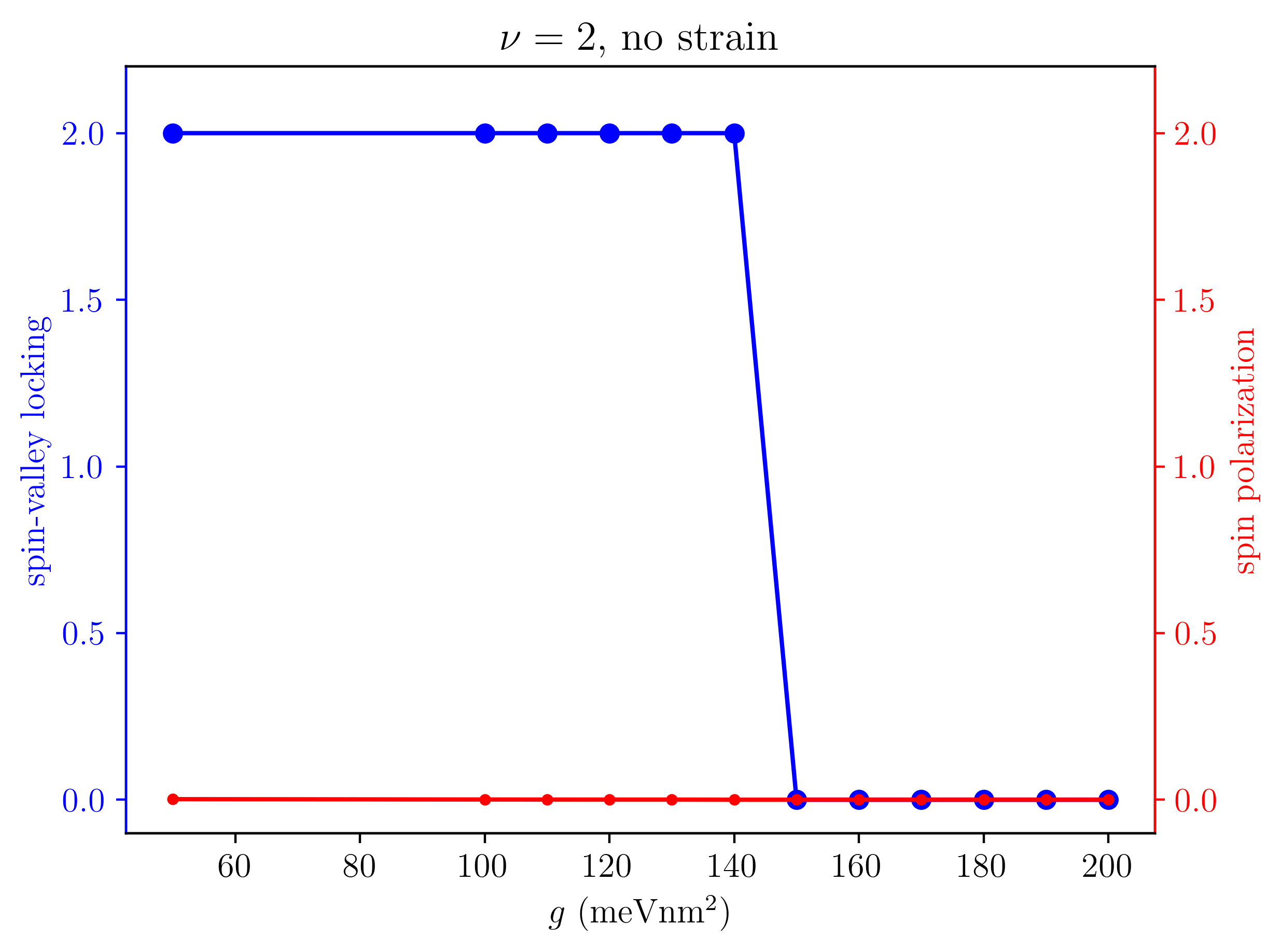}
    \caption{Spin polarization and spin-valley locking at $\nu = 2$ without strain, computed using $10 \times 10$ HF at $\theta = 1.1^\circ$ and $w_{AA} = 60$~meV using the SW method. This is consistent with a transition from KIVC to TIVC-QSH/IVC-QAH as the electron-phonon coupling is increased.}
    \label{fig:no_strain}
\end{figure}

In addition, for the case without strain where the strong-coupling insulators form the low-energy manifold~\cite{Lian2021,Bultinck2020}, Ref.~\cite{Kwan2024} predicts a transition from the KIVC state to TIVC-QSH/IVC-QAH state at a realistic electron-phonon coupling strength for $|\nu|=2$. Using the SW transformation, we are able to reproduce this transition (Fig.~\ref{fig:no_strain}), albeit at a higher threshold. 

\section{Hartree-Fock-Bogoliubov}\label{app:HFB}

\subsection{Generalities}
Consider a general system with an $M$-dimensional basis set indexed by $a,b$. Consider the Hamiltonian
\begin{equation}
    \hat{H}=\sum_{ab}^M h^0_{ab}\ha ^\dagger_a \ha_b+\frac{1}{2}\sum_{abcd}^MV_{abcd}\ha^\dagger_a\ha^\dagger_b\ha_d\ha_c
\end{equation}
where $V_{abcd}=V_{badc}$. $V$ has also been antisymmetrized in its first two and last two indices. Group the operators into a $2M$ component vector
\begin{equation}
    \hbalpha=\begin{pmatrix}
    \hba\\\hba^\dagger
    \end{pmatrix}.
\end{equation}
We seek a state that is a generalized Slater determinant associated with $M$ new creation operators $\hb^\dagger$ indexed by $i,j$
\begin{gather}
    \ket{\Psi}\sim \hb^\dagger_1\hb^\dagger_2\ldots\hb^\dagger_M\ket{0}\\
    \hb^\dagger_i=\sum_p^{2M}\halpha_p^\dagger\Phi_{pi},\quad i=1,\ldots,M\\
    \Phi\in\mathbb{C}^{2M\times M},
\end{gather}
where $\Phi^\dagger\Phi=\mathbb{1}_M$.
The normal and anomalous density matrices of size $M\times M$ are
\begin{gather}
    \rho_{ab}=\langle\ha_b^\dagger\ha_a\rangle,\quad  \rho^\dagger=\rho\\
    \kappa_{ab}=\langle\ha_b\ha_a\rangle,\quad \kappa^T=-\kappa.
\end{gather}
We can define a generalized $2M\times 2M$ density matrix $R$ with the following properties
\begin{gather}
    R_{pq}=\langle\halpha^\dagger_q\halpha_p\rangle\\
    R=\begin{pmatrix}
    \rho & \kappa \\ -\kappa^* & I-\rho^*
    \end{pmatrix} = R^\dagger=\Phi\Phi^\dagger\\
    \text{Tr}R=M,\quad R^2=R,\quad \text{Tr}\rho=N
\end{gather}
where $N$ is some target average particle number. 

\subsection{Conditions for co-existence of superconductivity and normal state order}
We now consider the case relevant to TBG, where we pair electrons with opposite momenta into Cooper pairs. Then 
\begin{equation}
    \hbalpha(\bk)=\begin{pmatrix}
    \hba(\bk)\\\hba^\dagger(-\bk)
    \end{pmatrix},
\end{equation}
and
\begin{equation}
        R(\bk)=\begin{pmatrix}
    \rho(\bk) & \kappa(\bk) \\
    -\kappa^*(\bk) & I-\rho^*(-\bk)
    \end{pmatrix}.
\end{equation}
We have $\sum_{\bk}\textrm{Tr}(\rho)=N$ and $R(\bk)^2=R(\bk)$. We also have $\rho(\bk)^\dagger=\rho(\bk)$ and $\kappa^T(\bk)=-\kappa(-\bk)$. This leads to the two equations
\begin{align}
    \rho(\bk)^2-\kappa(\bk)\kappa^*(-\bk)&=\rho(\bk)\\
    \kappa(\bk)\rho^*(\bk)&=\rho(-\bk)\kappa(\bk).
    \label{Eq:kappa_rho}
\end{align}
Let us focus on the situation with two spinless bands (the spin degree of freedom could for example be frozen out due to spin-valley locking). We assume the Hartree-Fock energy scale is much larger than the superconducting energy scale and we work in the Hartree-Fock band basis. Assuming we are at a filling such that one Hartree-Fock band is partially filled, the density matrix takes the form
\begin{equation}
    \rho(\bk)=\frac{1+\sigma_z}{2}n(\bk)
\end{equation}
where $\sigma_z$ is a Pauli matrix in the Hartree-Fock band basis. The solution for the anomalous correlator is 
\begin{equation}
    \kappa(\bk)=\sqrt{n(\bk)}\sqrt{1-n(\bk)}\frac{1+\sigma_z}{2}g(\bk)
\end{equation}
where $0\leq n(\bm{k})\leq 1$ is the occupation number. $|g(\bk)|=1$ and $g(-\bk)=-g(\bk)$ is required for fermionic antisymmetry (since we have a single occupied band, the SC must be odd parity). From this we can see that the superconductivity is purely intra-band. Finally, we note that Eq.~\eqref{Eq:kappa_rho} is reminiscent of the condition for superconducting fitness \cite{Ramires2016,Ramires2018,Ramires2022}, which also shows that intra-band pairing is more robust. 

\end{document}